\documentclass[times]{elsarticle}
\usepackage{moreverb} 
\usepackage[colorlinks,bookmarksopen,bookmarksnumbered,citecolor=red,urlcolor=red,pdfauthor=author]{hyperref}
\usepackage{amsfonts,amsmath,amssymb}
\usepackage{psfrag,pstricks,pst-node}

\usepackage[section]{placeins}
\usepackage{subfigure} 
\usepackage{mathrsfs}
\usepackage{graphicx}
\usepackage[ruled,vlined]{algorithm2e}
\usepackage{fancyhdr}  
\usepackage{psfrag} 
\usepackage[export]{adjustbox} 
\usepackage{enumerate} 
\usepackage{comment}
\usepackage[margin=2.5cm]{geometry}
\usepackage{algorithmic}
\usepackage[section]{placeins}

\usepackage{hyperref} 
\hypersetup{hidelinks,
	colorlinks=true,
	allcolors=black,
	pdfstartview=Fit,
	breaklinks=true}

\usepackage{bm}
\usepackage{booktabs}
\usepackage{threeparttable}
\graphicspath{{./figs/}{./model structure/}}
\DeclareGraphicsExtensions{.pdf,.eps}

\usepackage{pgfplots}
\pgfplotsset{compat=1.13}

\begin{document}
 
\begin{frontmatter}
\renewcommand{\thefootnote}{\fnsymbol{footnotemark}}

\fancypagestyle{plain}{%
\fancyhf{} 
\fancyhead[RO,RE]{\thepage} 
}

\title{Data discovery of low dimensional fluid dynamics of turbulent flows}
    \author[lab1,lab2]{X. Lin}
    \author[lab1,lab2]{D. Xiao\corref{cor1}} 
    \cortext[cor1]{Corresponding author}
    \ead{xiaodunhui@tongji.edu.cn}   
    \author[lab3]{F. Fang}
    \address[lab1]{School of Mathematical Sciences, Tongji University, Shanghai, P.R. China,200092}
    \address[lab2]{Key Laboratory of Intelligent Computing and Applications(Tongji University), Ministry of Education, China}
  \address[lab3]{Department of Earth Science and Engineering, Imperial College London, UK, SW7 2BP}

\begin{abstract}
Discovering governing equations from data, in particular high dimensional data, is challenging in various fields of science and engineering, and it has potential to 
revolutionise the science and technology in this big data era. This paper combines sparse identification and deep learning with non-linear fluid dynamics, in particular the turbulent 
flows, to discover governing equations of nonlinear fluid dynamics in the lower nonlinear manifold space. The autoencoder deep neural network is used to project the high dimensional space
into a lower dimensional nonlinear manifold space. The Proper Orthogonal Decomposition (POD) is then used to stabilise the nonlinear manifold space in order to guarantee a stable manifold space for pattern or equations discovery for the highly nonlinear problems such as turbulent flows. 
Sparse regression is then used to discover the lower dimensional governing equations of fluid dynamics in the lower dimensional nonlinear manifold space. What distinguishes this approach is its ability to discover a lower dimensional governing equations of fluid dynamics in the nonlinear manifold space. We demonstrate this method on a number of high-dimensional fluid dynamic systems such as lock exchange, flow past one and two cylinders. The results demonstrate that the resulting method is capable of discovering lower dimensional governing equations that took researchers in this community many decades years to resolve. In addition, this model discovers dynamics in a lower dimensional manifold space, thus leading to great computational efficiency, model complexity and avoiding overfitting. It also provides a new insight for our understanding of sciences such as turbulent flows.

\end{abstract}

\end{frontmatter}

\section{Introduction}
The governing equations in scientific disciplines are crucial for understanding physical processes, leading to technological advancements. Traditionally, these equations are derived from first principles or universal laws, for example, newton’s laws, symmetries, conservation laws, and kepler's laws\cite{ gonzalez2018deep, watters2017visual, langley1981data, zhang2018robust, crutchfield1987equations, zanna2020data,  emmertstreib2020introductory,carlberg2019recovering}. However, in this big data era, the traditional first-principles derivations for dynamic systems may not be practical or justifiable as the governing equations are unknown or partially known and the emerging various and massive data. Therefore, researchers are trying to find a new paradigm to discover sciences in this big data era\cite{schmidt2009distilling,rowley2009spectral,franz2014interpolation, minh2018covariances, morton2018deep, chen2014big, zhang2018survey, bevington1993data, korotcov2017comparison}. Discovering governing equations from data can be one of these new paradigms. However, the data driven equations discovery could be computational intensive as the system might be high dimensional. 

This leads to significant efforts in developing low-dimensional models\cite{floryan2022data, brunton2021modern, noack2003hierarchy, ozolins2013compressed, deng2020low, liu2020deep},  which represent the physical dynamics in a lower dimensional manifold space. One popular linear manifold method is Proper Orthogonal Decomposition (POD)\cite{mendez2017pod, rowley2004model, berkooz1993proper, kerschen2005method, liu2011development}. However, it restricts the evolution of the physical system within a linear space and have limitations to learn non-linear manifolds\cite{lee2020model, brunton2020machine, parente2009identification}. In order to represent more non-linear dynamics in nature, several non-linear manifold methods are presented\cite{tenenbaum2000global,roweis2000nonlinear, chen1998development}. In this work, a autoencoder deep learning method\cite{omata2019novel, wang2014generalized, zabalza2016novel, qi2017stacked, 2018stacked-ae} is used as a non-linear manifold method to represent the high dimensional physical dynamics.
After the lower dimensional non-linear manifold latent space is obtained, method of discovering equations in this latent space is required. 
Recently, a number of novel work has been presented to identify 
inherent structures of nonlinear dynamical systems from data\cite{bongard2007automated,2016SINDy,floryan2022data, zhang2019convergence, brunton2016sindyc, kaiser2018sparse, yang2020data-driven}. 

One popular method is to use the symbolic regression method to discover the dynamics\cite{kim2020integration, lacava2021contemporary, kutz2016dynamic, murata2020nonlinear}. However, it is computational expensive and encounters challenges in scaling effectively to large systems of interest and may be susceptible to overfitting. Therefore, the Sparse Identification of Nonlinear Dynamics (SINDy)\cite{2016SINDy, rudy2017data-driven, desilva2020pysindy, fasel2021sindy, fasel2022ensemble, kaheman2020sindy-pi} is presented to identify the minimum number of terms from a library of possible candidate functions necessary to capture and model the dynamics. An new regression method: adaptive least absolute shrinkage and selection operator (Adalasso)\cite{zou2006Alasso, zou2008one, champion2019data-driven, tibshirani1996regression}, combined with least angle regression (LARS) algorithm\cite{efron2004least}, is utilised in the computational process of SINDy.
 
Turbulent flows are characterised by chaotic and unpredictable fluid motions\cite{taira2017modal, mccomb1990physics, rogallo1984numerical, launder1983numerical}. The discovery of governing equations for turbulent flows has immense significance across various fields of science and engineering such as atmospheric dynamics, ocean currents, aircraft design and combustion\cite{fukami2023grasping, patil2001local, reichstein2019deep, zanna2020data-driven, byington2004data-driven, ihme2022combustion, zhou2022machine}.
In this work, we introduce a novel approach for deriving compact dynamical turbulent models in the lower dimensional nonlinear manifold latent space from data. This is achieved by non-linear dimensionality reduction deep learning method, proper orthogonal decomposition (POD) and sparse regression method. What distinguishes this approach is the ability to discover a lower dimensional nonlinear manifold governing equations of a complicated fluid dynamics problem: turbulent flows. This is achieved by using autoencoder deep learning, proper orthogonal decomposition (POD) and SINDy.


\section{Results}\label{sec:result}

\subsection{\textbf{Lock exchange}}\label{case:le}
The lock-exchange phenomenon refers to a fluid dynamics event that occurs when two fluids of different densities and temperatures are initially separated by a barrier (lock) and then released. When the lock is removed, the two fluids begin to exchange positions. In this process, the lighter fluid typically moves over the denser fluid, leading to a sharp interface between the two. The lock exchange is relevant to various natural phenomena, such as the mixing of air masses with different temperatures and the behavior of fluids in geological and environmental contexts\cite{Hiester2011,Simpson1999,Haertel1999vorticity}.

The computational domain is a 2D rectangular area with $0\leq x\leq 0.8 m, 0\leq y\leq 0.1 m$. In the left half of the domain, where $x< 0.4 m$, the tank is filled with dense and cold water of $T_1=-0.5^{\circ}$C, and the other right half, where $x\geq 0.4m$, is occupied by light and warm water of  $T_2=0.5^{\circ}$C. The physical parameters for lock-exchange is given in Table \ref{table:lock-exchange}. At $t=0 s$, the velocity $u$ is set to be zero throughout the computational domain. The bottom boundary enforces a no-slip condition, while the top and side walls is set to be a free-silp condition. The total time period is 40 seconds and the time interval is $\Delta t =0.05 s$, 800 time snapshots are generated via Fluidity, which is an open source unstructured adaptive mesh finite element computational fluid dynamic model \cite{amcg_2015}. The total number of nodes in the compuational mesh is 1491. 

\begin{figure}[ht]
\centering
\begin{tabular}{c}
\begin{minipage}{0.8 \linewidth}
\includegraphics[width = \linewidth,angle=0,clip=true]{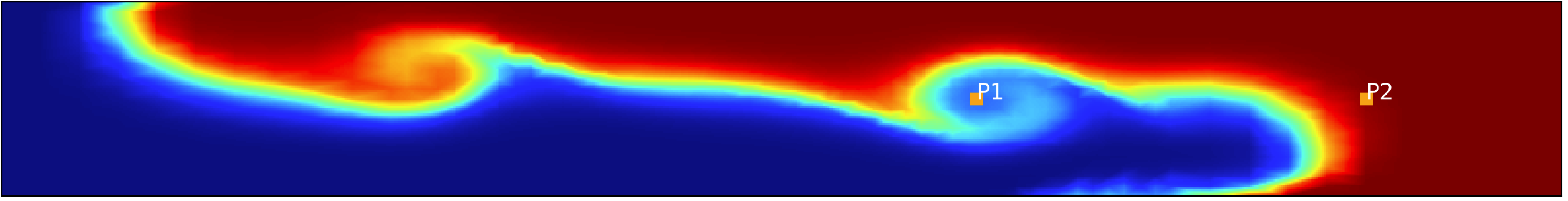} 
\end{minipage}
\\
\\
\begin{minipage}{0.8\linewidth}
\includegraphics[width = \linewidth,angle=0,clip=true]{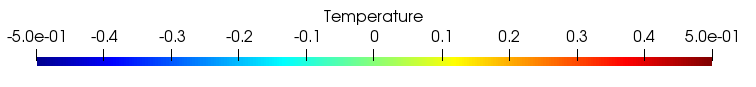} 
\end{minipage}
 \\
\\
\end{tabular}
\caption{\textbf{Lock exchange}: Location of t2o particular points: $P_1$ and $P_2$ in the case of lock exchange.}
\label{fig:le p1p2}
\end{figure}


\begin{table}[htbp!]\label{table:lock-exchange}
  \centering
  \begin{threeparttable}
    \begin{tabular}{c c c c c c c c c} 
      \hline
      Physical parameters    & label & value \\ 
      \hline
      gravitational acceleration & $g$  & 10$ms^{-2}$ \\ 
      kinematic viscosity & $\nu$ & $10^{-6}m^2s^{-1}$\\
      thermal diffusivity & $\kappa$& 0 $m^2s^{-1}$ \\
      thermal expansion coefficient & $\lambda$ & ${10^{-3}}^{\circ}C^{-1}$ \\
      buoyancy velocity & $u_b=\sqrt{g'H}$ & $\sqrt{10^{-3}}$ \\
      \hline
    \end{tabular}
    \end{threeparttable}    
    \caption{Physical parameters for lock-exchange}
\end{table}

\begin{figure}[htbp]
\centering 
\begin{minipage}{1.0 \linewidth}
\includegraphics[width = \linewidth,angle=0,clip=true]{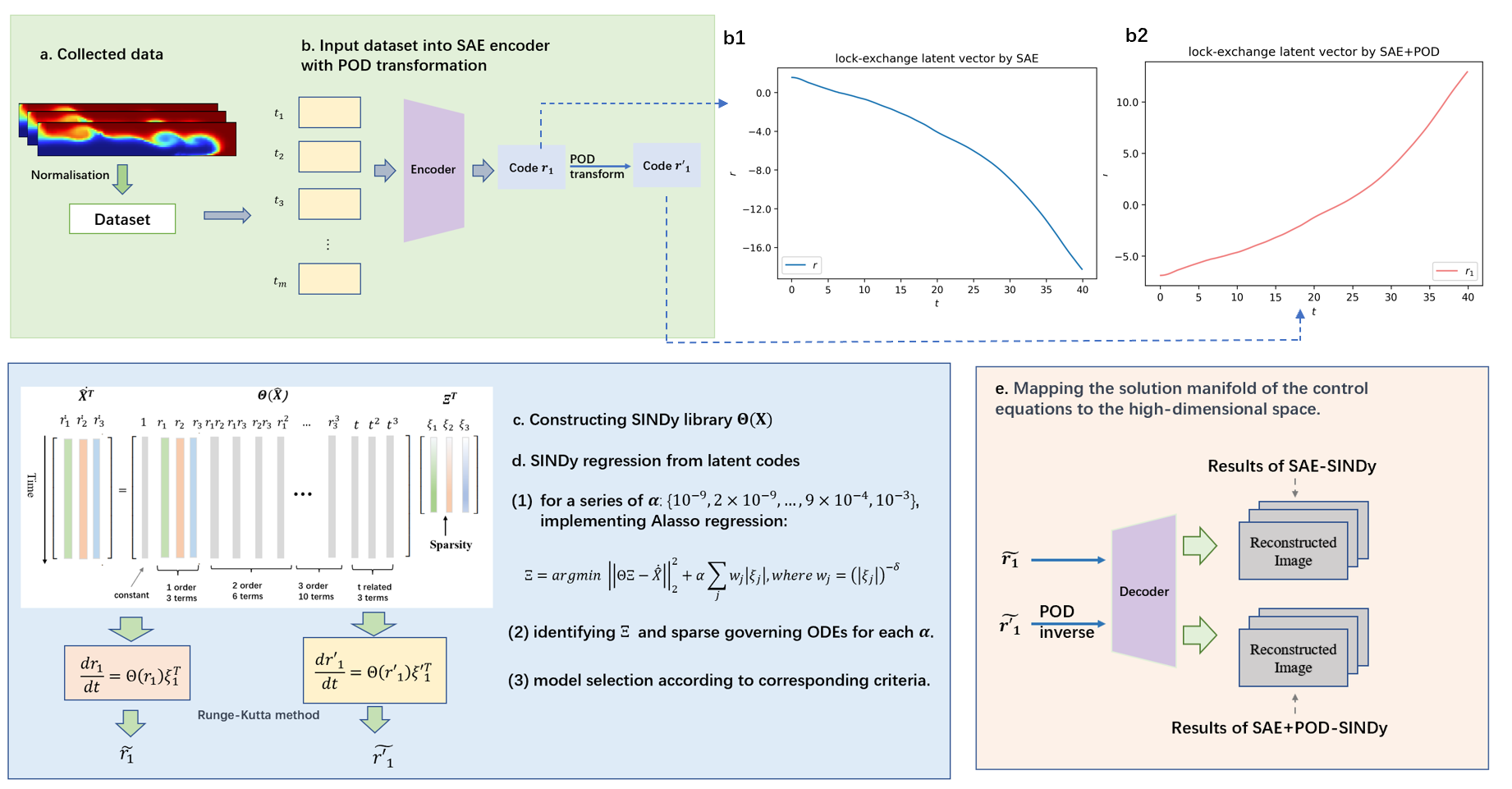} 
\end{minipage}
\caption{\textbf{The framework of proposed method.}
\textbf{(a)} Projecting the high-dimensional data into the lower dimensional manifold space via SAE and POD method.
\textbf{(b)} Code values $r_1$ (shown in \textbf{(b1)} in the SAE manifold space. POD coefficients $r'_1$ (shown in \textbf{(b2)}), which are generated by performing POD process to the code values in the SAE network.
\textbf{(c)} Constructing the SINDy library for $r_1$ and $r'_1$. The detailed process can be found in Section \ref{sec:SINDy}.
\textbf{(d)} Sparse regression process: Select a series of $\lambda$ as parameters. These parameters are then sequentially incorporated into Adalasso equation for sparse regression, resulting in a series of models (i.e., ODEs). Finally, model selection is performed.
\textbf{(e)} Projecting $r_1$ and $r'_1$, which are obtained from solving the reduced governing equations, back into the high-dimensional space.}
\label{Model:process1}
\end{figure}

Figure \ref{Model:process1} shows the procedure of proposed method. In the method, the high dimensional data is projected into a manifold space with a dimensional size of one. The projection is conducted via non-linear dimensionality reduction method: Autoencoder deep neural network. The reconstruction rate of this SAE network is 98.9562\%, which means that SAE accurately capture most of the fluid dynamics using only one dimension. This is superior over the traditional dimensionality reduction method such as POD. After the manifold space of SAE is obtained, a POD process is performed in the manifold space of SAE with the same dimensional size with SAE. In our numerical examples, we found that the POD method is capable of stabilising the manifold space of SAE. The stabilising is easier for SINDy to discover equations. 

The code values (values in the manifold latent space) is given in the Figure \ref{Model:process1}(b1) and the POD coefficients is shown in Figure \ref{Model:process1}(b2). In Supplementary materials, we prove that, under transformations of the same order, POD incurs no loss in energy, except for rounding error. Hence, errors in POD are negligible in this context.
It can be observed that the average of the upper and lower bounds of the values in the manifold space is more close to zero if POD method is performed for the SAE. This probably leads to be more stable for the SINDy discovery.

The process of discovering governing equations in the manifold space can be found in \ref{sec:SINDy}. The SINDy's model selection process involves the adjustment of various threshold parameter $\lambda$. In STLS of Brunton's work \cite{2016SINDy}, $\lambda$ functions as a hard threshold parameter, which will be removed if it is less than a certain value. In this paper, a soft threshold is used, which use a penalty function to control it\cite{zou2006Alasso}. As the computational cost of the candidate function terms in the SINDy library increases with the growth of unknown variables, we reduce the inputs of SINDy in order to reduce the computational cost and lead to be easier for SINDy discovery.  
We input a series of parameter values $\lambda$ into SINDy regression and compute them sequentially to generate distinct models, which are the governing equations embedded in the manifold space. Then, a model section process is performed, which essentially balance the errors and parsimony. The model selection process relies on the criteria of Pearson correlation coefficient and RMSE in the manifold space. 

It is evident that, as $\lambda$ increases, the complexity of the system of ordinary differential equations decreases, and the error exhibits a fluctuating upward trend. $\lambda=10^{-8}$ is chosen in our methods. The models we discovered using the SINDy in the manifold space are:
\begin{equation}
    \begin{aligned}
    \frac{d r_1}{dt}=f_1(r_1,t)=-0.389+0.0975r_1-0.00384r_1^{2}-0.0002599r_1^{3}+0.02211t \\
    \end{aligned}
    \label{equ:le-sae}
\end{equation}
\begin{equation}
    \begin{aligned}
    \frac{d r'_1}{dt}=f_2(r'_1,t)=-0.4932 \\
    \end{aligned}
    \label{equ:le-sp}
\end{equation}

 The Equation \ref{equ:le-sae} is the equation which is disvovered via SAE+SINDy method whereas the Equation \ref{equ:le-sp} is the equation obtained from SAE+POD+SINDy method.  In the SAE+SINDy approach, the equation $f_1$ contains a total of 5 terms, while in the SAE+POD-SINDy approach, there is only 1 constant term in $f_2$.
 


\begin{figure}
\centering
\begin{tabular}{ccc}
\begin{minipage}{0.3\linewidth}
\includegraphics[width = \linewidth,angle=0,clip=true]{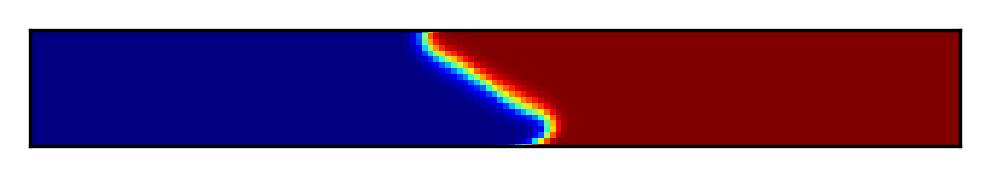}  
\end{minipage}
&
\begin{minipage}{0.3\linewidth}
\includegraphics[width = \linewidth,angle=0,clip=true]{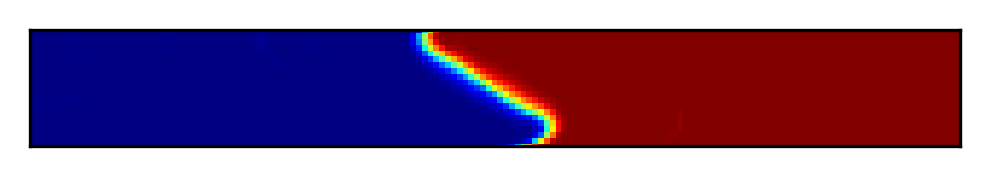}  
\end{minipage} 
&
\begin{minipage}{0.3\linewidth}
\includegraphics[width = \linewidth,angle=0,clip=true]{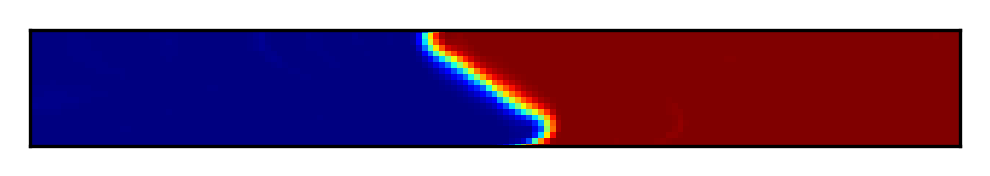}  
\end{minipage} 
 \\
(a-1) {\small Full Model, t = 5s}&
(a-2) {\small SAE, t=5s}&
(a-3) {\small SAE+POD, t=5s}
\\

\begin{minipage}{0.3\linewidth}
\includegraphics[width = \linewidth,angle=0,clip=true]{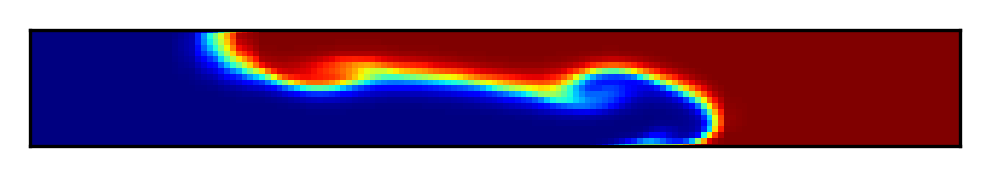}  
\end{minipage}
&
\begin{minipage}{0.3\linewidth}
\includegraphics[width = \linewidth,angle=0,clip=true]{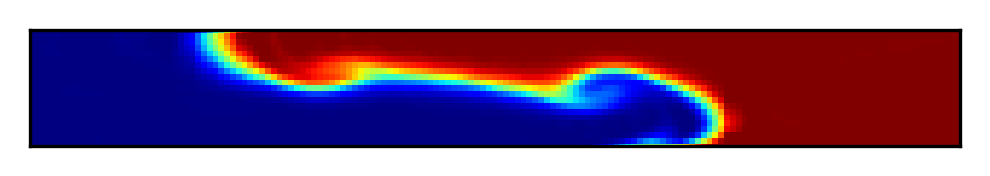}  
\end{minipage} 
&
\begin{minipage}{0.3\linewidth}
\includegraphics[width = \linewidth,angle=0,clip=true]{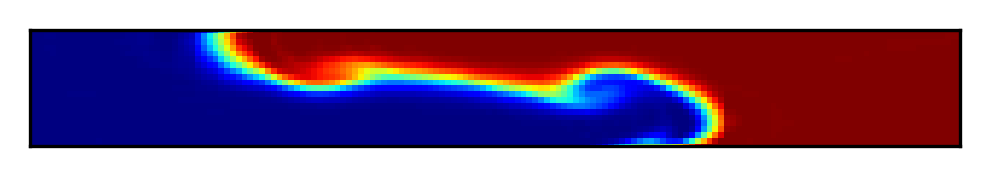}  
\end{minipage} 
 \\
(b-1) {\small Full Model, t = 17.5s}&
(b-2) {\small SAE, t = 17.5s}&
(b-3) {\small SAE+POD, t = 17.5s}
\\

\begin{minipage}{0.3\linewidth}
\includegraphics[width = \linewidth,angle=0,clip=true]{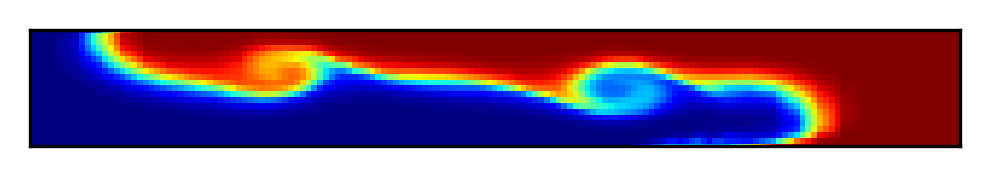}  
\end{minipage}
&
\begin{minipage}{0.3\linewidth}
\includegraphics[width = \linewidth,angle=0,clip=true]{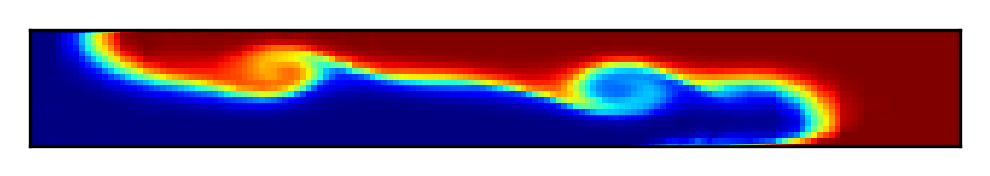}  
\end{minipage} 
&
\begin{minipage}{0.3\linewidth}
\includegraphics[width = \linewidth,angle=0,clip=true]{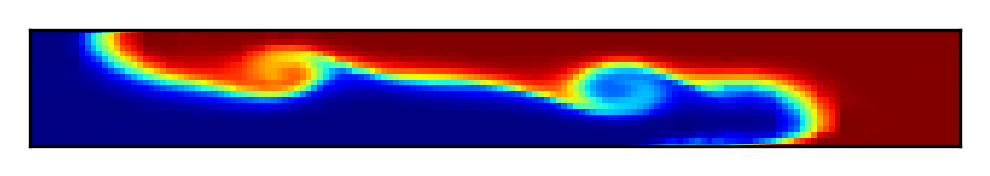}  
\end{minipage} 
 \\
(c-1) {\small Full Model, t = 25s}&
(c-2) {\small SAE, t = 25s}&
(c-3) {\small SAE+POD, t = 25s}
\\

\begin{minipage}{0.3\linewidth}
\includegraphics[width = \linewidth,angle=0,clip=true]{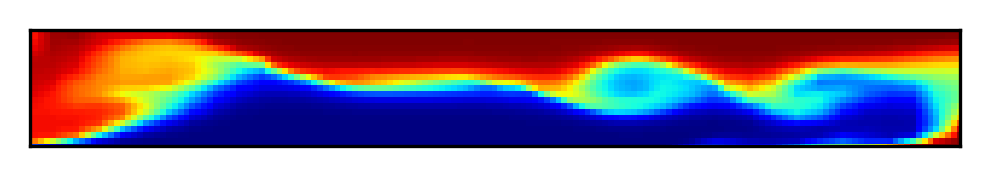}  
\end{minipage}
&
\begin{minipage}{0.3\linewidth}
\includegraphics[width = \linewidth,angle=0,clip=true]{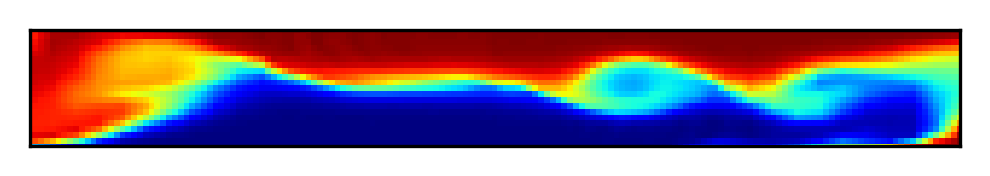}  
\end{minipage} 
&
\begin{minipage}{0.3\linewidth}
\includegraphics[width = \linewidth,angle=0,clip=true]{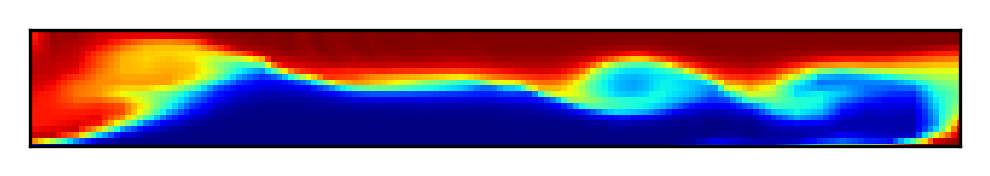}  
\end{minipage} 
 \\
(d-1) {\small Full Model, t = 37.5s}&
(d-2) {\small SAE, t = 37.5s}&
(d-3) {\small SAE+POD, t = 37.5s}
\\

\begin{minipage}{0.3\linewidth}
\includegraphics[width = \linewidth,angle=0,clip=true]{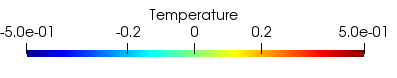} 
\end{minipage}
&
\begin{minipage}{0.3\linewidth}
\includegraphics[width = \linewidth,angle=0,clip=true]{results/lock_exchange/le_grid3.png} 
\end{minipage}
&
\begin{minipage}{0.3\linewidth}
\includegraphics[width = \linewidth,angle=0,clip=true]{results/lock_exchange/le_grid3.png} 
\end{minipage}
 \\
\\
\end{tabular}
\caption{\textbf{Lock exchange}: velocity(u) solutions comparison between the high fidelity model, SAE+SINDy and SAE+POD+SINDy model}
\label{fig:le_solution}
\end{figure}

\begin{figure}
\centering
\begin{tabular}{cc}
\begin{minipage}{0.4\linewidth}
\includegraphics[width = \linewidth,angle=0,clip=true]{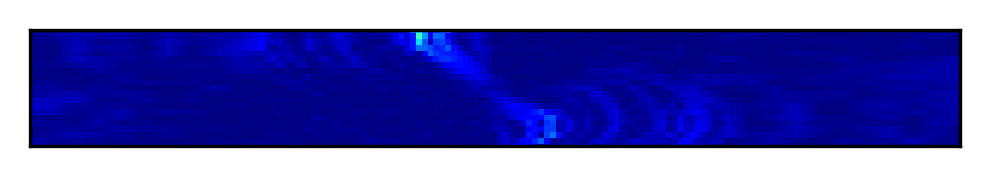}  
\end{minipage} 
&
\begin{minipage}{0.4\linewidth}
\includegraphics[width = \linewidth,angle=0,clip=true]{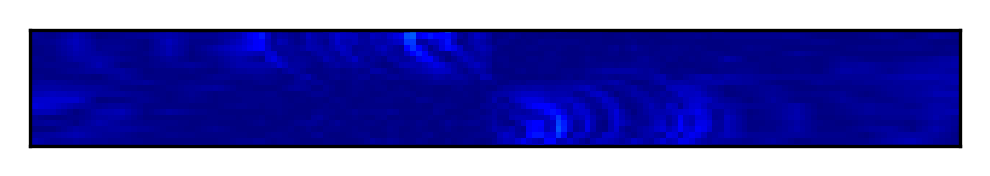}  
\end{minipage} 
 \\
(a-4) {\small SAE error, t=5s}&
(a-5) {\small SAE+POD error, t=5s}
\\

\begin{minipage}{0.4\linewidth}
\includegraphics[width = \linewidth,angle=0,clip=true]{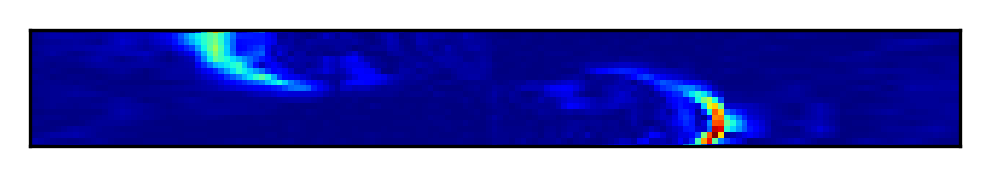}  
\end{minipage} 
&
\begin{minipage}{0.4\linewidth}
\includegraphics[width = \linewidth,angle=0,clip=true]{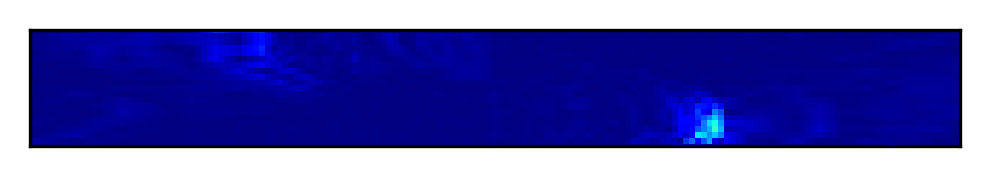}  
\end{minipage} 
 \\
(b-4) {\small SAE error, t=17.5s}&
(b-5) {\small SAE+POD error, t=17.5s}
\\

\begin{minipage}{0.4\linewidth}
\includegraphics[width = \linewidth,angle=0,clip=true]{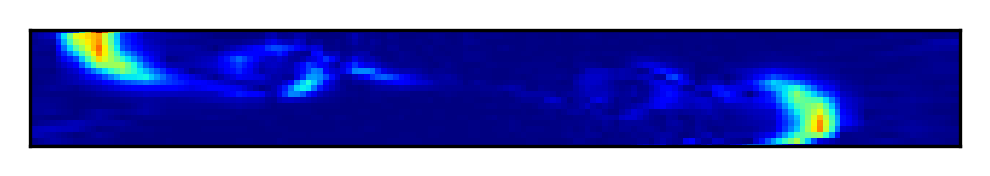}  
\end{minipage} 
&
\begin{minipage}{0.4\linewidth}
\includegraphics[width = \linewidth,angle=0,clip=true]{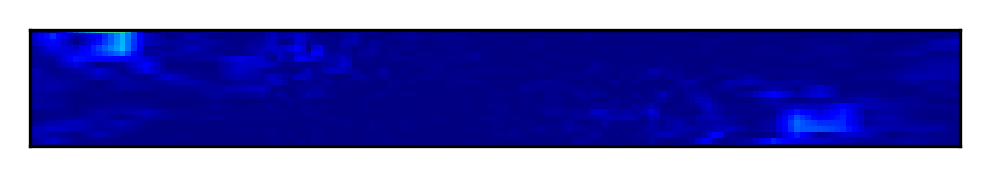}  
\end{minipage} 
 \\
(c-4) {\small SAE error, t=25s}&
(c-5) {\small SAE+POD error, t=25s}
\\

\begin{minipage}{0.4\linewidth}
\includegraphics[width = \linewidth,angle=0,clip=true]{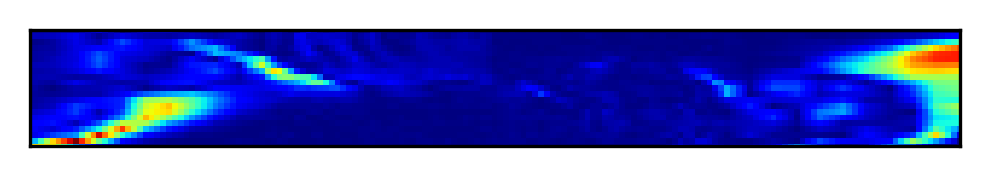}  
\end{minipage} 
&
\begin{minipage}{0.4\linewidth}
\includegraphics[width = \linewidth,angle=0,clip=true]{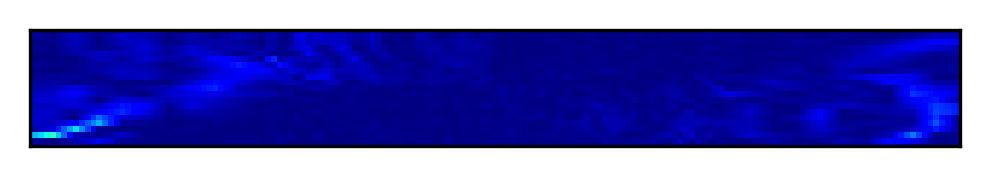}  
\end{minipage} 
 \\
(d-4) {\small SAE error, t=37.5s}&
(d-5) {\small SAE+POD error, t=37.5s}
\\

\begin{minipage}{0.4\linewidth}
\includegraphics[width = \linewidth,angle=0,clip=true]{results/lock_exchange/le_grid3.png} 
\end{minipage}
&
\begin{minipage}{0.4\linewidth}
\includegraphics[width = \linewidth,angle=0,clip=true]{results/lock_exchange/le_grid3.png} 
\end{minipage}
\\
\end{tabular}
\caption{The lock-exchange. The velocity(u) errors of full-model, SAE-SINDy and SAE+POD-SINDy}
\label{fig:le_error}
\end{figure}

After obtaining the manifold version of governing equations \ref{equ:le-sae} and \ref{equ:le-sp}, the solutions of those equations in the manifold space can be projected back to a high-dimensional space. This is conducted via the decoding process in the autoencoder network. 
Figure \ref{fig:le_solution} shows the velocity solutions from the high fidelity model and both of the discovered models: SAE+SINDy and SAE+POD+SINDy at time levels: 5s, 17.5s and 37.5s. As we can see in the figure that both of the discovered governing equations in the manifold space are close to the high fidelity model. In order to see the differences of the two discovered models, the errors are presented in the Figure \ref{fig:le_error}, which shows that the SAE+POD+SINDy method is more accurate and powerful in capturing the fluid dynamics.


\begin{figure}[ht]
\centering
\begin{tabular}{cc}
 \\
\textbf{a} &
\textbf{b}
\\
\begin{minipage}{0.4 \linewidth}
\includegraphics[width = \linewidth,angle=0,clip=true]{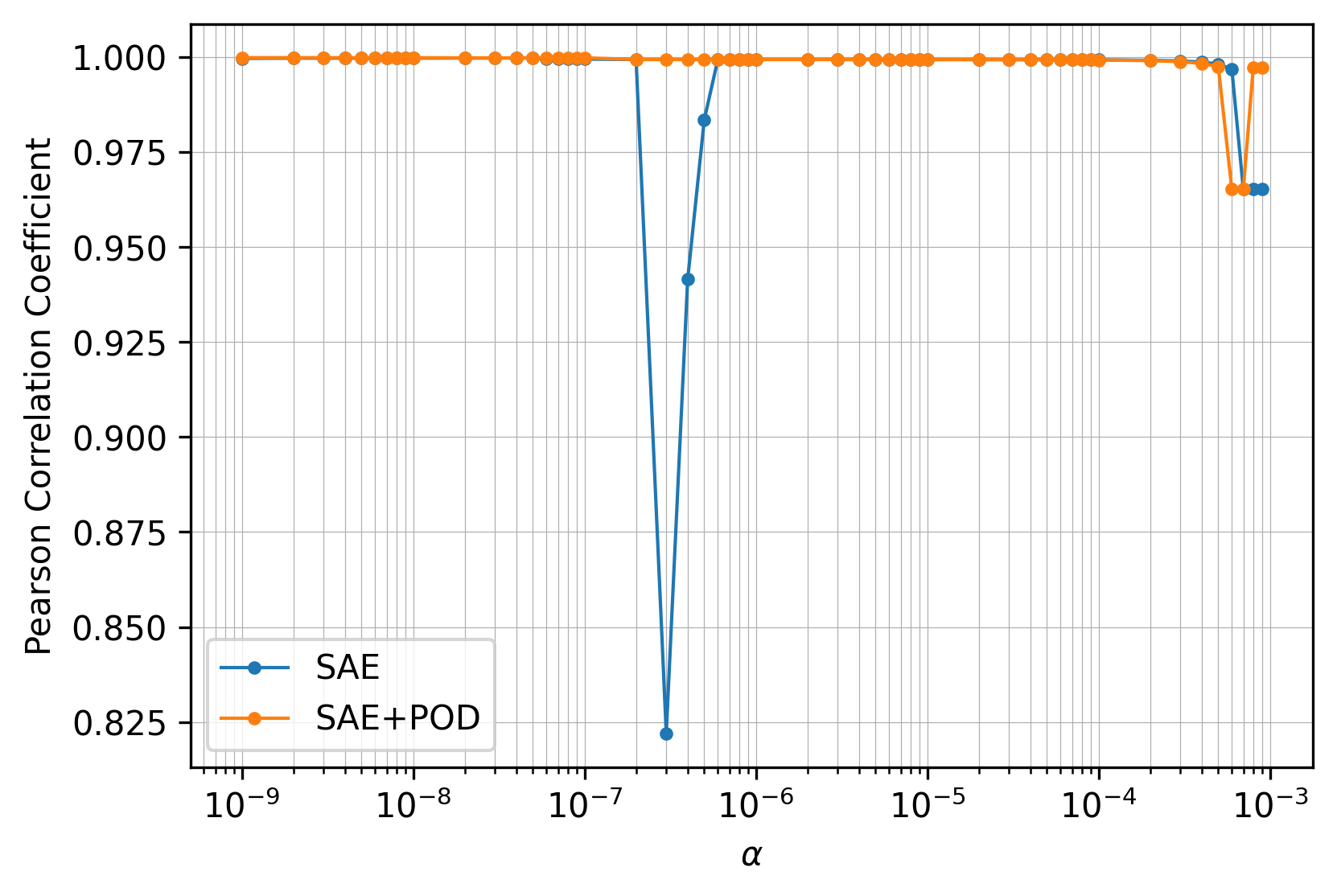} 
\end{minipage}
&
\begin{minipage}{0.4 \linewidth}
\includegraphics[width = \linewidth,angle=0,clip=true]{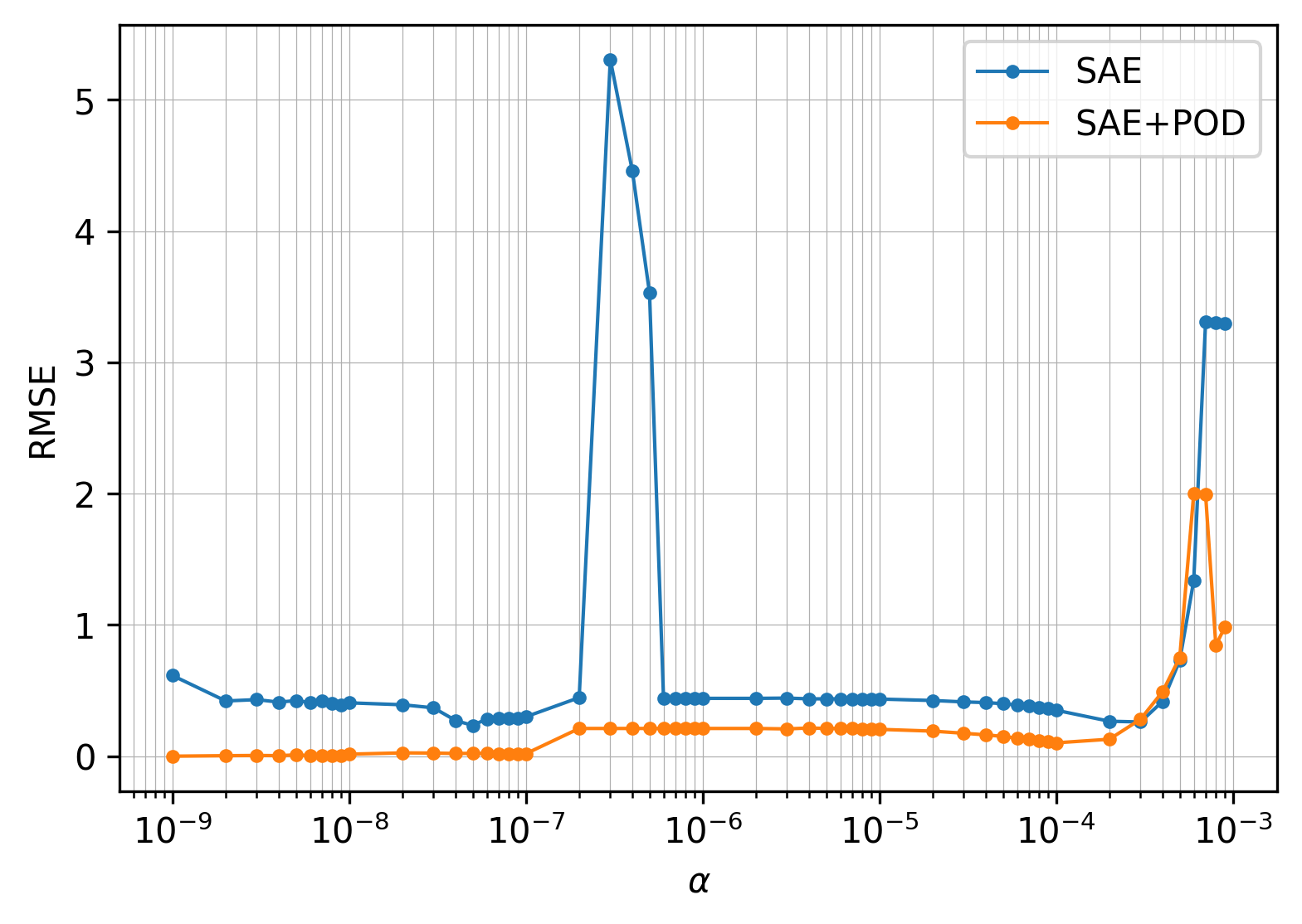} 
\end{minipage}

 \\
\textbf{c} &
\textbf{d}
\\
\begin{minipage}{0.4 \linewidth}
\includegraphics[width = \linewidth,angle=0,clip=true]{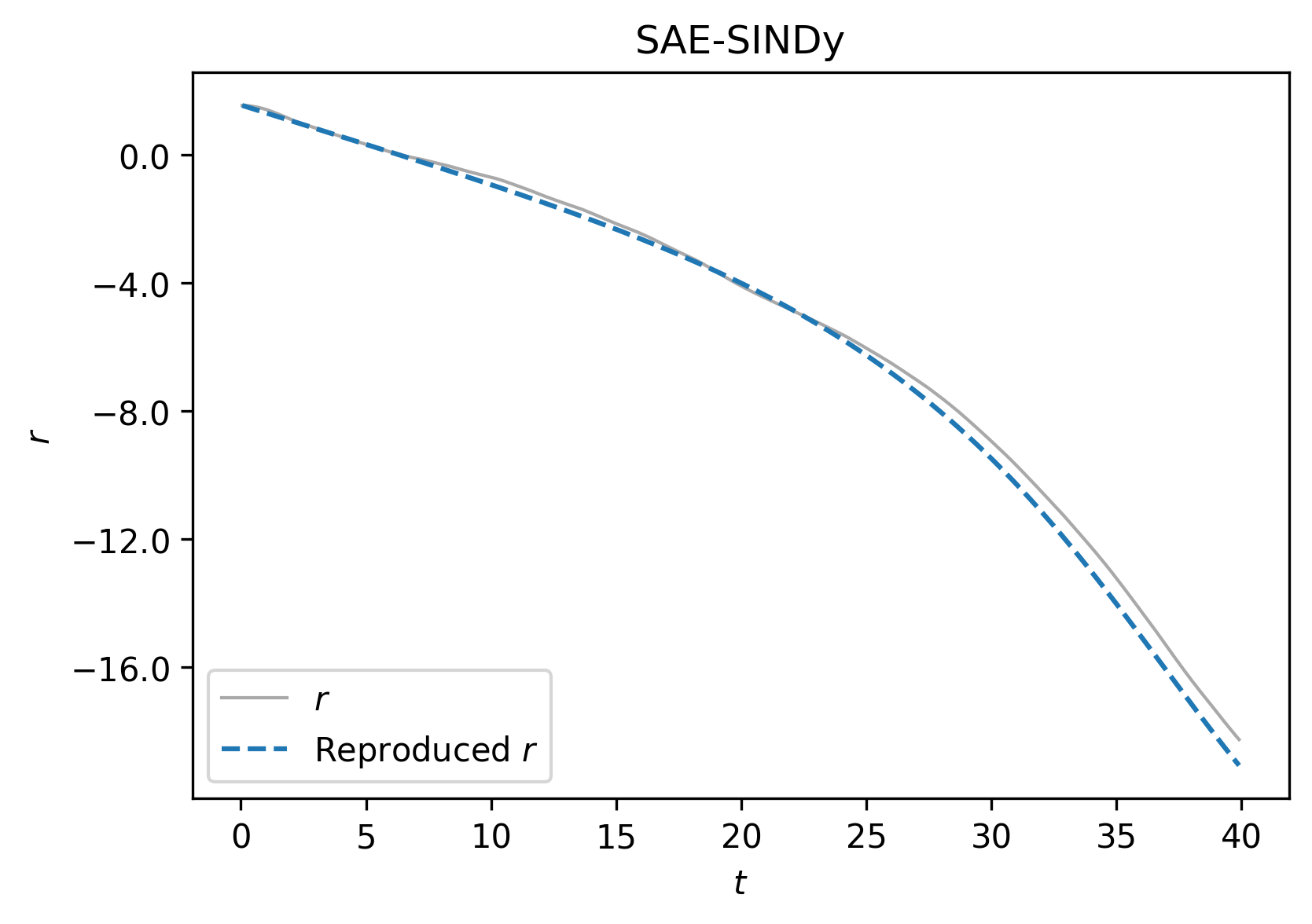} 
\end{minipage}
&
\begin{minipage}{0.4 \linewidth}
\includegraphics[width = \linewidth,angle=0,clip=true]{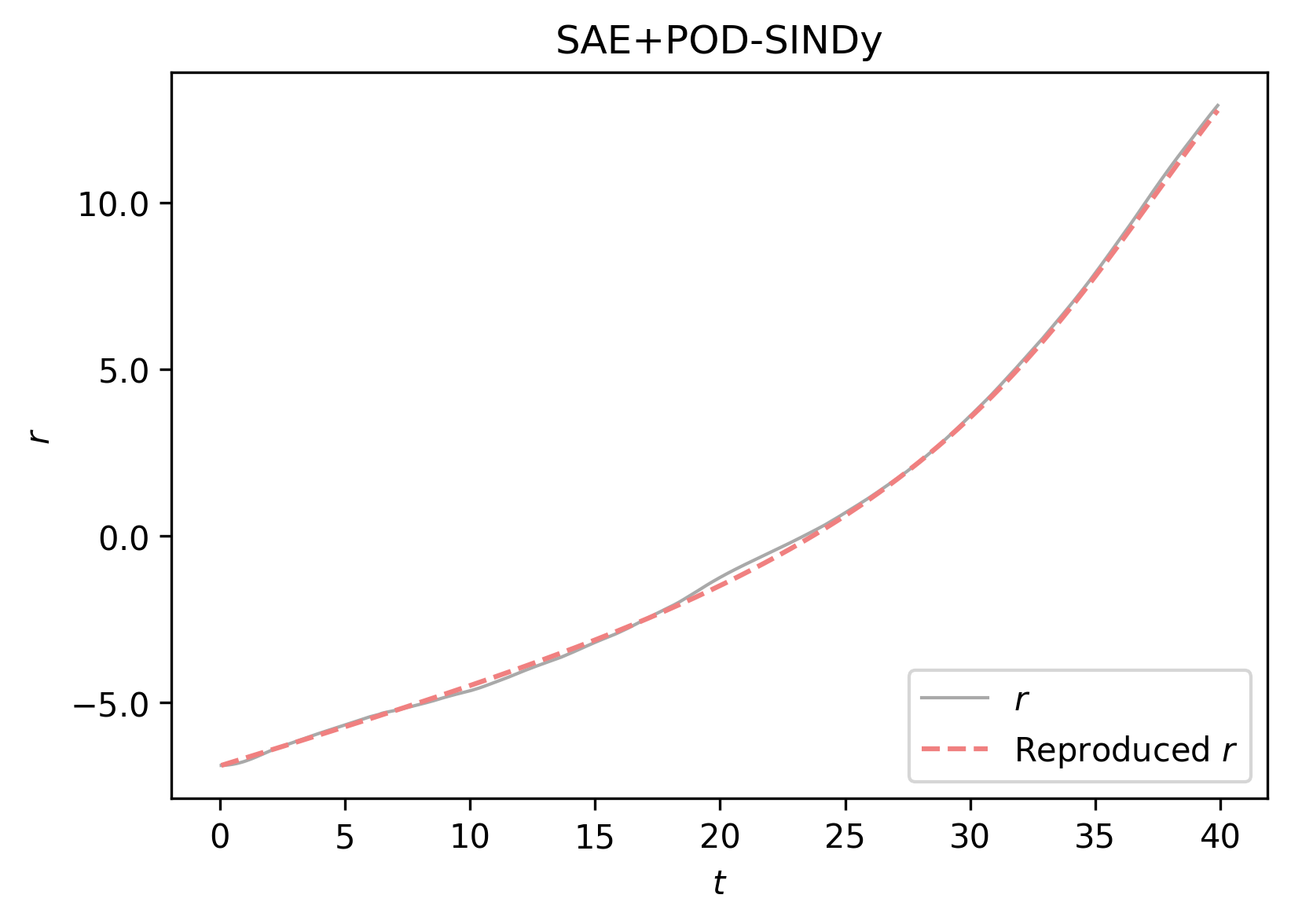} 
\end{minipage}
\end{tabular}
\caption{\textbf{Lock exchange case:}
\textbf{(a)} and \textbf{(b)} correlation coefficients and RMSE comparison between SAE and SAE+POD respectively.   
\textbf{(c)} latent code values comparison between original values and the ones generated by SAE+SINDy in the manifold non-linear space. \textbf{(d)} POD coefficients comparisons between the original values and ones generated by SAE+POD+SINDy.}
\label{fig:le parameter search + sindy result}
\end{figure}

 In the figure\ref{fig:le parameter search + sindy result}, \textbf{(a)} and \textbf{(b)} show the correlation coefficients comparison between the SAE and SAE+POD respectively. As we can see from the figure, SAE will be more accurate and stable if POD process is used to update the SAE method. In addition, it is easier for the SINDy to discover lower dimensional equation as it is more stable after the POD transformation, thus leading to more accurate results. This can be seen from the figure \ref{fig:le parameter search + sindy result} \textbf{(c)} and \textbf{(d)}.

In addition, in order to see clearly the accuracy of both of the reduced version of governing equations in the manifold space, temperature solutions from the high fidelity model and both of the discovered SAE+POD and SAE+POD+SINDy models at two particular points P1 and P2 in the computational domain are given in the Figure \ref{fig:le p1p2 solution + pcc,rmse}.

\begin{figure}[ht]
\centering
\begin{tabular}{cc}
\\
\textbf{a} &
\textbf{b}
\\
\begin{minipage}{0.4 \linewidth}
\includegraphics[width = \linewidth,angle=0,clip=true]{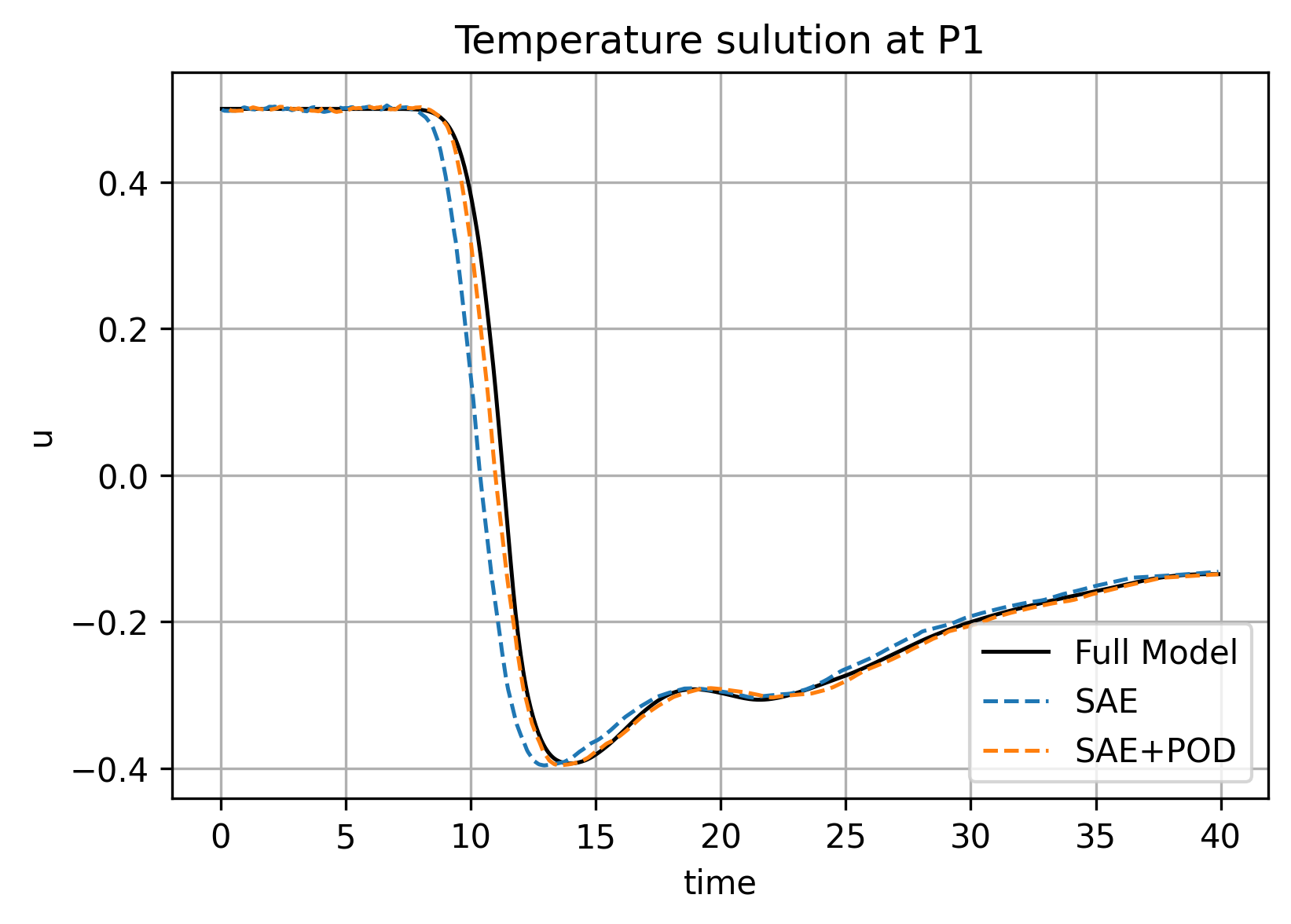} 
\end{minipage}
&
\begin{minipage}{0.4\linewidth}
\includegraphics[width = \linewidth,angle=0,clip=true]{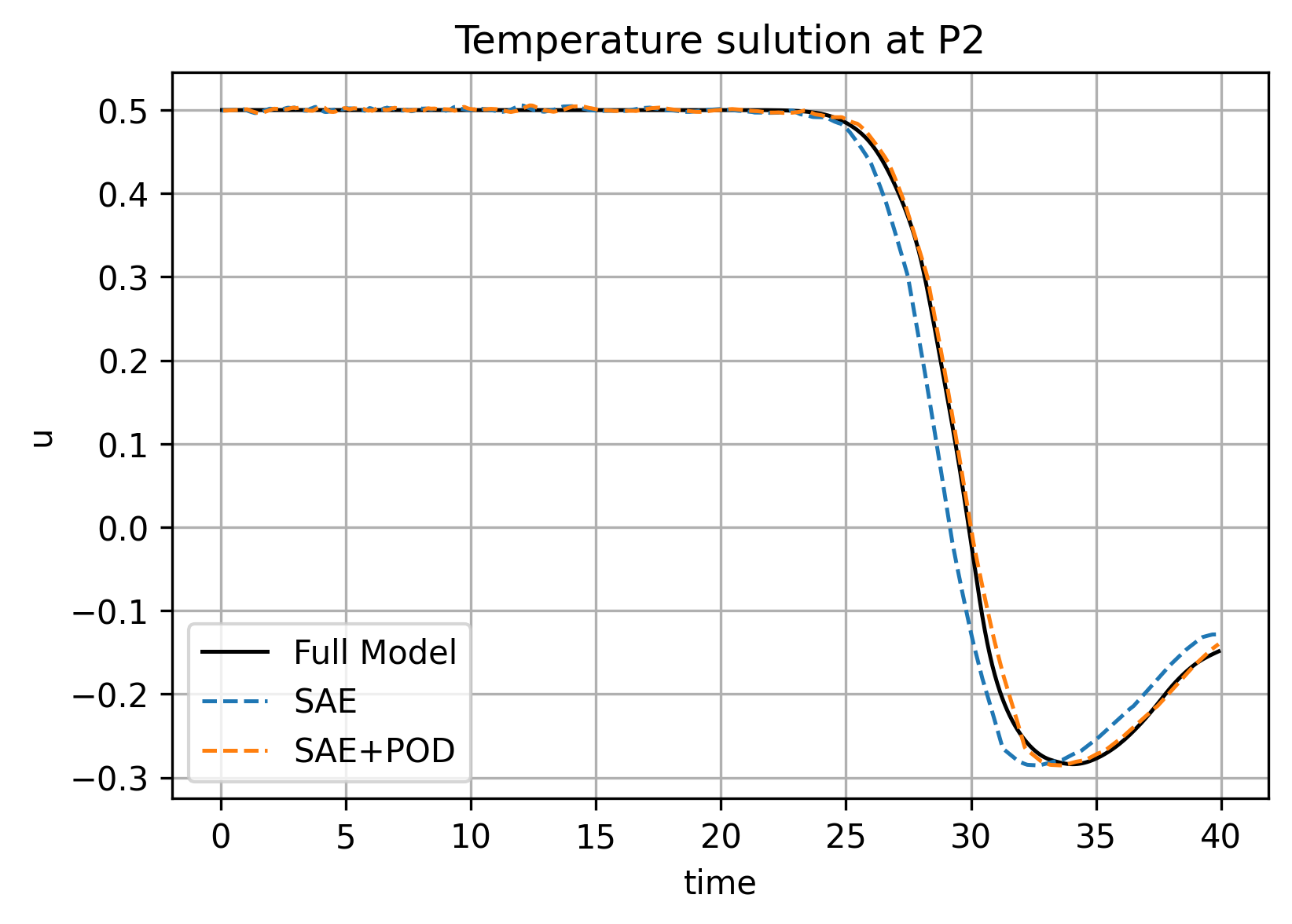} 
\end{minipage}

\end{tabular}
\caption{\textbf{lock-exchange}: Temperature solutions comparisons between the high fidelity model and both of the discovered SAE+POD and SAE+POD+SINDy models at two particular points P1 P2 in the computational domain.}
\label{fig:le p1p2 solution + pcc,rmse}
\end{figure}

\pagebreak
\clearpage

\subsection{\textbf{Flow past a cylinder}}\label{case:1c}
In order to test our novel method to discover complex partial differential equation
(PDE), a flow past a cylinder case is considered. The governing equation of this problem is Navier-Stokes equations, which are given as:
\begin{equation}
    \nabla \cdot \bm{u} = 0
\end{equation}
\begin{equation}
    \frac{{\partial \bm{u}}}{{\partial t}} + (\bm{u} \cdot \nabla) \bm{u} = - \nabla p + \frac{1}{Re_D} \nabla^2 \bm{u}
\end{equation}
where $\bm{u}$, $p$, and $Re_D$ represent velocity vector, pressure, and Reynolds number respectively. The Reynolds number $Re_D=3000$ is chosen, which means that this case is a turbulence phenomenon.  The centre of the inlet boundary is defined as the origin of the computational domain.

\begin{figure}[ht]
\centering
\begin{tabular}{c}
\begin{minipage}{0.4 \linewidth}
\includegraphics[width = \linewidth,angle=0,clip=true]{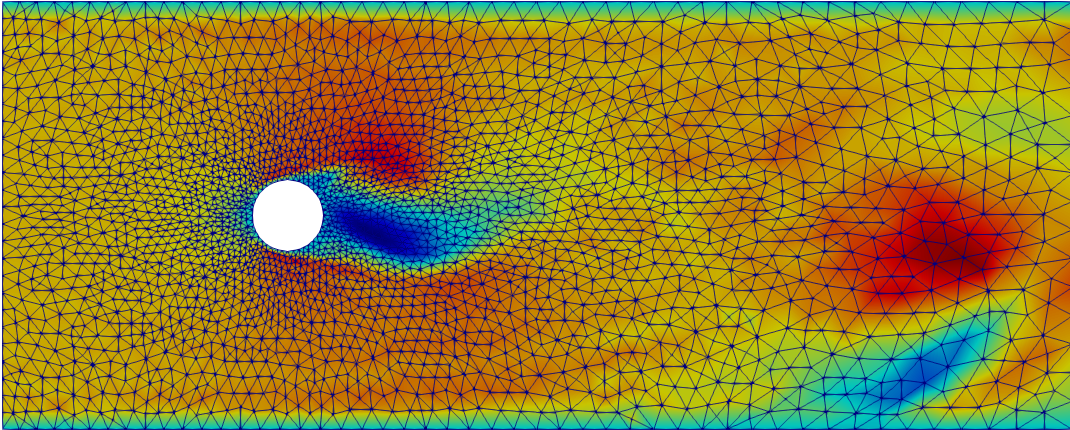} 
\end{minipage}
\\
\\
\begin{minipage}{0.4\linewidth}
\includegraphics[width = \linewidth,angle=0,clip=true]{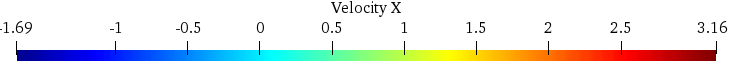} 
\end{minipage}
 \\
\\
\end{tabular}
\caption{\textbf{Flow past a circle cylinder case:} computational domain}
\label{fig:1c-mesh}
\end{figure}

No slip and zero outward boundary conditions are applied to the lower and upper edges. The circle is located at the location of $(x=9, y=0)$. The computational domain has a horizontal length of $0 \leq x \leq 33$ and a vertical length of $-5 \leq y \leq 5$. The simulation data is generated via Fluidity with a time interval of $\Delta t = 0.05 s$ within the time period [$0-100s$]. There are 21812 nodes in the computational domain and 2000 time levels results are generated. The computational domain is shown in Figure \ref{fig:1c-mesh}. 

\begin{figure}[htbp]
\centering 
\begin{minipage}{1.0 \linewidth}
\includegraphics[width = \linewidth,angle=0,clip=true]{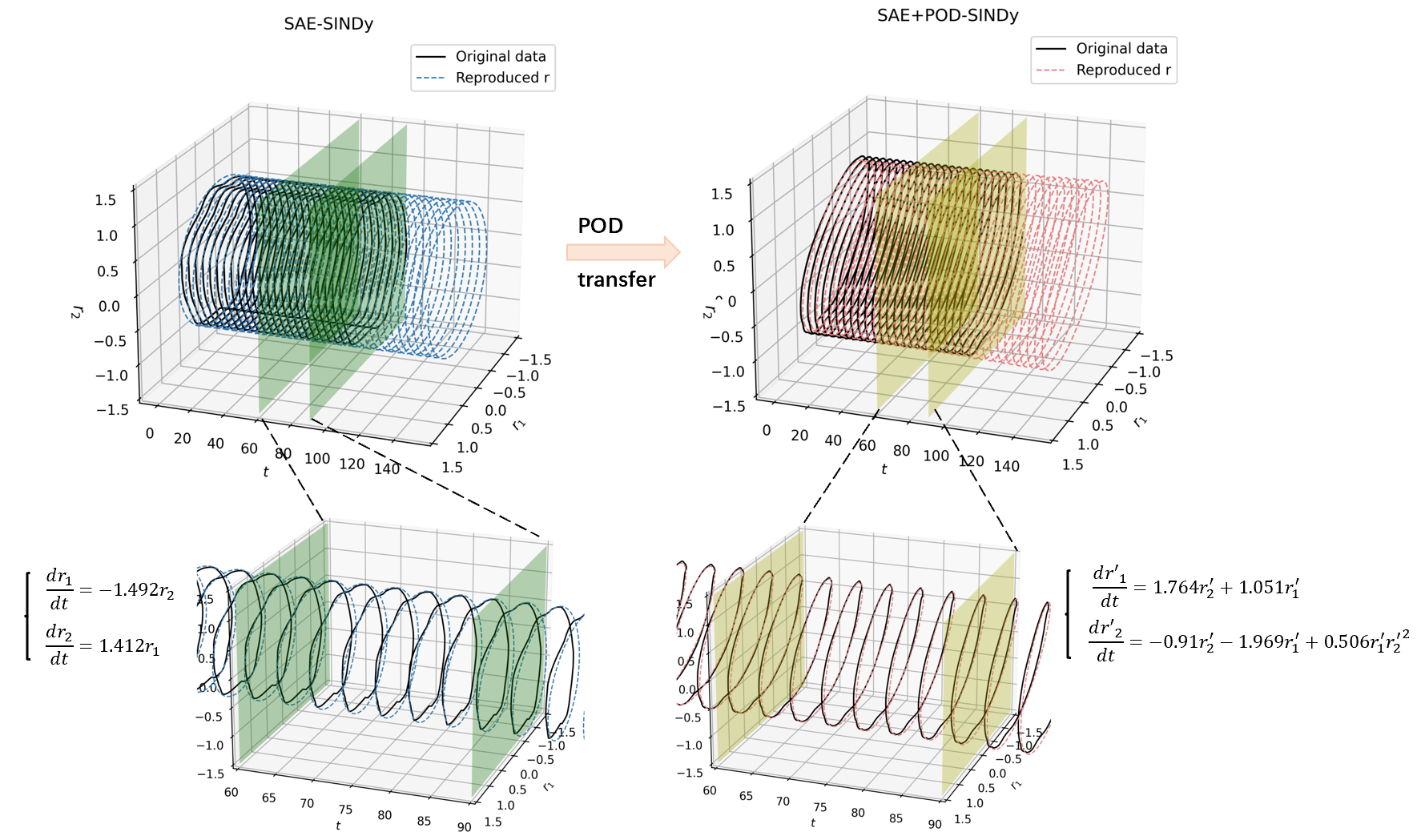} 
\end{minipage}
\caption{\textbf{Flow around a cylinder case: }
\textbf{a}. the left figure presents the results of solutions in the SAE latent space while the right one shows the results of SAE+POD. 
\textbf{b} demonstrates the results of SAE-SINDy during the time period [60s-90s] and the relevant ODEs discovered by SINDy.
\textbf{c} shows the results of SAE+POD-SINDy in the same time period [60s-90s] and the relevant ODEs discovered by SINDy.}
\label{fig:1c sae sp sindy}
\end{figure}

In the latent space,  there are only two dimensions, which capture 99.50\% of the total energy. The SAE structure is listed in Table \ref{table:1c-sae-network}.
The results in the non-linear manifold space obtained by both of the SAE and SAE+POD methods are given in Figure \ref{fig:1c sae sp sindy}. As we can see from the figure, the SAE+POD is more close to the original results than the SAE. 

\begin{table}[htbp!]\label{table:1c-sae-network}
\centering
\begin{tabular}{llll}
\hline
\multicolumn{2}{c}{Encoder}           & \multicolumn{2}{c}{Decoder}      \\
\hline
Layer                   & Output Size & Layer              & Output Size \\
Input: original data    & (16384,1)    & Input: latent code & (2,1)       \\
1st (Dense)              & (1024,1)     & 6st (Dense)         & (15,1)       \\
2nd (Dense)              & (128,1)     & 7st (Dense)         & (40,1)      \\
3nd (Dense)              & (40,1)      & 8st (Dense)         & (128,1)     \\
4nd (Dense)              & (15,1)       & 9st(Dense)         & (1024,1)     \\
5nd (Dense): latent code & (2,1)       & 10st(Dense)        & (16384,1)    \\
\hline
\end{tabular}
\caption{Stacked autoencoder structure on flow past a cylinder case.}
\end{table}

SINDy is then employed to further discover the governing equations in the manifold space. The frequency and root mean square(RMS) is used to evaluate the error of quasi-periodic nature of the one-cylinder wake in this case. We balance the accuracy and total number of terms in the equations, which aims to obtain accurate results using the smallest number of terms in the equations. The optimisation process will lead to an optimal parameter $\lambda=0.0007$ in this case. 


 
Equation \ref{equ:1c-sae-sindy} and Equation \ref{equ:1c-sp-sindy} show the equations discovered by SAE-SINDy and SAE+POD-SINDy, respectively:
\begin{equation}
    \begin{aligned}
    \frac{d r_1}{dt} & =f_1(r_1,r_2,t) = -1.492r_2 \\
    \frac{d r_2}{dt} & =f_2(r_1,r_2,t) = 1.412r_1 \\
    \end{aligned}
    \label{equ:1c-sae-sindy}
\end{equation}
\begin{equation}
    \begin{aligned}
    \frac{d r'_1}{dt} & =f'_1(r'_1,r'_2,t) = 1.764r'_2+1.051r'_1  \\
    \frac{d r'_2}{dt} & =f'_2(r'_1,r'_2,t) = -0.91r'_2-1.969r'_1+0.506r'_1 {r'}_2^2  \\
    \end{aligned}
    \label{equ:1c-sp-sindy}
\end{equation}



\begin{figure}[htbp]
\centering 
\begin{minipage}{0.8 \linewidth}
\includegraphics[width = \linewidth,angle=0,clip=true]{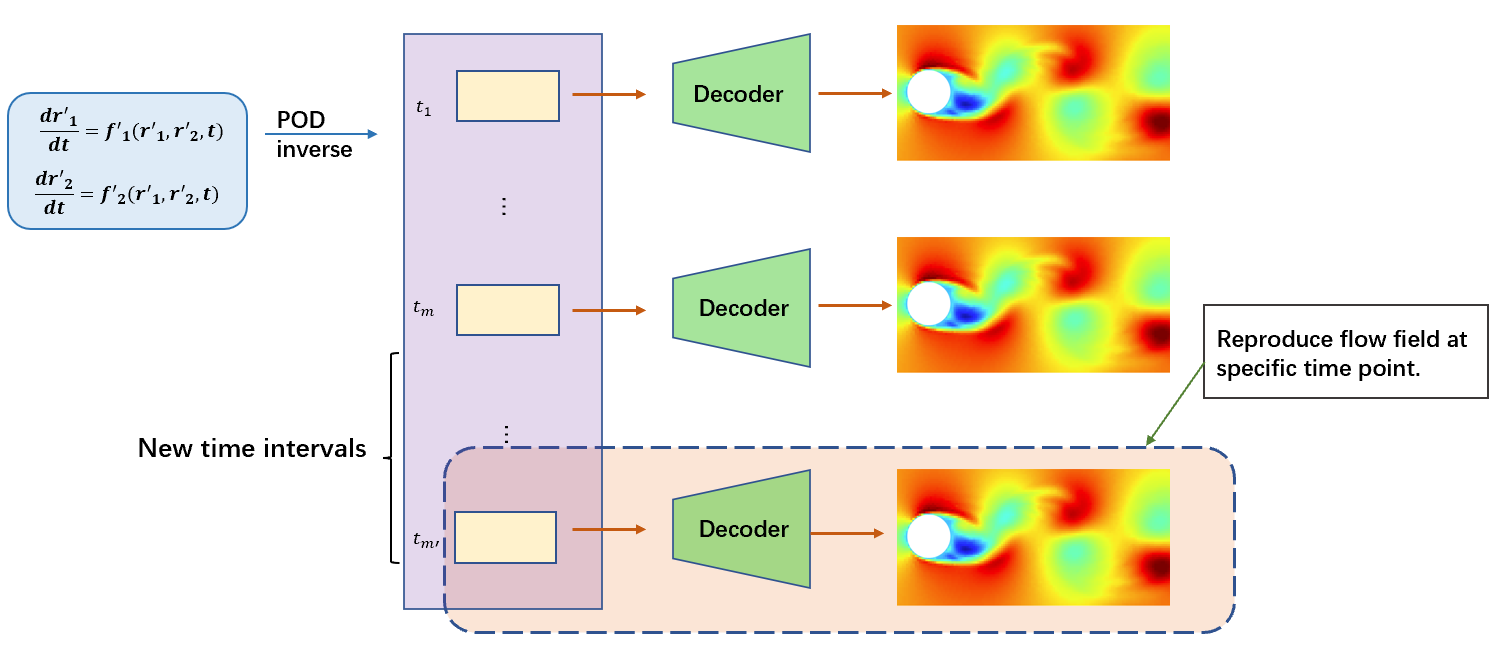} 
\end{minipage}
\caption{\textbf{Flow past a circle cylinder case: Back projection process.}
After solving the reduced version of equations in the manifold space, the solutions are performed via reverse POD process and then projected back into the high dimensionality space via decoder. }
\label{fig: 1c online process}
\end{figure}

The Navier-Stokes equations are simplified into a Ordinary Equations \ref{equ:1c-sae-sindy} and \ref{equ:1c-sp-sindy} via the dimensionality reduction and sparse regression methods, which is much easier to solve. Runge-Kutta method is used to solve the reduced version of equations discovered in the manifold space.  
After solving the ODEs in the manifold space, the results can be projected back into high-dimensional space. The process of projecting back into the high dimensional space is shown in figure \ref{fig: 1c online process}.


After identifying the govern equation f through SINDy, the low-dimensional system can be formulated as 
\begin{equation}
    \dot{\bm{r}'}=\Theta(\bm{r}')\bm{\Xi}
\end{equation}
where $\Theta(\bm{r}')$ stands for SINDy library of $\bm{r}'$ (in which $\bm{r}'$ is a vector of $d$ latent variables),  and $\bm{\Xi}$ is the sparse coefficient vector. 
Utilising the Runge-Kutta method to approximate the manifold solution, the low-dimensional solution $\bm{r}'_{test}$ can be computed at any given time point $t_{test}$. 
The final reduced variables $\bm{r}$ are obtained after an inverse POD transformation $\mathcal{P}^{-1}$ as follow:
\begin{equation}
    \bm{r}' = \mathcal{P}^{-1}(\bm{r}')
\end{equation}

\begin{figure}
\centering
\begin{tabular}{ccc}
 \\
(a-1) {\small Full Model, t = 5s}&
(a-2) {\small SAE+SINDy, t=5s}&
(a-3) {\small SAE+POD+SINDy, t=5s}
\\
\begin{minipage}{0.25\linewidth}
\includegraphics[width = \linewidth,angle=0,clip=true]{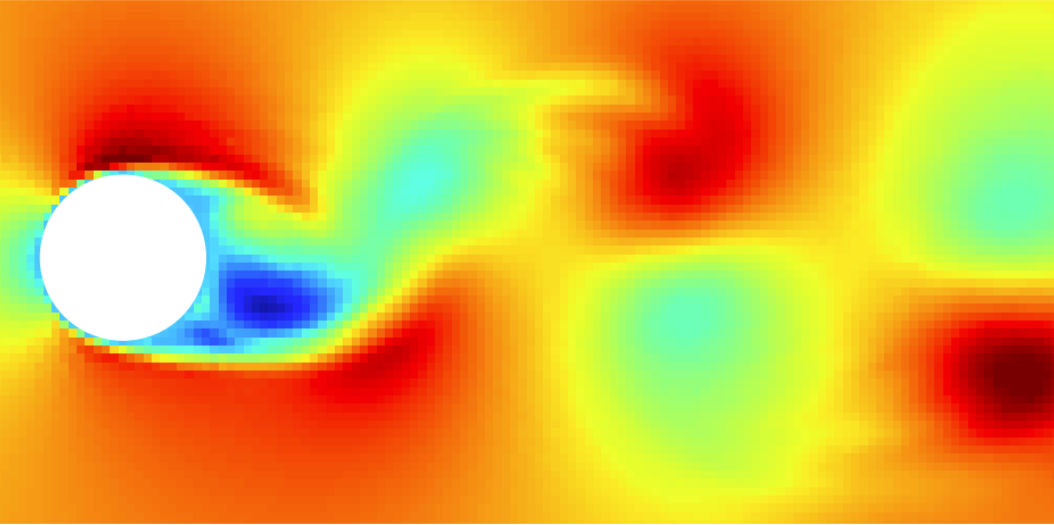}  
\end{minipage}
&
\begin{minipage}{0.25\linewidth}
\includegraphics[width = \linewidth,angle=0,clip=true]{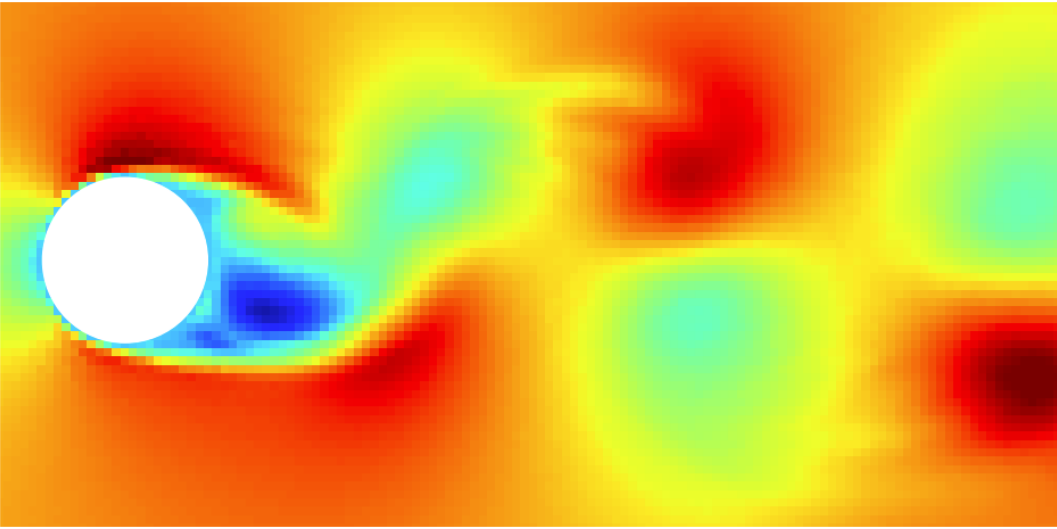}  
\end{minipage} 
&
\begin{minipage}{0.25\linewidth}
\includegraphics[width = \linewidth,angle=0,clip=true]{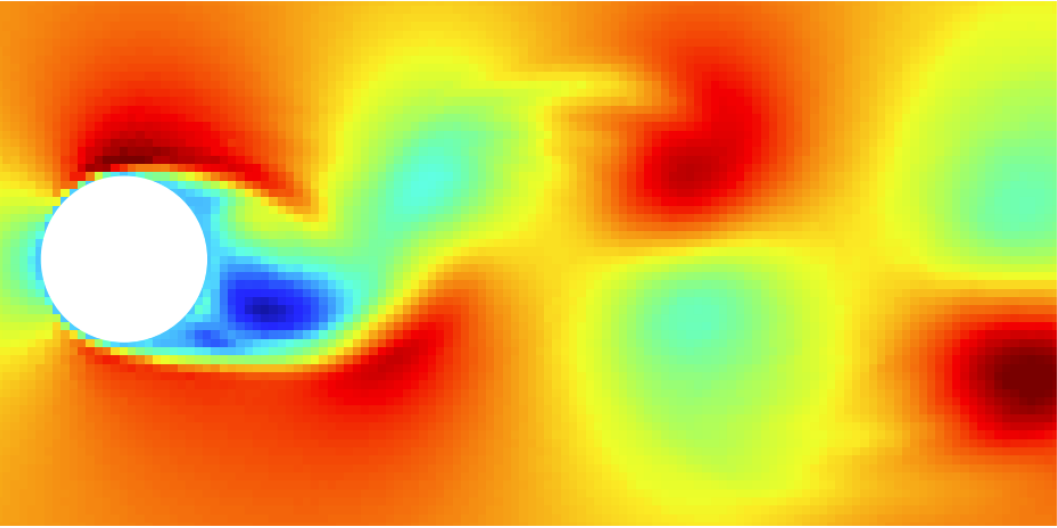}  
\end{minipage} 

\\
(b-1) {\small Full Model, t = 27.5s}&
(b-2) {\small SAE+SINDy, t = 27.5s}&
(b-3) {\small SAE+POD+SINDy, t = 27.5s}
\\
\begin{minipage}{0.25\linewidth}
\includegraphics[width = \linewidth,angle=0,clip=true]{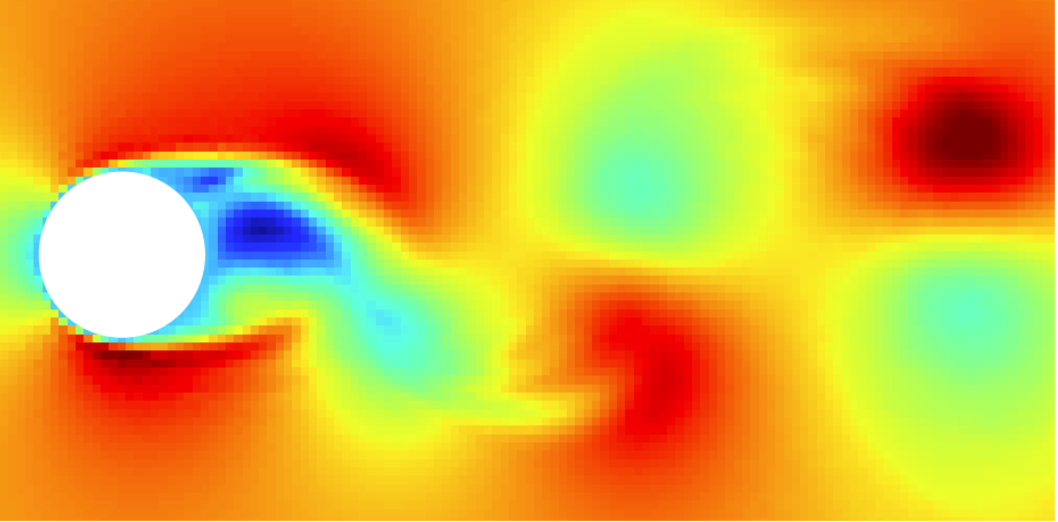}  
\end{minipage}
&
\begin{minipage}{0.25\linewidth}
\includegraphics[width = \linewidth,angle=0,clip=true]{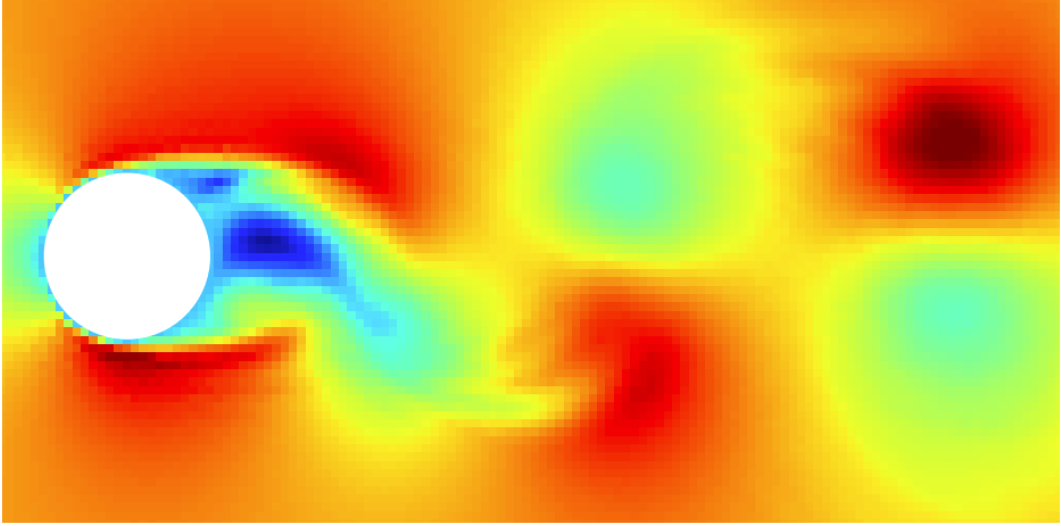}  
\end{minipage} 
&
\begin{minipage}{0.25\linewidth}
\includegraphics[width = \linewidth,angle=0,clip=true]{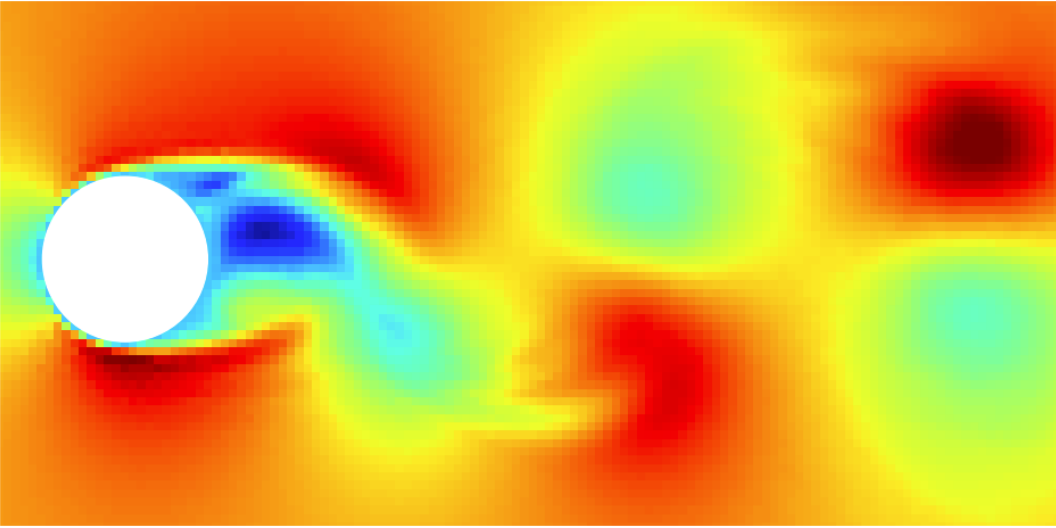}  
\end{minipage} 

 \\
(c-1) {\small Full Model, t = 75s}&
(c-2) {\small SAE+SINDy, t = 75s}&
(c-3) {\small SAE+POD+SINDy, t = 75s}
\\
\begin{minipage}{0.25\linewidth}
\includegraphics[width = \linewidth,angle=0,clip=true]{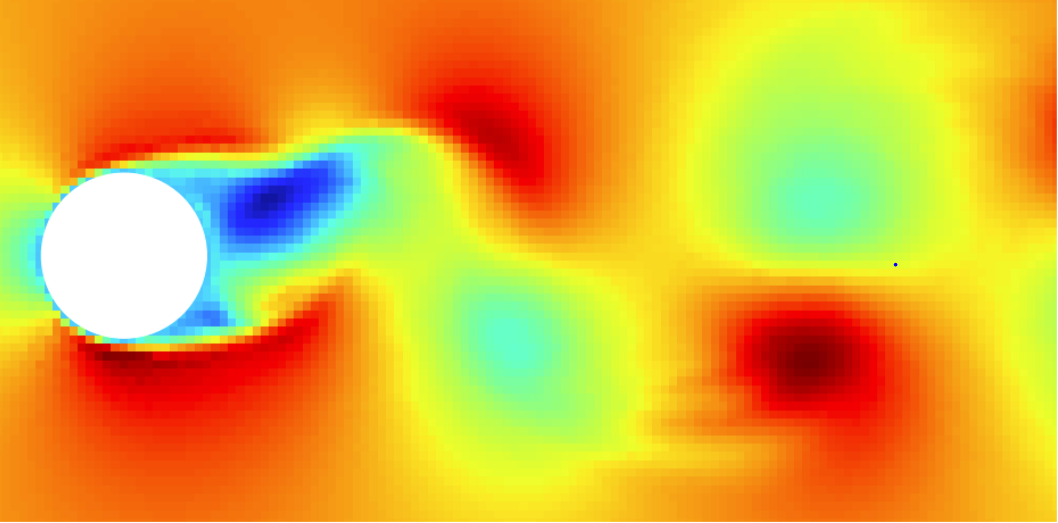}  
\end{minipage}
&
\begin{minipage}{0.25\linewidth}
\includegraphics[width = \linewidth,angle=0,clip=true]{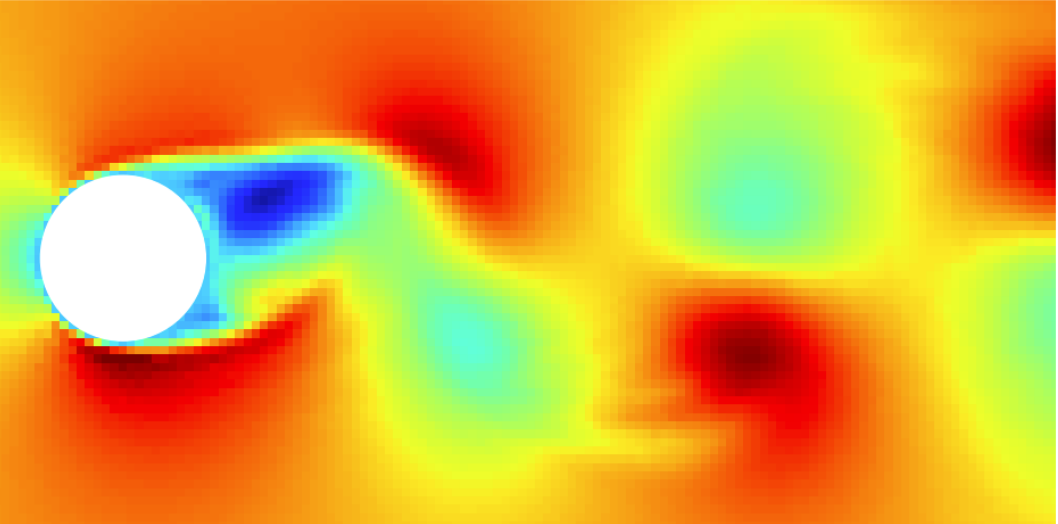}  
\end{minipage} 
&
\begin{minipage}{0.25\linewidth}
\includegraphics[width = \linewidth,angle=0,clip=true]{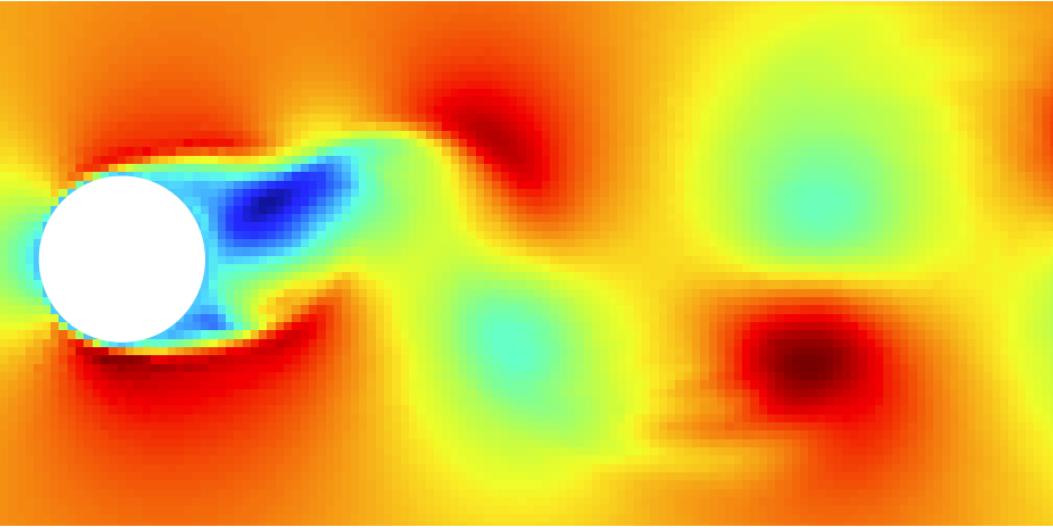}  
\end{minipage} 

\\
\\
\begin{minipage}{0.25\linewidth}
\includegraphics[width = \linewidth,angle=0,clip=true]{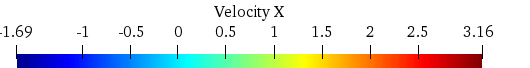} 
\end{minipage}
&
\begin{minipage}{0.25\linewidth}
\includegraphics[width = \linewidth,angle=0,clip=true]{results/1cylinder/1c_grid2.png} 
\end{minipage}
&
\begin{minipage}{0.25\linewidth}
\includegraphics[width = \linewidth,angle=0,clip=true]{results/1cylinder/1c_grid2.png} 
\end{minipage}
\end{tabular}
\caption{\textbf{Flow past a cylinder}. The velocity solutions obtained from the full high fidelity model, SAE+SINDy model and SAE+POD+SINDy model at time levels $t=0s$, $t=125s$, and $t=20s$.}
\label{fig:1c_solution}
\end{figure}

\begin{figure}
\centering
\begin{tabular}{cc}
\begin{minipage}{0.25\linewidth}
\includegraphics[width = \linewidth,angle=0,clip=true]{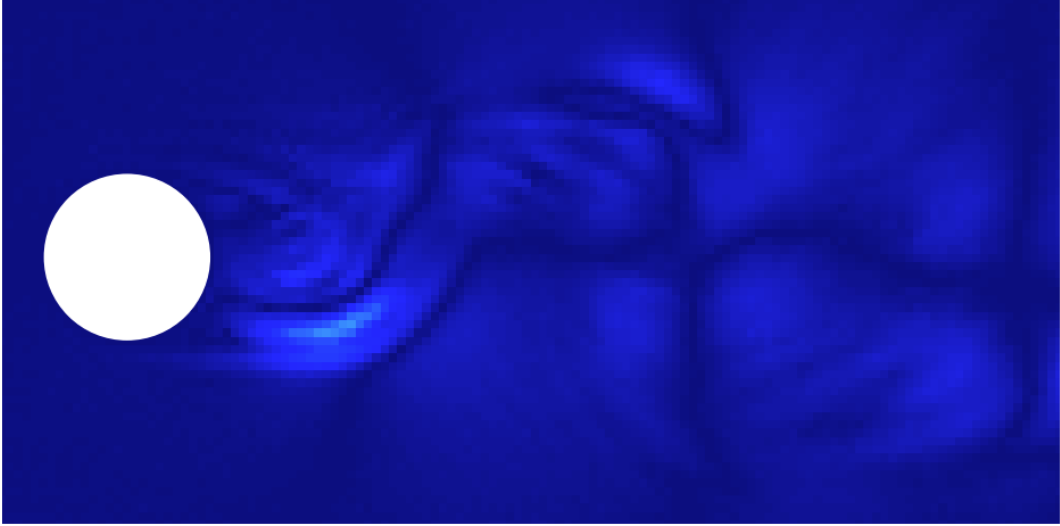}  
\end{minipage} 
&
\begin{minipage}{0.25\linewidth}
\includegraphics[width = \linewidth,angle=0,clip=true]{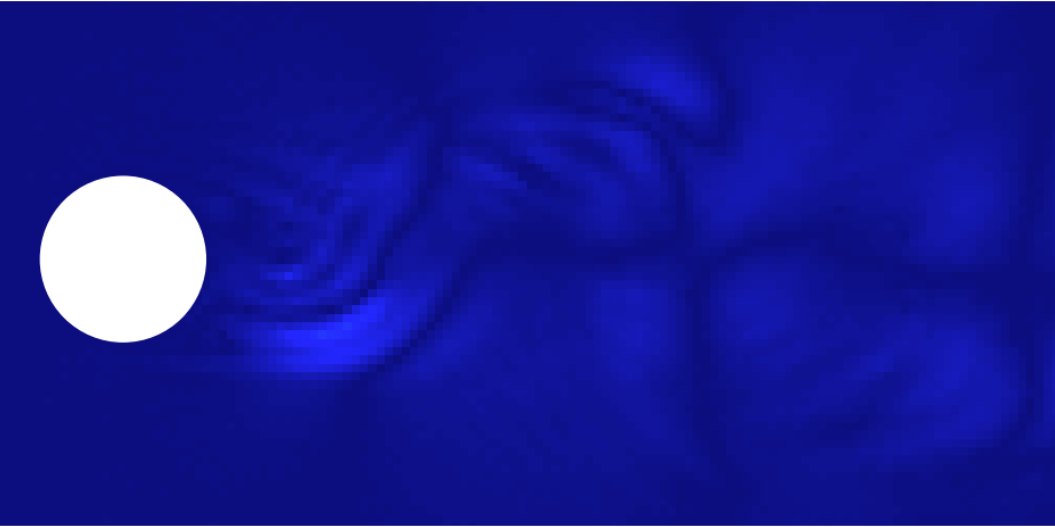}  
\end{minipage} 
 \\
(a-4) {\small SAE error, t=5s}&
(a-5) {\small SAE+POD error, t=5s}
\\

\begin{minipage}{0.25\linewidth}
\includegraphics[width = \linewidth,angle=0,clip=true]{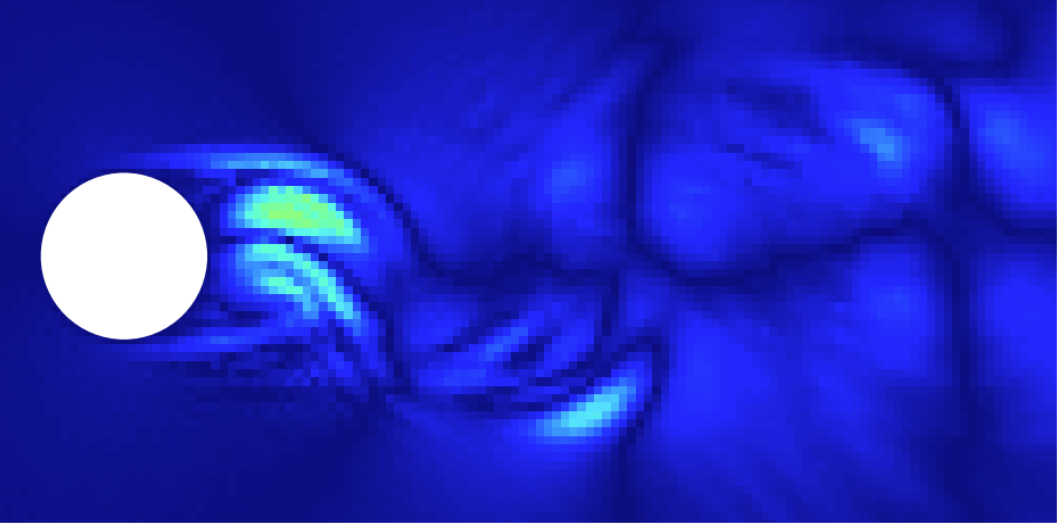}  
\end{minipage} 
&
\begin{minipage}{0.25\linewidth}
\includegraphics[width = \linewidth,angle=0,clip=true]{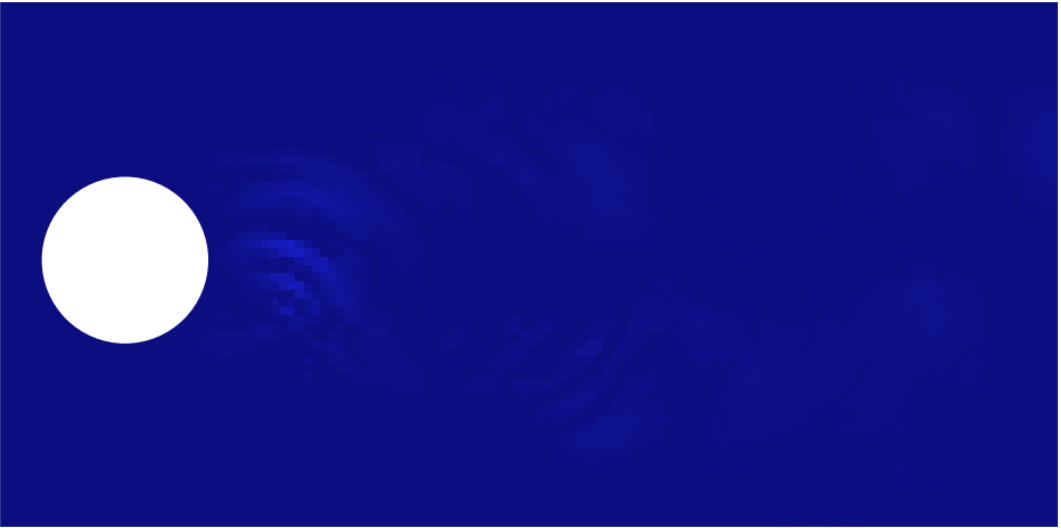}  
\end{minipage} 
 \\
(b-4) {\small SAE error, t=27.5s}&
(b-5) {\small SAE+POD error, t=27.5s}
\\

\begin{minipage}{0.25\linewidth}
\includegraphics[width = \linewidth,angle=0,clip=true]{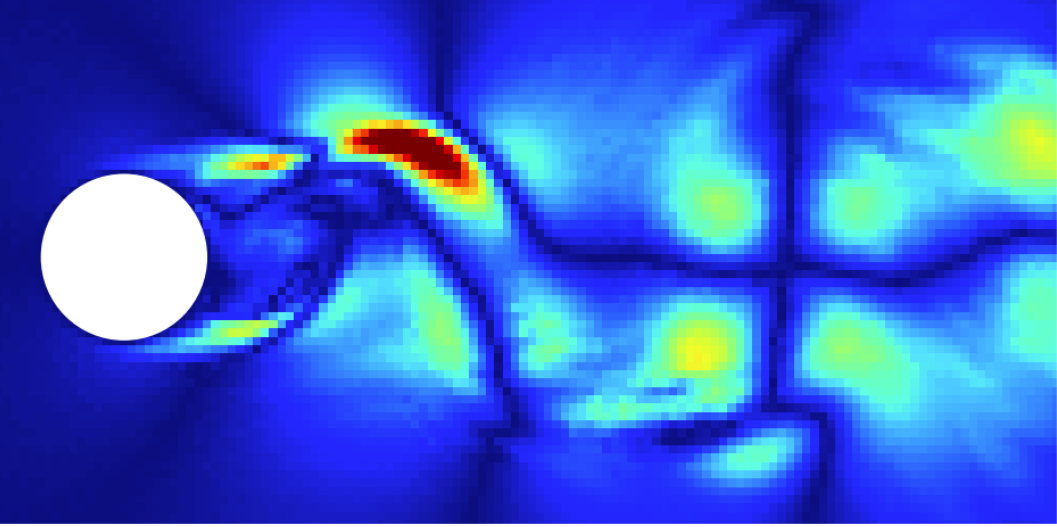}  
\end{minipage} 
&
\begin{minipage}{0.25\linewidth}
\includegraphics[width = \linewidth,angle=0,clip=true]{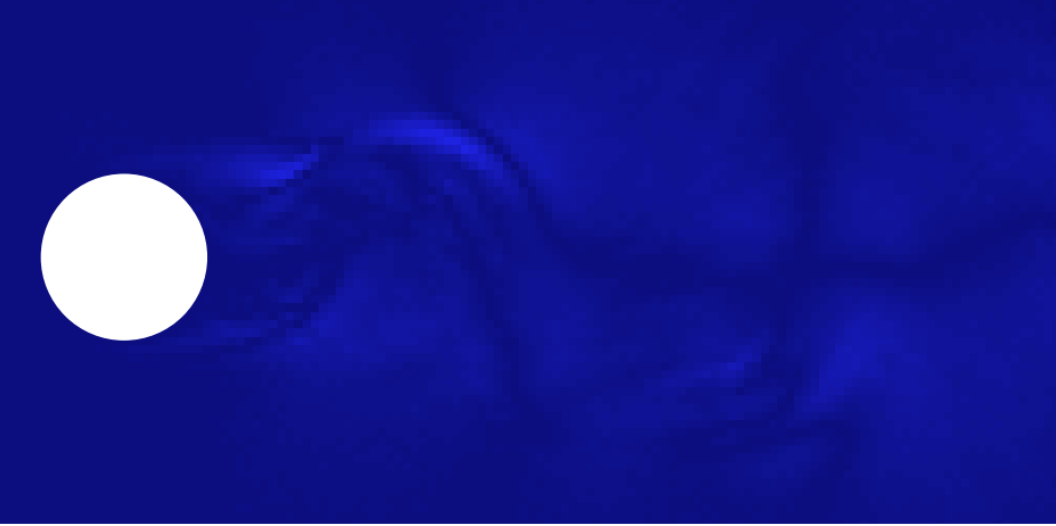}  
\end{minipage} 
 \\
(c-4) {\small SAE error, t=75s}&
(c-5) {\small SAE+POD error, t=75s}
\\

\begin{minipage}{0.25\linewidth}
\includegraphics[width = \linewidth,angle=0,clip=true]{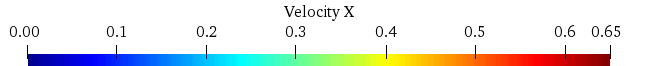} 
\end{minipage}
&
\begin{minipage}{0.25\linewidth}
\includegraphics[width = \linewidth,angle=0,clip=true]{results/1cylinder/1c_grid_error.png} 
\end{minipage}
 \\
\\
\end{tabular}
\caption{\textbf{Flow past a cylinder}. The errors of the SAE+SINDy model and SAE+POD+SINDy model at time levels $t=0s$, $t=125s$, and $t=20s$.}
\label{fig:1c_error}
\end{figure}

Figure \ref{fig:1c_solution} shows the velocity solutions obtained from the full high fidelity model, SAE+SINDy model and SAE+POD+SINDy model at time levels $t=5s$, $t=27.5s$, and $t=75s$.  Figure \ref{fig:1c_error} shows the errors of two discovered simplified equations at time levels $t=5s$, $t=27.5s$, and $t=75s$.  

In order to see clearly the differences between two discovered simplified version of NS equations in the manifold space for this flow past a cylinder case, the solutions at two particular points are given in the Figure \ref{fig:1c p1p2 + pcc,rmse} $\textbf{a}$ and $\textbf{b}$. The correlation coefficients and RMSE are given in \ref{fig:1c p1p2 + pcc,rmse} $\textbf{c}$ and $\textbf{d}$. which shows that the solutions of equations discovered by SAE+POD is more close to the that of the high fidelity full model. 

\begin{figure}[ht]
\centering
\begin{tabular}{c}
\begin{minipage}{0.4 \linewidth}
\includegraphics[width = \linewidth,angle=0,clip=true]{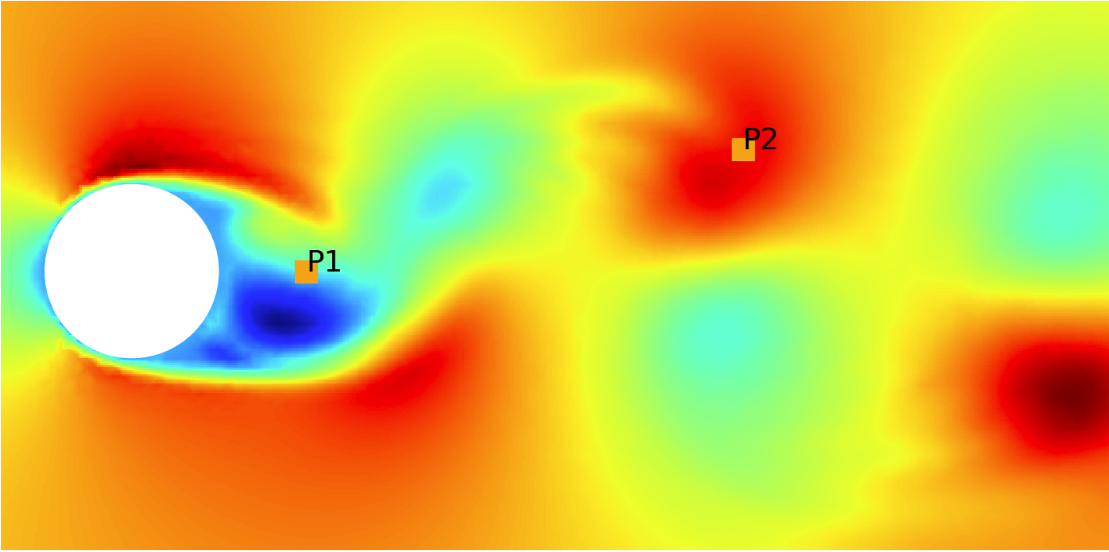} 
\end{minipage}
\\
\\
\begin{minipage}{0.4\linewidth}
\includegraphics[width = \linewidth,angle=0,clip=true]{results/1cylinder/1c_grid1.png} 
\end{minipage}
 \\
\\
\end{tabular}
\caption{\textbf{Flow past a cylinder}: locations of two particular points: P1 and P2}
\label{fig:1c p1p2 location}
\end{figure}

\begin{figure}[ht]
\centering
\begin{tabular}{cc}
\\
\textbf{a} &
\textbf{b}
\\
\begin{minipage}{0.5 \linewidth}
\includegraphics[width = \linewidth,angle=0,clip=true]{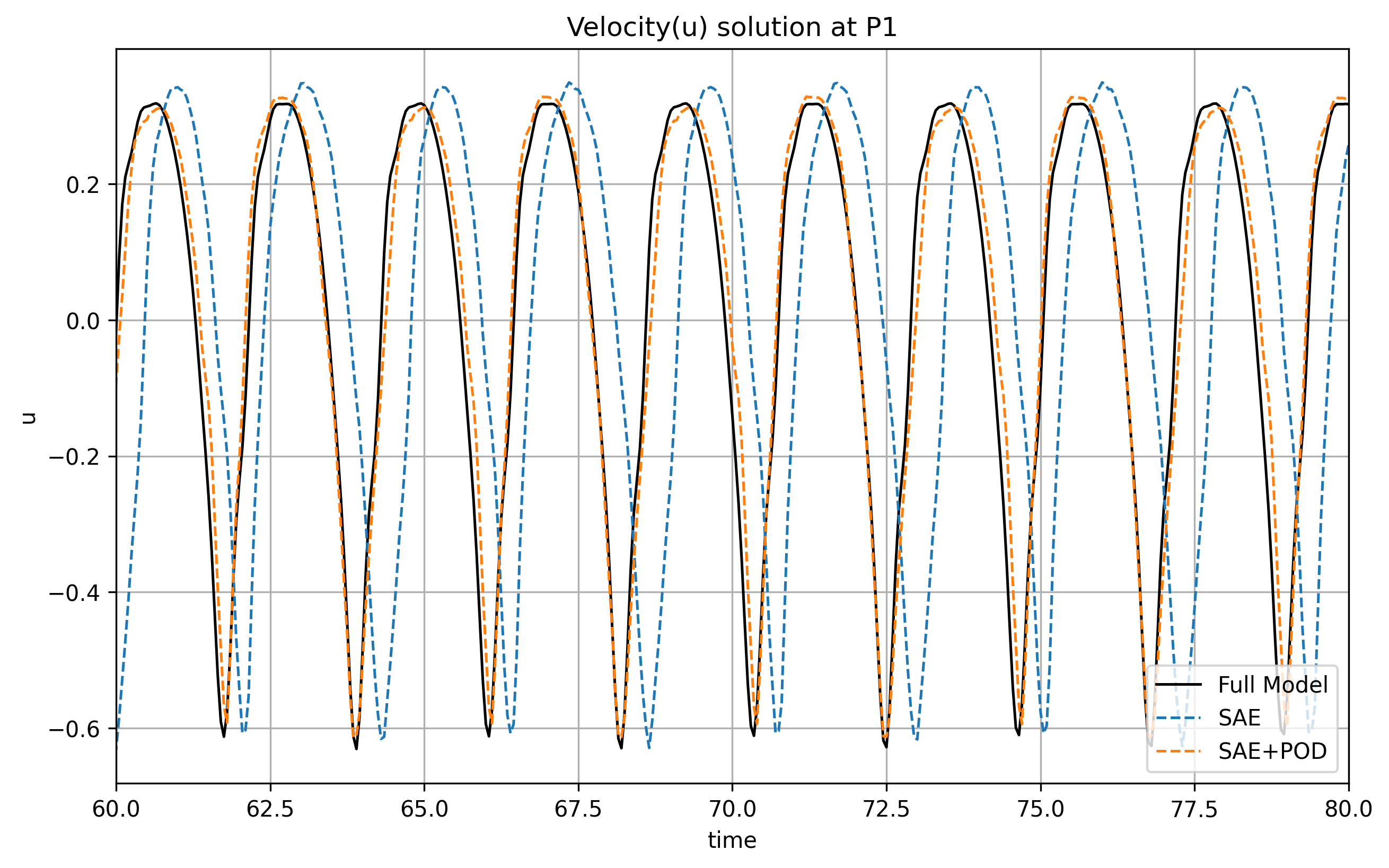} 
\end{minipage}
&
\begin{minipage}{0.5\linewidth}
\includegraphics[width = \linewidth,angle=0,clip=true]{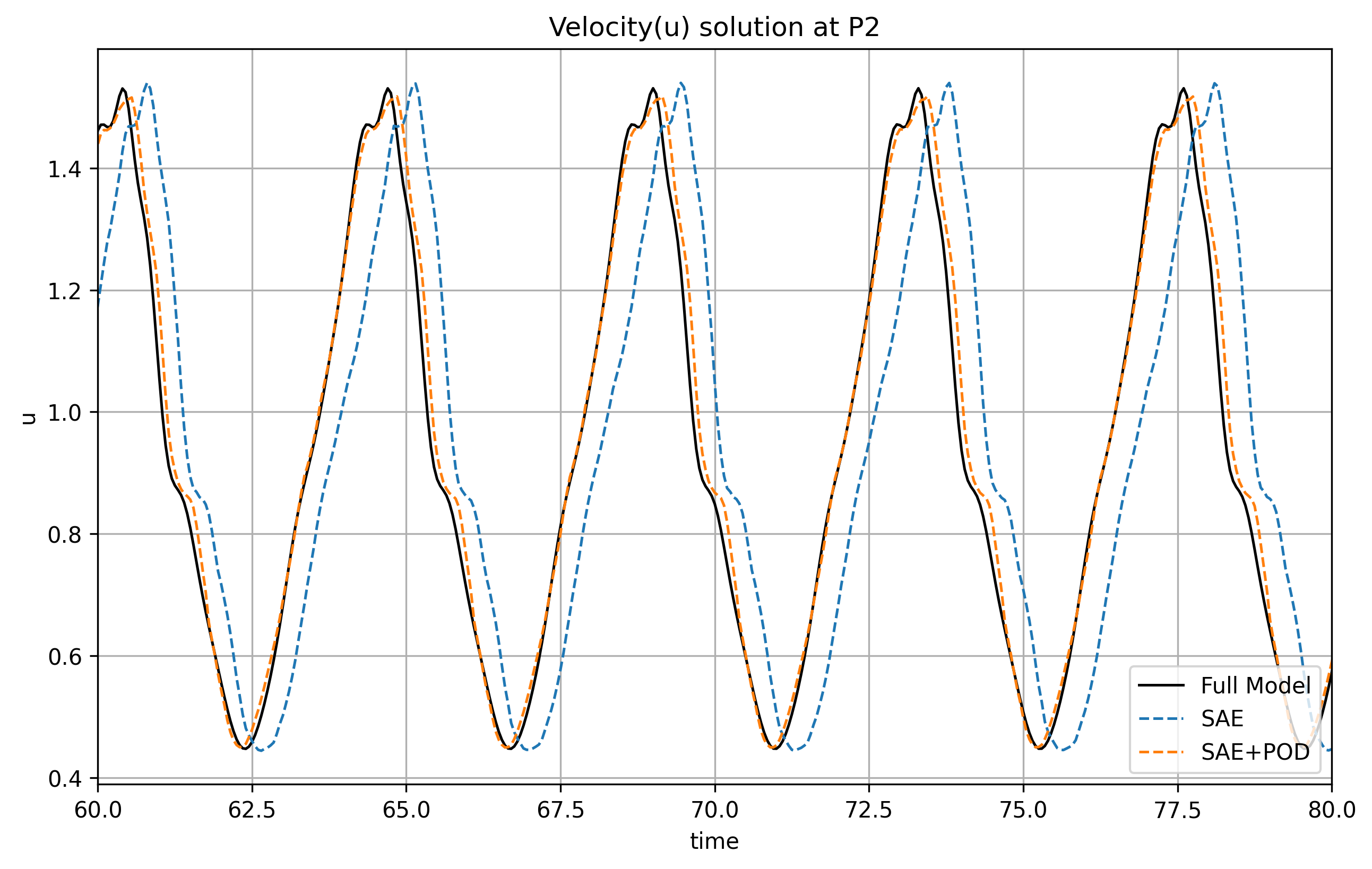} 
\end{minipage}

\\
\textbf{c} &
\textbf{d}
\\
\begin{minipage}{0.5 \linewidth}
\includegraphics[width = \linewidth,angle=0,clip=true]{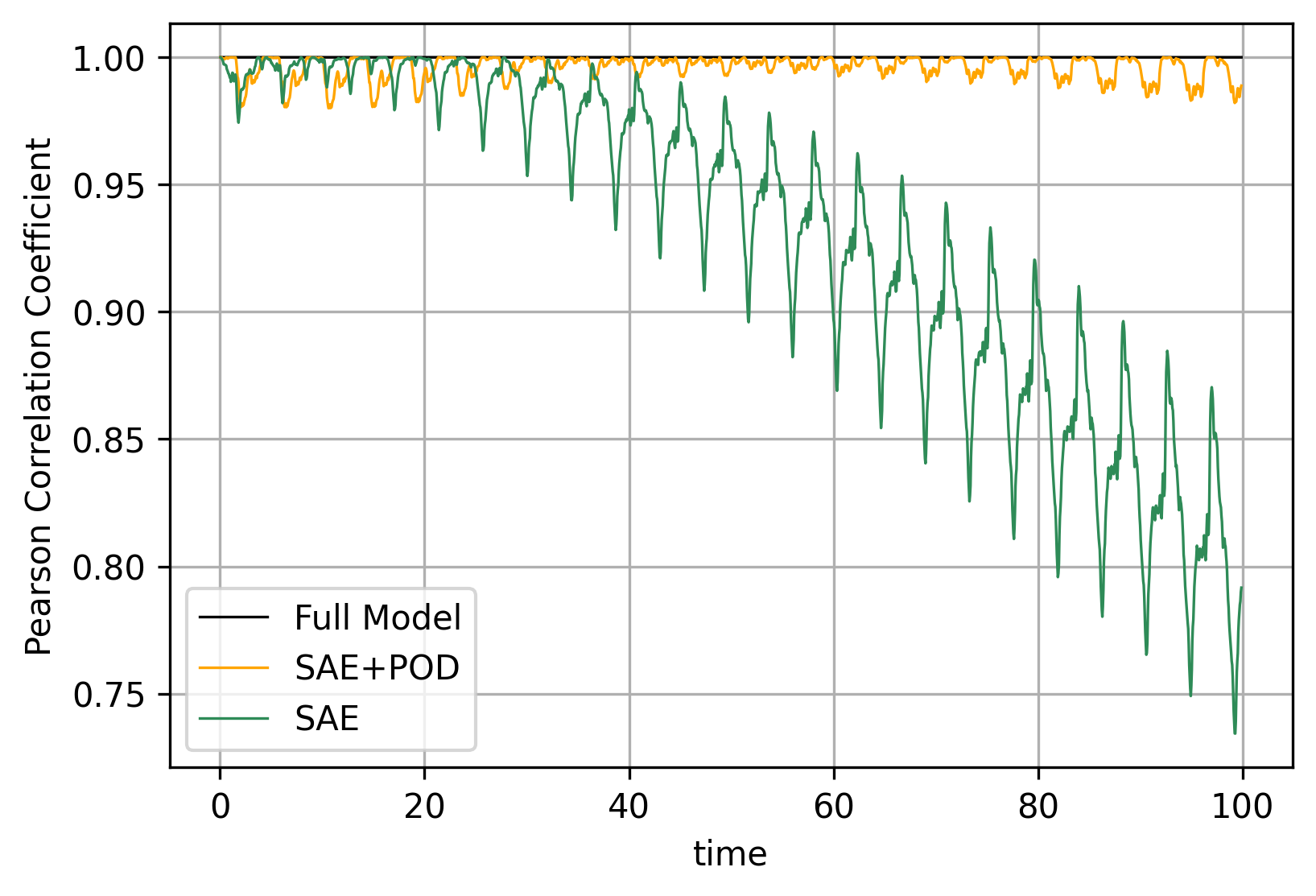} 
\end{minipage}
&
\begin{minipage}{0.5 \linewidth}
\includegraphics[width = \linewidth,angle=0,clip=true]{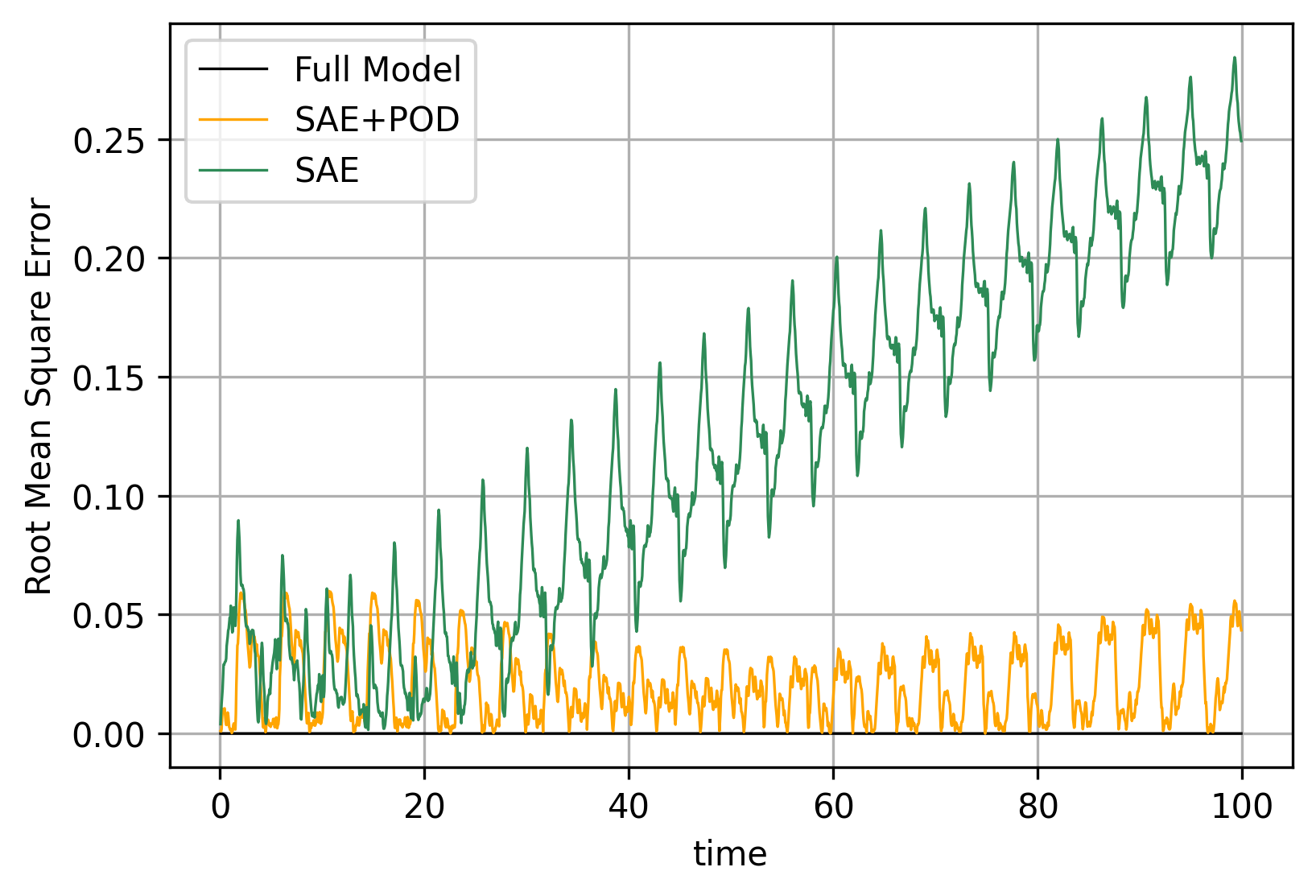} 
\end{minipage}
\end{tabular}
\caption{\textbf{Flow around a cylinder.}
\textbf{a} and\textbf{b}: velocity solutions comparison between the full high fidelity model, SAE+SINDy, SAE+POD+SINDy models at points P1 and P2 respectively shown in figure \ref{fig:1c p1p2 location}. 
\textbf{c} and \textbf{d}: Correlation coefficient and RMSE between the full high fidelity model and SAE+SINDy, SAE+POD+SINDy models.}
\label{fig:1c p1p2 + pcc,rmse}
\end{figure}

\pagebreak

\clearpage

\subsection{\textbf{Flow past two cylinders}}\label{case:2c}
\begin{figure}[htbp]
\centering 
\begin{minipage}{0.5 \linewidth}
\includegraphics[width = \linewidth,angle=0,clip=true]{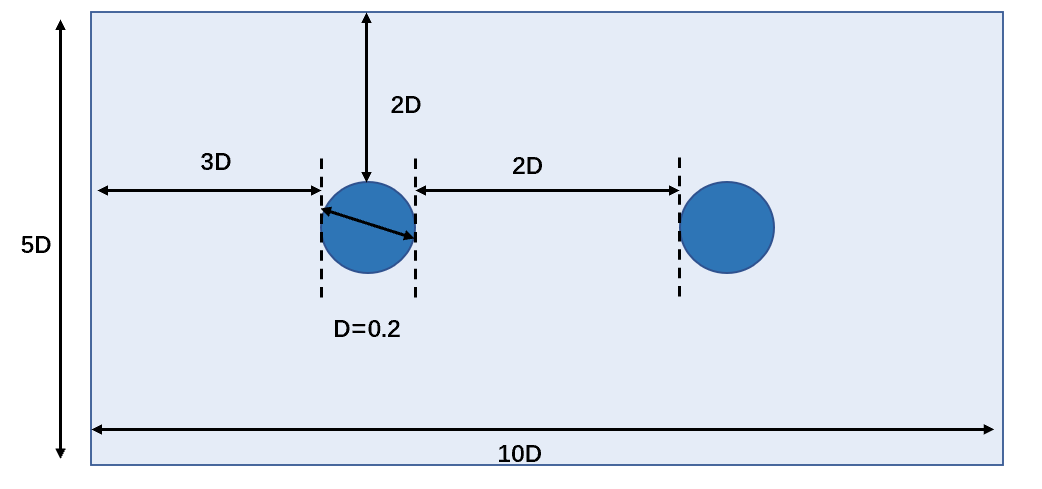} 
\end{minipage}
\caption{\textbf{Schematic of flow around two cylinders.} }
\label{fig: 2c gmsh}
\end{figure}
In the third case, we applied our method to discover governing equations of flow past two cylinders in the manifold space. The Reynolds number $Re_D$ is $3000$ and the vortex shedding can be found in the case. The simulation data was generated via Fluidity.

Figure \ref{fig: 2c gmsh} shows the schematic of this test case. Two cylinders with a diameter of $D=0.2$ are in the domain. The distance between the two cylinders is set to be $2D$. The total length of the computational domain is $10D$. 
1500 time levels' simulation data with time interval of $\Delta t=0.01$ was used.  
The SAE network structure is illustrated in Figure \ref{table:2c-sae-network}.

\begin{table}[htbp!]\label{table:2c-sae-network}
\centering
\begin{tabular}{llll}
\hline
\multicolumn{2}{c}{Encoder}           & \multicolumn{2}{c}{Decoder}      \\
\hline
Layer                   & Output Size & Layer              & Output Size \\
Input: original data    & (5217,1)    & Input: latent code & (2,1)       \\
1st (Dense)              & (1500,1)     & 6st (Dense)         & (20,1)       \\
2nd (Dense)              & (500,1)     & 7st (Dense)         & (100,1)      \\
3nd (Dense)              & (100,1)      & 8st (Dense)         & (500,1)     \\
4nd (Dense)              & (20,1)       & 9st(Dense)         & (1500,1)     \\
5nd (Dense): latent code & (2,1)       & 10st(Dense)        & (5217,1)    \\
\hline
\end{tabular}
\caption{Flow past two cylinders: stacked autoencoder structure.}
\end{table}


\begin{figure}[htbp]
\centering 
\begin{minipage}{1.0 \linewidth}
\includegraphics[width = \linewidth,angle=0,clip=true]{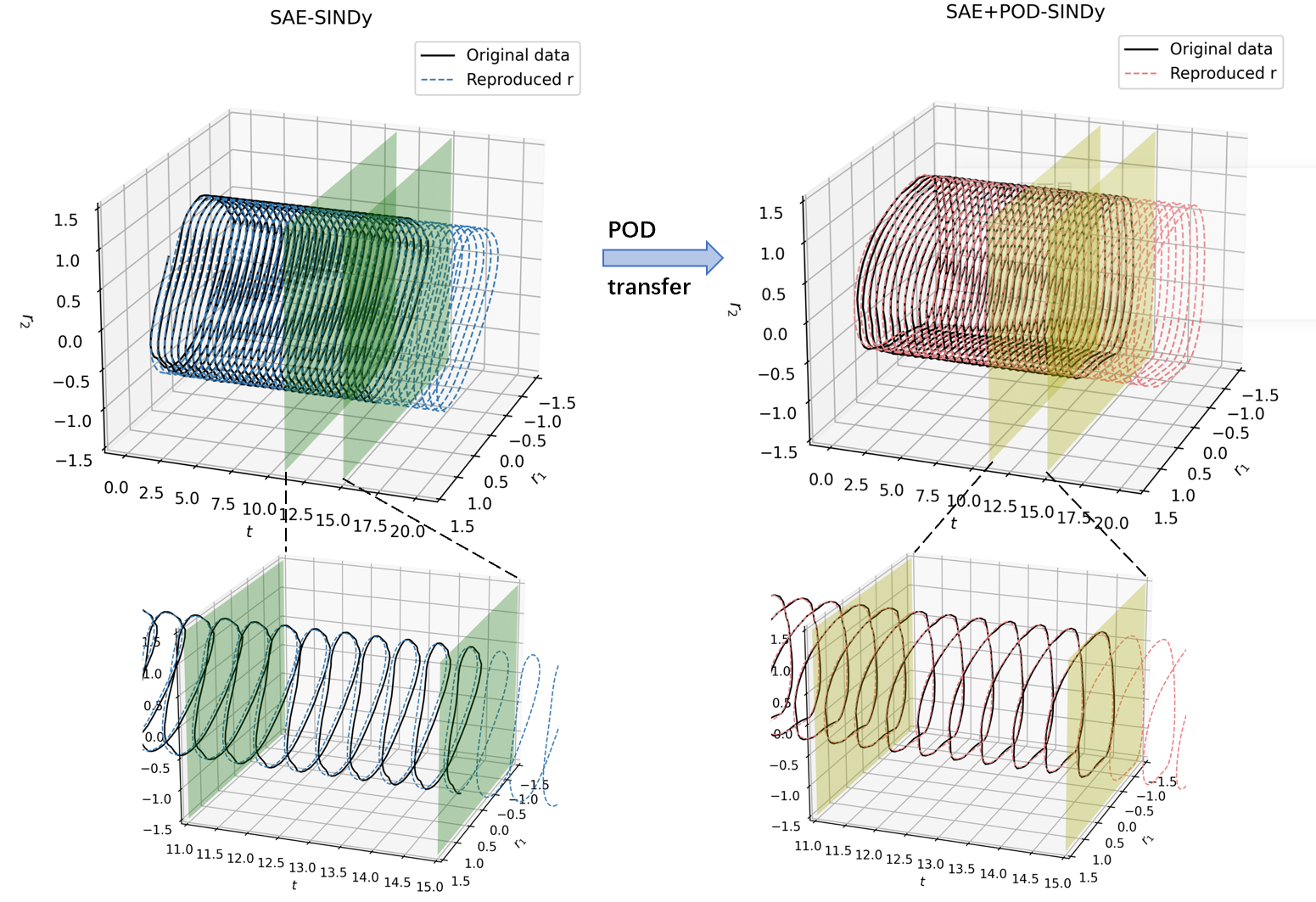} 
\end{minipage}
\caption{\textbf{Flow past two cylinders case}.
\textbf{a} shows the images after dimension reduction to a two-dimensional space using the SAE encoder.
\textbf{b} represents the results obtained from \textbf{a} after applying the POD transformation.
\textbf{c} and \textbf{d} correspond to the outcomes when utilising SINDy algorithm to solve governing equations \ref{equ:2c sae sindy} and \ref{equ:2c sp sindy}, respectively.
Both SAE-SINDy and SAE+POD-SINDy are applied to all training data. }
\label{fig: 2c sindy}
\end{figure}

\begin{figure}
\centering
\begin{tabular}{ccc}
\begin{minipage}{0.25\linewidth}
\includegraphics[width = \linewidth,angle=0,clip=true]{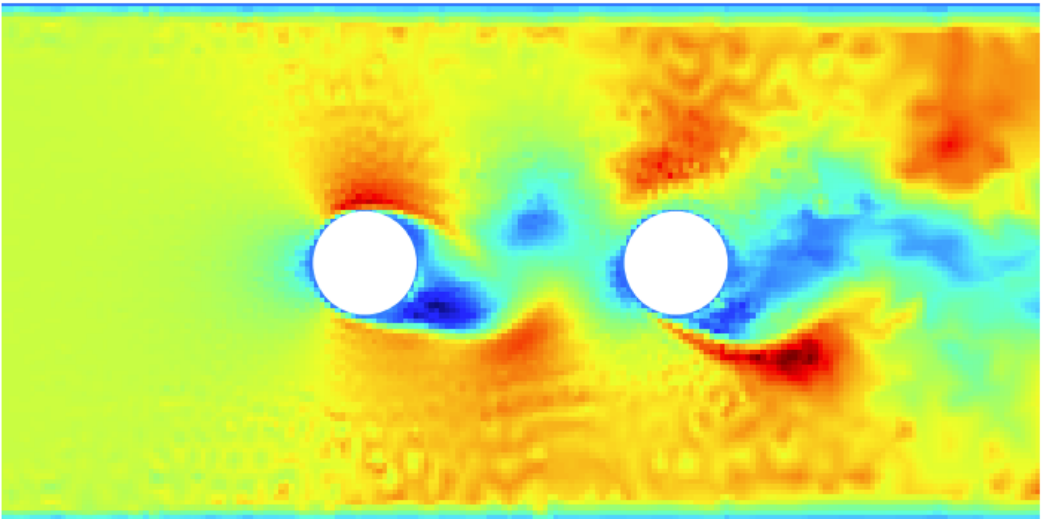}  
\end{minipage}
&
\begin{minipage}{0.25\linewidth}
\includegraphics[width = \linewidth,angle=0,clip=true]{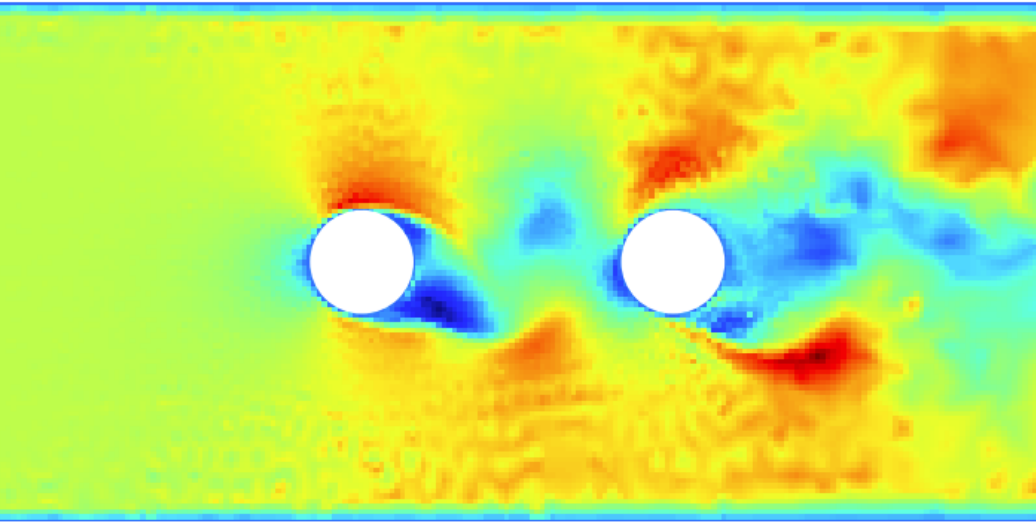}  
\end{minipage} 
&
\begin{minipage}{0.25\linewidth}
\includegraphics[width = \linewidth,angle=0,clip=true]{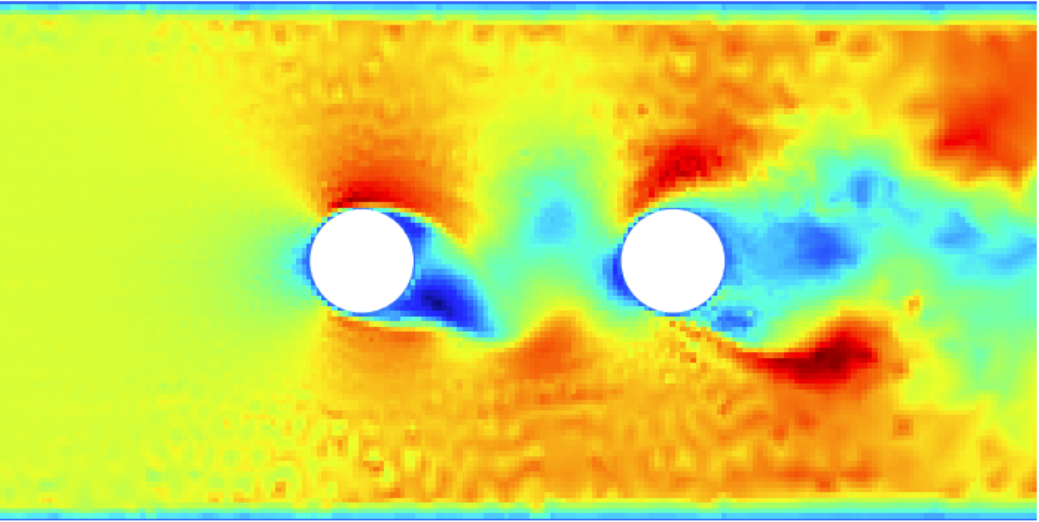}  
\end{minipage} 
 \\
(a-1) {\small Full Model, t = 3s}&
(a-2) {\small SAE, t=5=3s}&
(a-3) {\small SAE+POD, t=3s}
\\

\begin{minipage}{0.25\linewidth}
\includegraphics[width = \linewidth,angle=0,clip=true]{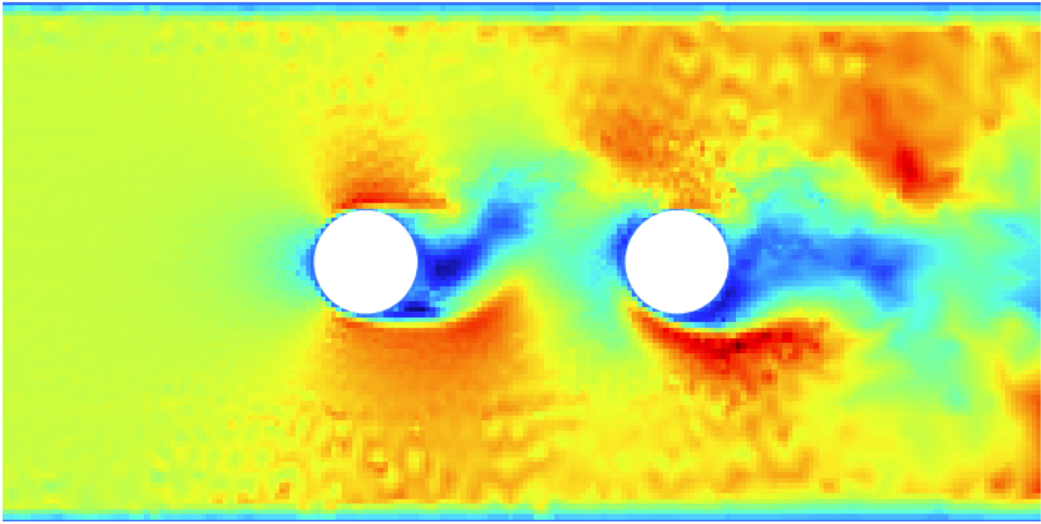}  
\end{minipage}
&
\begin{minipage}{0.25\linewidth}
\includegraphics[width = \linewidth,angle=0,clip=true]{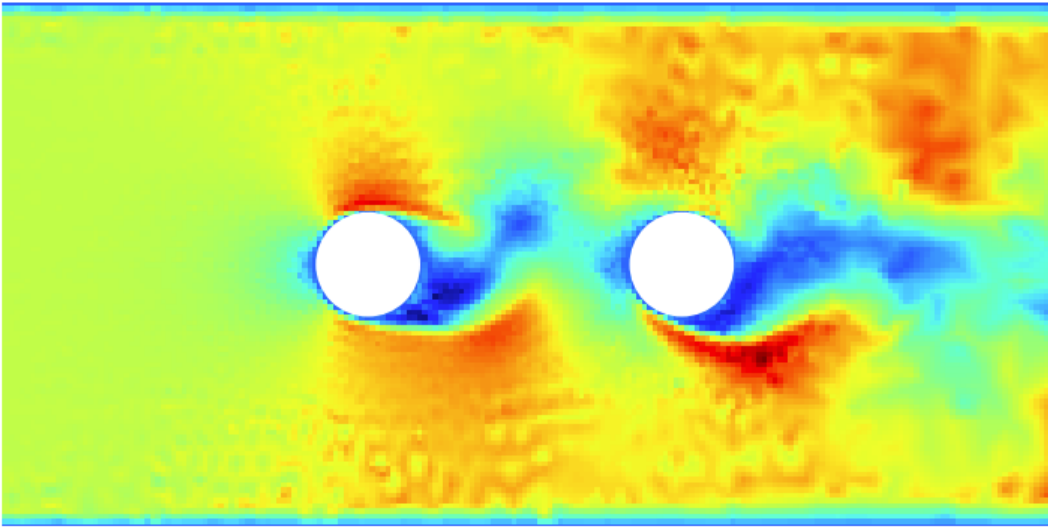}  
\end{minipage} 
&
\begin{minipage}{0.25\linewidth}
\includegraphics[width = \linewidth,angle=0,clip=true]{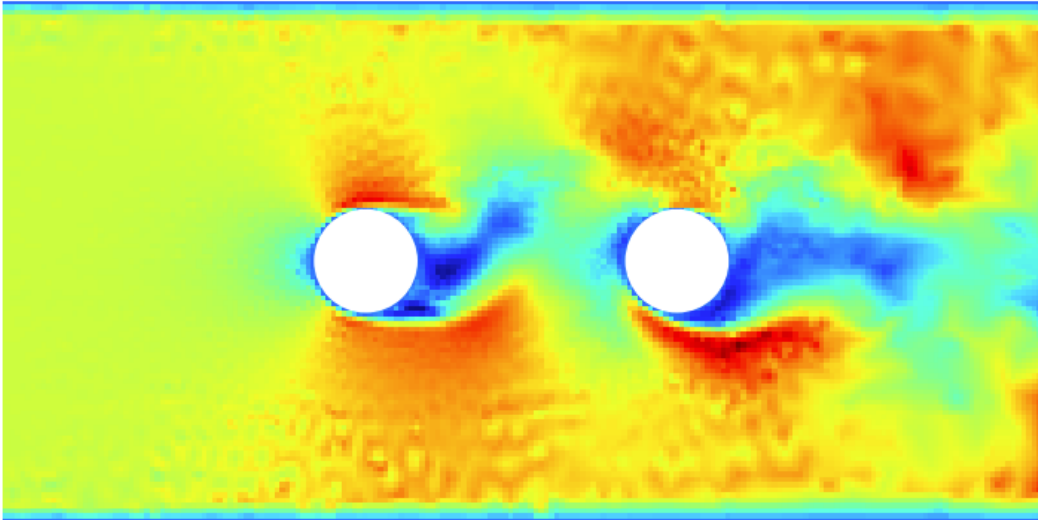}  
\end{minipage} 
 \\
(b-1) {\small Full Model, t = 8s}&
(b-2) {\small SAE, t = 8s}&
(b-3) {\small SAE+POD, t = 8s}
\\

\begin{minipage}{0.25\linewidth}
\includegraphics[width = \linewidth,angle=0,clip=true]{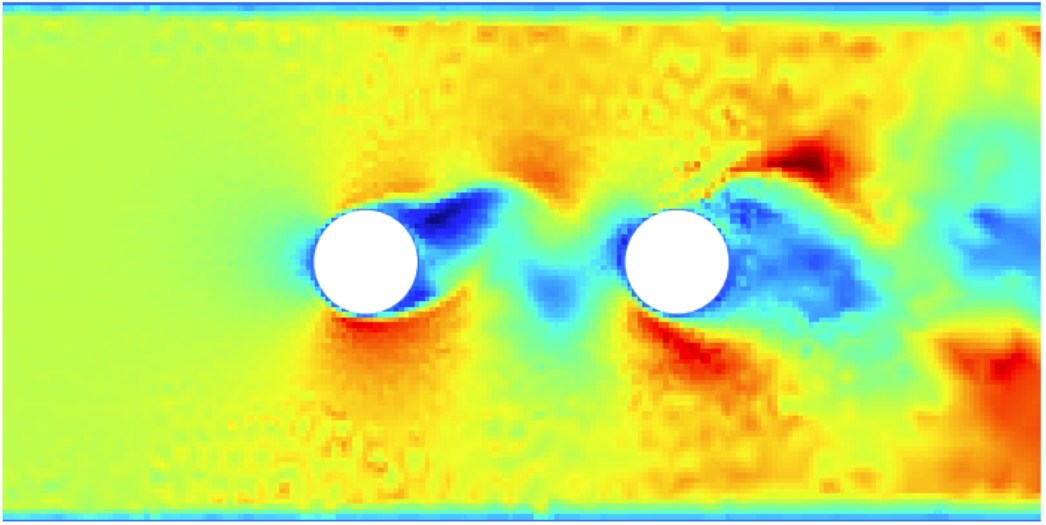}  
\end{minipage}
&
\begin{minipage}{0.25\linewidth}
\includegraphics[width = \linewidth,angle=0,clip=true]{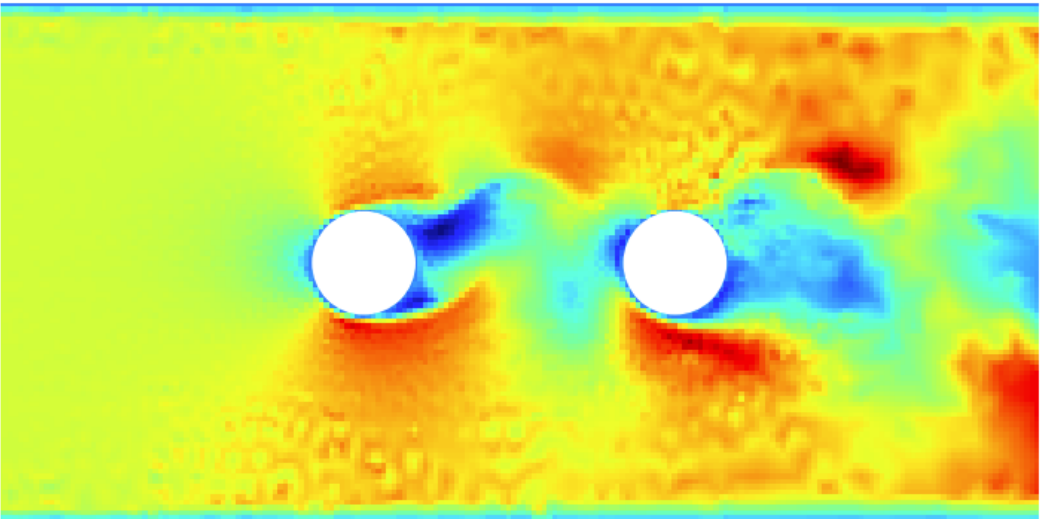}  
\end{minipage} 
&
\begin{minipage}{0.25\linewidth}
\includegraphics[width = \linewidth,angle=0,clip=true]{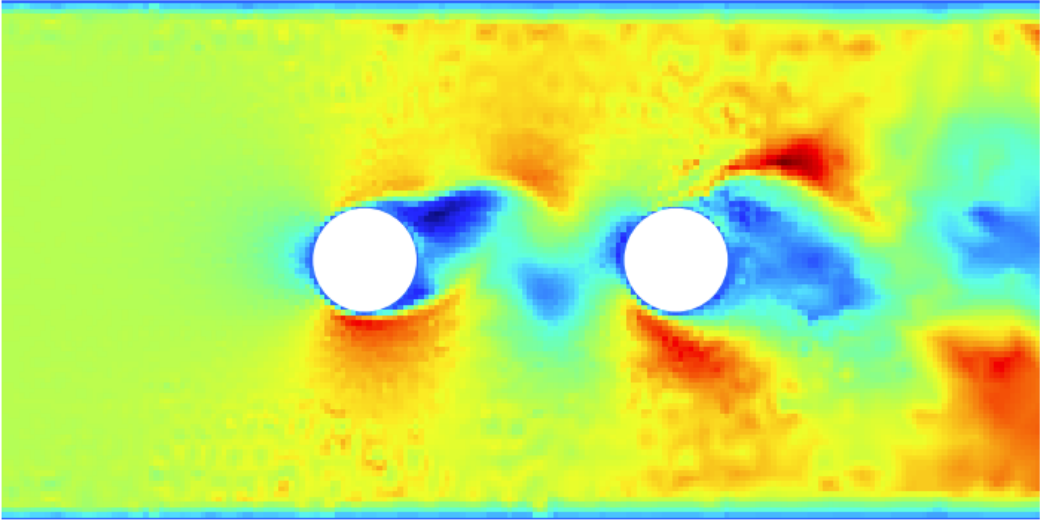}  
\end{minipage} 
 \\
(c-1) {\small Full Model, t = 14s}&
(c-2) {\small SAE, t = 14s}&
(c-3) {\small SAE+POD, t = 14s}
\\

\begin{minipage}{0.25\linewidth}
\includegraphics[width = \linewidth,angle=0,clip=true]{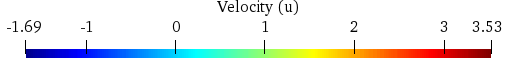} 
\end{minipage}
&
\begin{minipage}{0.25\linewidth}
\includegraphics[width = \linewidth,angle=0,clip=true]{results/2cylinder/2c_grid2.png} 
\end{minipage}
&
\begin{minipage}{0.25\linewidth}
\includegraphics[width = \linewidth,angle=0,clip=true]{results/2cylinder/2c_grid2.png} 
\end{minipage}
 \\
\\
\end{tabular}
\caption{\textbf{Flow past two cylinders}. The velocity($u$) solutions of full-model, SAE and SAE+POD}
\label{fig:2c solution}
\end{figure}
\begin{figure}
\centering
\begin{tabular}{cc}
\begin{minipage}{0.25\linewidth}
\includegraphics[width = \linewidth,angle=0,clip=true]{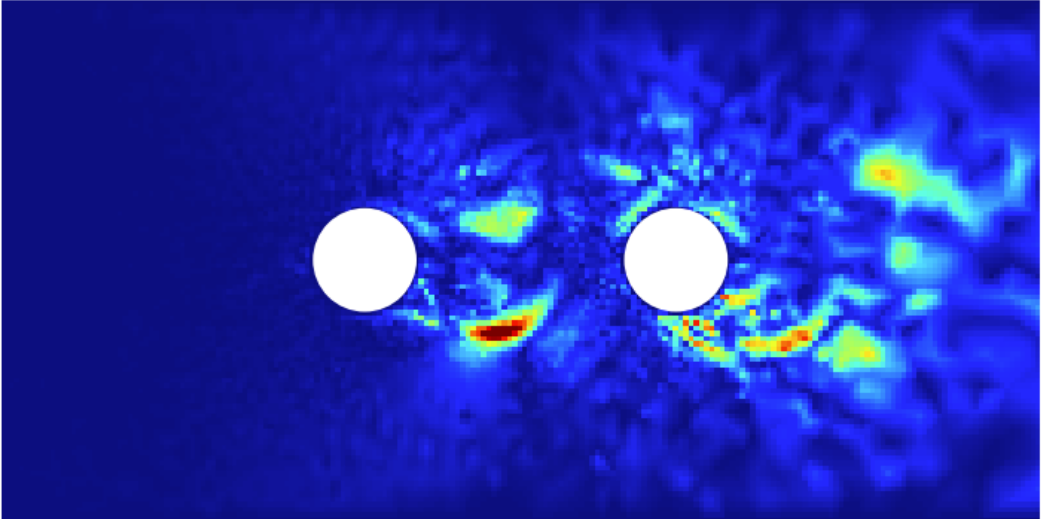}  
\end{minipage} 
&
\begin{minipage}{0.25\linewidth}
\includegraphics[width = \linewidth,angle=0,clip=true]{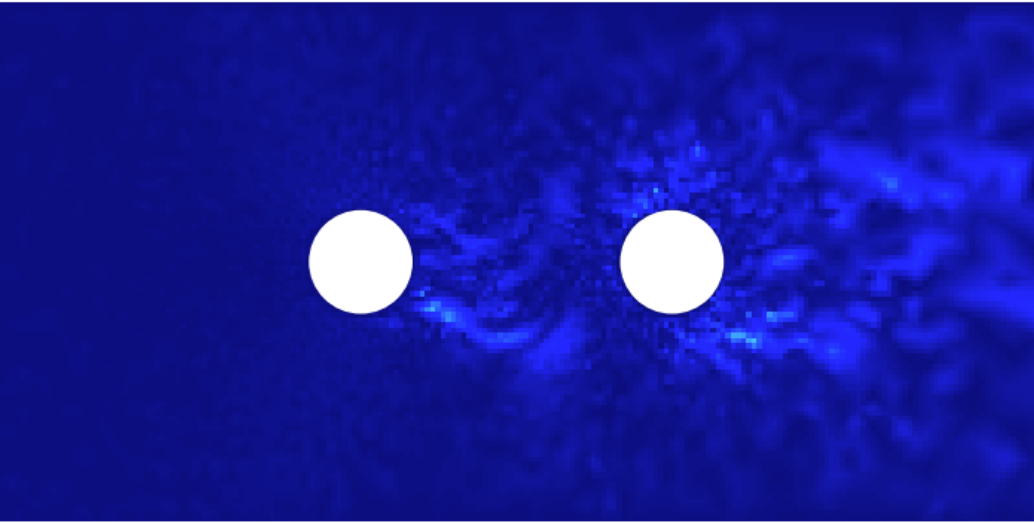}  
\end{minipage} 
 \\
(a-4) {\small SAE error, t=3s}&
(a-5) {\small SAE+POD error, t=3s}
\\

\begin{minipage}{0.25\linewidth}
\includegraphics[width = \linewidth,angle=0,clip=true]{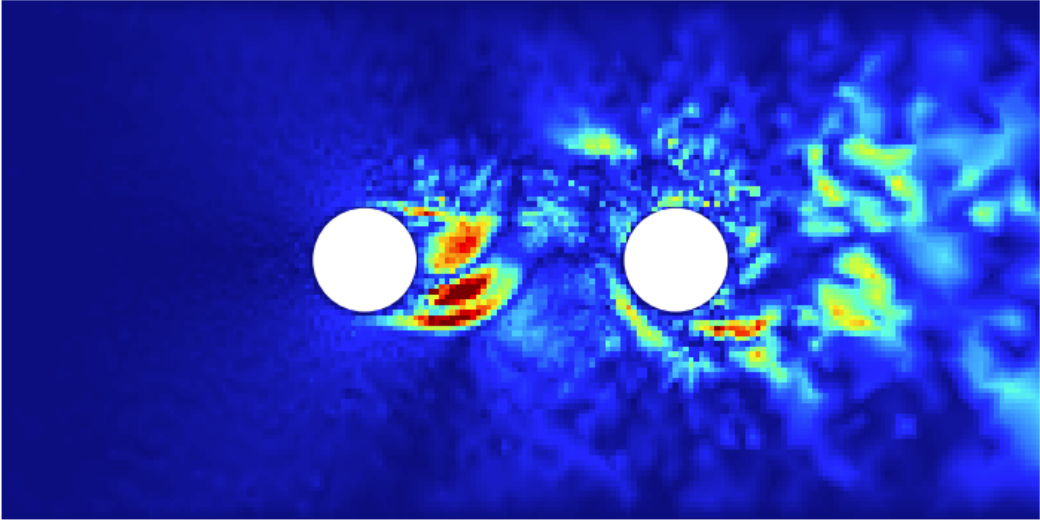}  
\end{minipage} 
&
\begin{minipage}{0.25\linewidth}
\includegraphics[width = \linewidth,angle=0,clip=true]{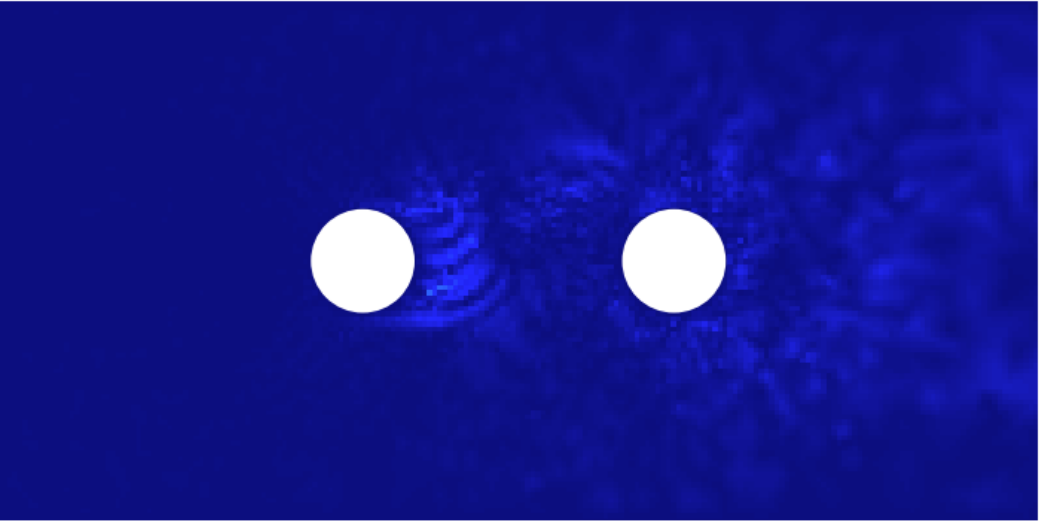}  
\end{minipage} 
 \\
(b-4) {\small SAE error, t=8s}&
(b-5) {\small SAE+POD error, t=8s}
\\

\begin{minipage}{0.25\linewidth}
\includegraphics[width = \linewidth,angle=0,clip=true]{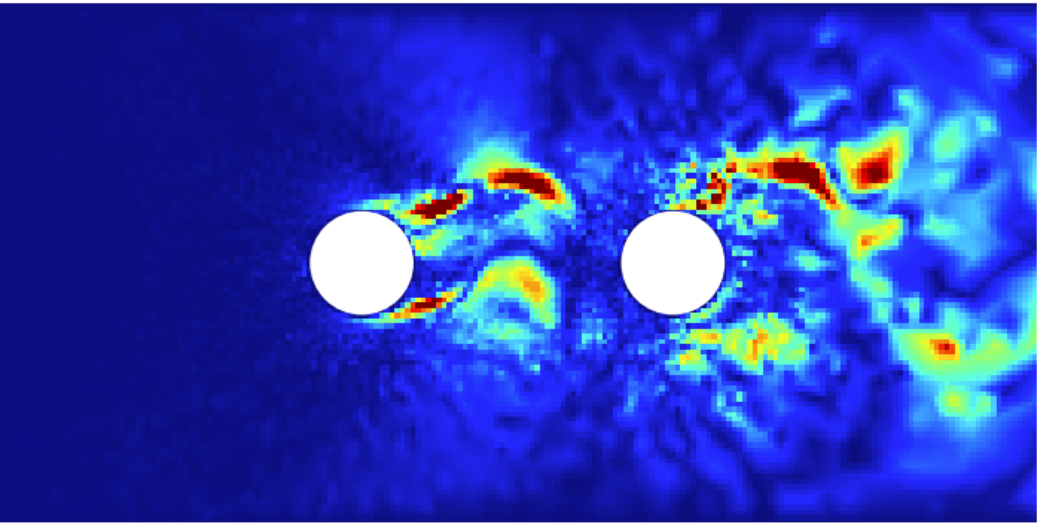}  
\end{minipage} 
&
\begin{minipage}{0.25\linewidth}
\includegraphics[width = \linewidth,angle=0,clip=true]{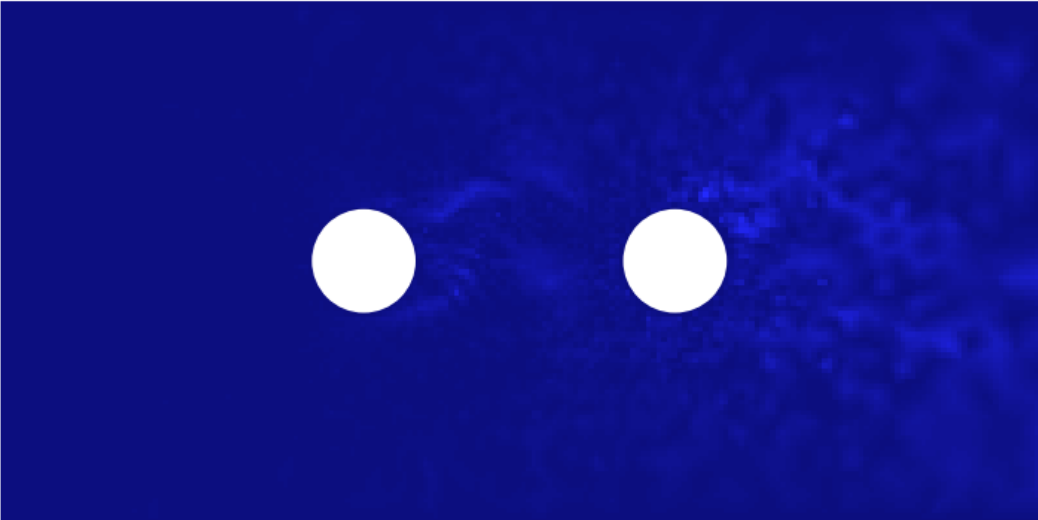}  
\end{minipage} 
 \\
(c-4) {\small SAE error, t=14s}&
(c-5) {\small SAE+POD error, t=14s}
\\

\begin{minipage}{0.25\linewidth}
\includegraphics[width = \linewidth,angle=0,clip=true]{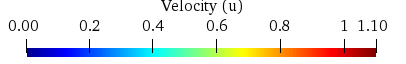} 
\end{minipage}
&
\begin{minipage}{0.25\linewidth}
\includegraphics[width = \linewidth,angle=0,clip=true]{results/2cylinder/2c_grid_error.png} 
\end{minipage}
 \\
\\
\end{tabular}
\caption{\textbf{Flow past two cylinders}. The velocity($u$) solution errors of SAE and SAE+POD.}
\label{fig:2c error}
\end{figure}
\begin{figure}[ht]
\centering
\begin{tabular}{c}
\begin{minipage}{0.5 \linewidth}
\includegraphics[width = \linewidth,angle=0,clip=true]{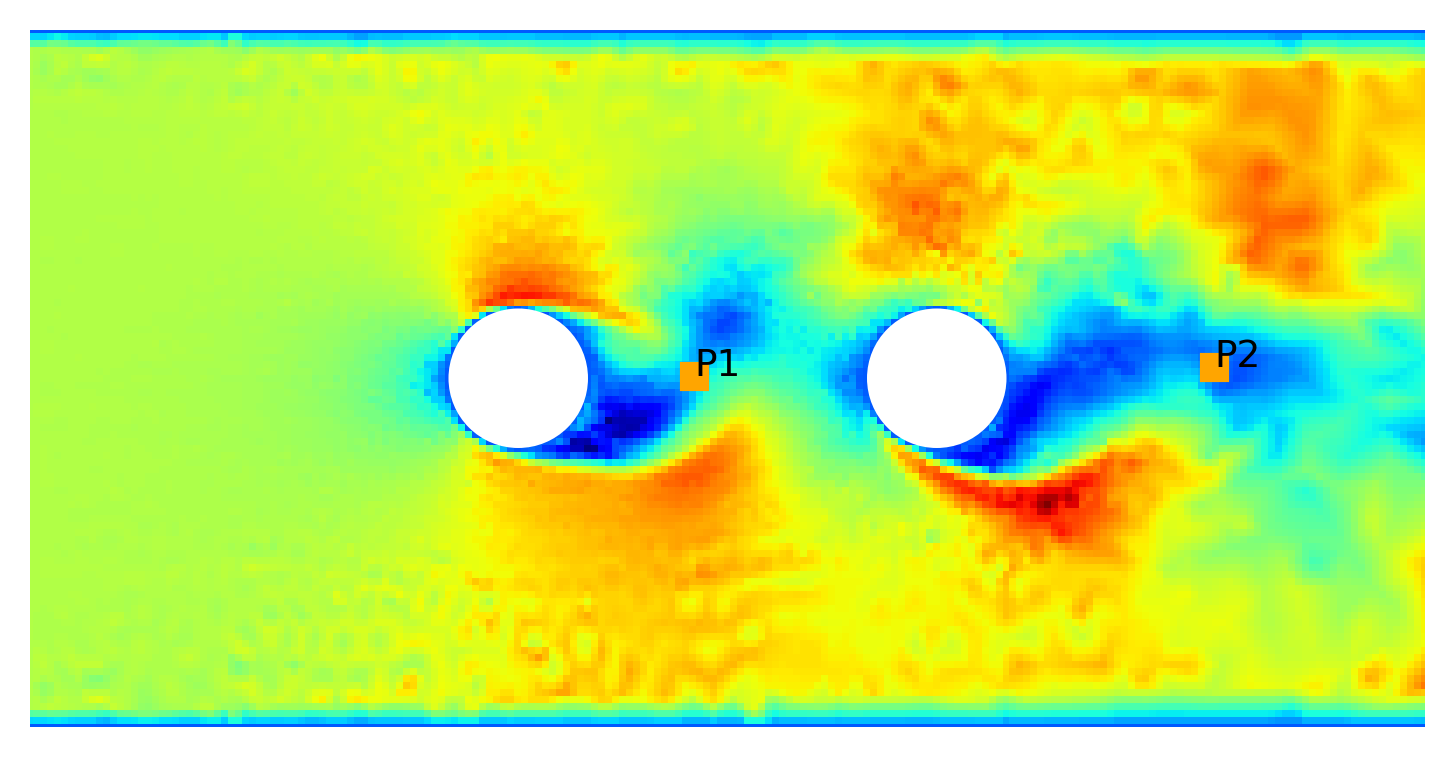} 
\end{minipage}
\\
\\
\begin{minipage}{0.5\linewidth}
\includegraphics[width = \linewidth,angle=0,clip=true]{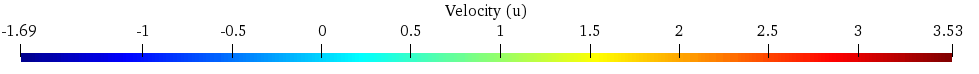} 
\end{minipage}
 \\
\\
\end{tabular}
\caption{\textbf{Flow past 2 cylinder}: locations of P1 and P2.}
\label{fig:2c p1p2}
\end{figure}
\begin{figure}[ht]
\centering
\begin{tabular}{cc}
\\
\textbf{a} &
\textbf{b}
\\
\begin{minipage}{0.5 \linewidth}
\includegraphics[width = \linewidth,angle=0,clip=true]{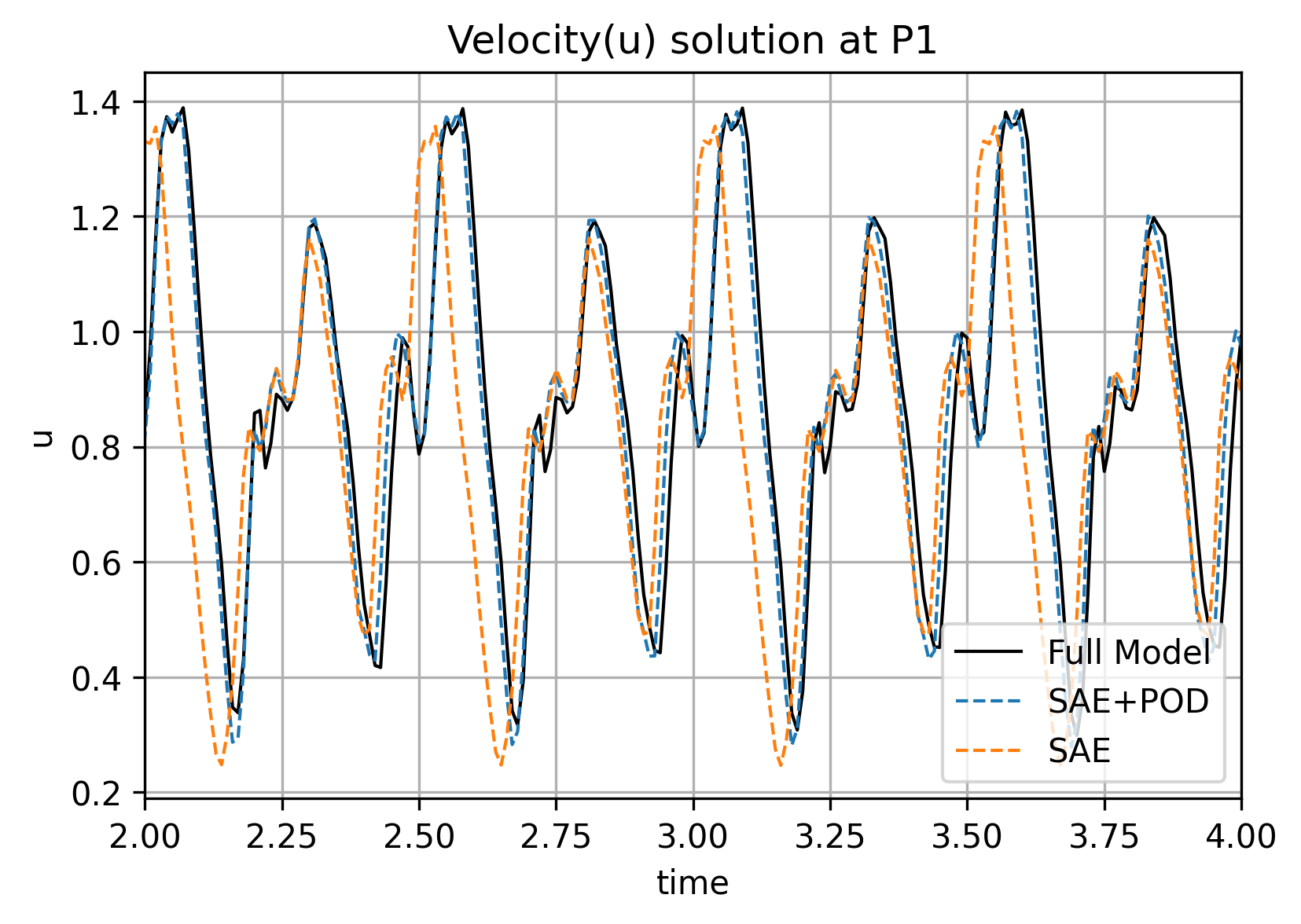} 
\end{minipage}
&
\begin{minipage}{0.5\linewidth}
\includegraphics[width = \linewidth,angle=0,clip=true]{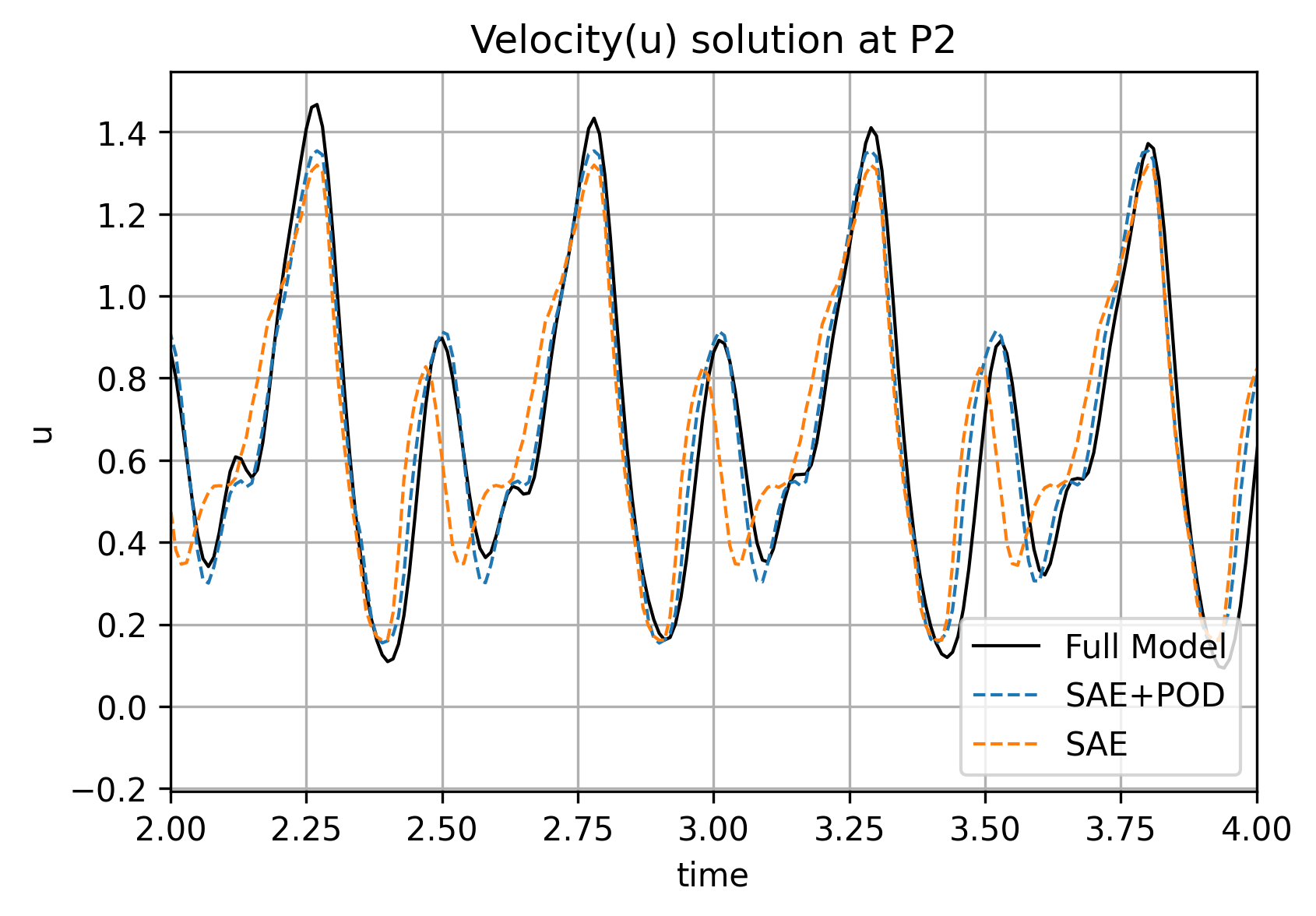} 
\end{minipage}
\end{tabular}
\caption{\textbf{Flow past two cylinders}: Velocity comparisons between full model, SAE and SAE+POD at two particular points: P1 and P2 which are located in Figure \ref{fig:2c p1p2}.}
\label{fig2cp1p2}
\end{figure}

In this example, 2 codes in the manifold space are used, which represent 97.78\% of the total energy. 
Figure \ref{fig: 2c sindy} shows the results in the manifold space which are obtained by solving equations \ref{equ:2c sae sindy} and \ref{equ:2c sp sindy}. 
In order to describe accurate fluid dynamics, the candidate functions are used up to third order in the SINDy. Equation \ref{equ:2c sae sindy} and Equation \ref{equ:2c sp sindy} are governing equations for this problem in the manifold space using SAE+SINDy and SAE+POD+SINDy respectively.  Figure \ref{fig: 2c sindy} compares the solutions in the manifold space of the two methods. 
\begin{equation}
    \begin{aligned}
    \frac{d r_1}{dt} =& -2.664+1.693r_2+11.01r_2^{3}+17.81r_1-9.23r_1r_2-10.99r_1r_2^{2}+2.81r_1^{2}+11.14r_1^{2}r_2-12.81r_1^{3}  \\
    \frac{d r_2}{dt} =& -4.67r_2+4.254r_2^{2}-10.14r_1+3.94r_1r_2-10.69r_1r_2^{2}-11.99r_1^{2}r_2-6.836r_1^{3}  \\
    \end{aligned}
    \label{equ:2c sae sindy}
\end{equation}

\begin{equation}
    \begin{aligned}
    \frac{d r'_1}{dt} =& -11.14r'_2-2.861{r'}_2^{2}-4.293{r'}_2^{3}+7.91r'_1-5.105r'_1{r'}_2-4.246r'_1{r'}_2{2}+1.271{r'}_1^{2}r'_2-8.75{r'}_1^{3}  \\
    \frac{d r'_2}{dt} =& +3.396+7.14r'_2-7.016{r'}_2^{3}-6.664r'_1-1.26r'_1r'_2+9.375r'_1{r'}_2^{2}-2.115{r'}_1^{2}-4.24{r'}_1^{2}r'_2+20.0{r'}_1^{3}  \\
    \end{aligned}
    \label{equ:2c sp sindy}
\end{equation}
Figure \ref{fig:2c solution} shows the velocity comparisons between the high fidelity full model, SAE+SINDy and SAE+POD+SINDy at time levels: t=3s, t=8s, and t=14s. Figure \ref{fig:2c error} shows the errors of SAE+SINDy and SAE+POD+SINDy at time levels: t=3s, t=8s, and t=14s. As shown in the figure, the solutions are more accurate if POD is used to stabilise the SAE values. In order to see clearly the differences, the solutions at two particular points $P_1$ and $P_2$ are given in the Figure \ref{fig2cp1p2}. The locations of two points $P_1$ and $P_2$ are shown in Figure \ref{fig:2c p1p2}. 
 

\section{Discussion}\label{sec:conclusion}
We have presented an approach that is capable of discovering governing equations in the manifold minimal dimensional space from high-dimensional data. In addition, this method has been used to discover governing equations of turbulent flows in the manifold space. Our method firstly projects the high dimensional dynamics into a lower dimensional manifold space, and then stabilised it. After that, it discovers the reduced version of governing equations that describe the dynamics on that stabilised low dimensional manifold. The projection is achieved via autoencoder neural network and the stabilisation is conducted via POD. The SINDy is used to discover the governing equations in the manifold space. 

In our examples, we demonstrated that our method is able to discover reduced version of governing equations for different test cases such as lock exchange, flow past one and two cylinders with unsteady vortex shedding. In addition, our method is outperforms the SAE+SINDy method. We speculate that this is because the POD stabilised the SAE values in the manifold space. The distribution of hidden patterns or dynamics in the SAE latent space is hard to be discovered via SINDy. This becomes easier if POD is performed to the SAE latent space. 

Stacked AE has strong capabilities on feature representation, compared to AE. This means that SAE is able to   extract more complex information or features and thus enhancing the model's robustness and generalisation. 

However, SAE, being a fully connected network, does not take into account coordinate information. In practical experiments, instability issues often arise in the encoding process, potentially leading to errors in SINDy's training outcomes and subsequently hindering the accurate generation of high-dimensional manifolds during online processes. Thus, as a dimensionality reduction method, the equidimensional POD transformation can enhance model discovery stability, thereby strengthening the performance of SINDy.

A question that arises in the method is: how many dimensional size in the manifold space must be used?  Unlike the traditional dimensionality reduction method such as POD, there is no formula or mathematical expression to calculate the energy captured. The minimal number of dimensional size in the reduced manifold space is similar to the problem raised by the Lusternik–Schnirelmann
category of the manifold $\mathcal{M}$. Unfortunately, this is not realistically to be computed \cite{cornea2003lusternik, floryan2022data1}. However, POD formulation of calculating the captured energy can still be a reference for choosing the number of codes in the autoencoder network as autoender needs much less number of dimansionality in the reduced manifold space. We can choose less number of codes than POD \cite{fu2023}. In addition, we can select dimensional size of two in the manifold latent space, which is an empirical approach as two codes exhibit good performance for SINDy discovery. Also, larger number of nodes in the autoencoder network can be used to capture the fluid dynamics if smaller number of codes are used. 



In addition to discover and predict a simplified dynamics of a system, our method is also able to unveil hidden behaviour or properties of a system. For complicated problems which the dynamics are difficult to be modelled, our method is able to project it into a reduced space, and  
hidden properties of a system are not obvious, the intrinsic variables
provided by our method would need to be translated to interpretable
physics, and we suggest this as a future research direction.

Finally, as our approach discovers reduced version of governing equations, it is well-suited for other dynamic systems and problems without explicit governing equations in the full high dimensional space.

\section{Methods} \label{sec:method}
 The innovations presented in this work include three key ideas: (i) Finding the low-dimensional representation of turbulent flows, (ii) Selecting a number of terms in the libraries of candidate functions to represent the fluid dynamics in the manifold space. (iii) Optimising the sparse coefficients of selected terms.

\subsection{Obtaining low dimensional manifolds with stabilisation}\label{sec:sae}

Stacked autoencoder (SAE) and proper orthogonal decomposition (POD) are combined and used for dimensionality reduction. 
In our framework, suppose the turbulence flow can be expressed as $\bm{u} \in \mathbb{R}^{N_x}$ and vectors in low dimensional space can be wrote as $\bm{r} \in \mathbb{R}^{N_r}$.
Consider a function to map coordinates onto manifolds as $\bm{u}(\bm{x};t): \Omega \times \mathbb{T} \rightarrow \mathbb{R}^{N_x}$, where $\mathbb{T}=[0,T]$ and $\Omega$ denote for time and the whole flow computational fields, respectively. $\bm{u}$ usually refers to properties of fluid dynamics, including velocity, pressure, vorticity, etc. of the fluid.


As a type of unsupervised feedforward neural network, the traditional autoencoder (AE) intends for reducing the dimensional size and identifies latent mapping. This is achieved via the encoder $h_{enc}$ and the decoder $h_{dec}$.
SAE is a combination of several (assuming $k$) autoencoders, which are stacked in the hidden layers and learned layer-wise. The encoder $g_{enc}$ and decoder $g_{dec}$ of SAE has a form of
\begin{equation}
    g_{enc}: \bm{u}  \rightarrow h_{enc}^{(k)}  \circ \dots \circ h_{enc}^{(1)} (\bm{u})=\bm{r},
\end{equation}
\begin{equation}
    g_{dec}: \bm{r}  \rightarrow h_{dec}^{(k)}  \circ \dots \circ h_{dec}^{(1)} (\bm{r})=\tilde{\bm{u}},
\end{equation}
where $h_{enc}^{(i)}:\mathbb{R}^{d_{i-1}} \rightarrow \mathbb{R}^{d_i},h_{dec}^{(i)}:\mathbb{R}^{d_{i-1}} \rightarrow \mathbb{R}^{d_i}, i=1,\dots,k$ denotes the mapping relation represented by the $i$-th hidden layer. The loss function of SAE can be expressed as
\begin{equation}
    \mathcal{L}_{SAE}=\min \frac{1}{k} \sum_{i=1}^k \mathcal{L}_{AE}^{(i)},
\end{equation}
where $\mathcal{L}_{AE}^{(i)}$ is defined by $L_2$ error and the rectified linear units activation function (ReLU) is used as the activation function.
Due to the inherent randomness in the training of SAE, the generated low-dimensional space may exhibit some degree of instability. This variability could potentially impact the performance of SINDy. Thus, the POD is introduced to improve the stabilisation of latent codes to obtain better performance of SINDy algorithm. The POD plays a role to enhance the smoothness of low-dimensional manifolds and facilitate the searching for approximate polynomial estimations. It is performed to SAE before using SINDy to search the governing equations in the manifold space. 

The purpose of POD is to compute the reduced bases of the latent flow $\bm{r}$ and project the low-dimensional dynamics into these bases, which should minimise the kneric energy $E(\cdot)$ of the flow field as
\begin{equation}
    \arg \min_{\{r_i,\phi_i\}} E(\bm{r}(\bm{x};t)-\sum_{i=1}^{d} r_i(t)\phi_i(t)),
\end{equation}
where $\{r'_i(t) , r'_i:[0,T]\rightarrow \mathbb{R}\}_{i=1}^{d}$ stands for POD coefficients and $\{\phi_i(t) , \phi_i: \Omega \rightarrow \mathbb{R}^{N_r} \}_{i=1}^{d}$. Here, $d$ is equal to $N_r$ and $\Omega := \mathbb{R}^{N_x}$. The $r(t_i)$ is defined as $r_i$, $i=1,2,\dots,N_t$.
When $d=N_r$, POD costs nearly no loss of energy, which is proved in Supplementary materials.

\subsection{Building function libraries and sparse regression}\label{sec:SINDy}
The state of a fluid dynamical system in the manifold space at different times can be denoted as $\bm{x}(t)=[x_1(t),x_2(t),\dots,x_n(t)]\in R^n$, where $t$ represents the time and $n$ is the dimensionality of input features. $\bm{x}(t)$ is represented in the form of the following system of differential equations:
\begin{equation}
    \dot{\bm{x}}(t) = f(\bm{x}(t)),
    \label{equ:xt=fxt}
\end{equation}
in which $f$ represents the fluid dynamic relationships in the manifold space. The primary training target of SINDy is to discover a $f$ considering both of the sparsity and accuracy. 
The system states $\bm{x}(t)$ and the time derivatives $\dot{\bm{x}}(t)$ are collected through second-order central difference scheme.
Subsequently, to solve the equation (\ref{equ:xt=fxt}), SINDy library $\Theta(\bm{X})$ is constructed to store the candidate functions associated with the state $\bm{x}(t)$. The library $\Theta:=\Theta(\bm{X})$ can be wrote as
\begin{equation}
    \Theta = 
    \begin{pmatrix}
    |  &  |  &  |  &  |  &  \cdots  &  |  &  |  &  \cdots  \\
    1  & \bm{X} & \bm{X}^P_2 & \bm{X}^P_3 &  \cdots  & \sin(\bm{X}) & \cos(\bm{X}) & \cdots \\
    |  &  |  &  |  &  |  &  \cdots  &  |  &  |  &  \cdots,
    \end{pmatrix}  
\end{equation}
where $\bm{X}^P_n$ stands for the n-order polynomials and $\sin(\bm{X}), \cos(\bm{X})$ stand for trigonometric terms, generated by $\bm{x}(t)$. Notice that $\Theta$ is data dependent. 
In SINDy, the function $f$ can be discovered via obtaining sparse coefficient matrix $\Xi=[\xi_1,\xi_2,\dots,\xi_p]$ and libraries with nonlinear terms:
\begin{equation}\label{sindy}
    \dot{\bm{X}}=\Theta(\bm{X})\Xi.
\end{equation}

\subsection{Adaptive least absolute shrinkage and selection operator regression}
Instead of using sequential thresholed least-squares (STLS) to determine the sparse coefficient matrix $\Xi$ in Equation \ref{sindy} proposed by Brunton et al. in \cite{2016SINDy}, 
we transform this problem into a optimisation problem and utilise adaptive least absolute shrinkage and selection operator\cite{zou2006Alasso, zou2008one} (Adalasso). It is a more widely used sparse regression algorithm compared to Lasso regression\cite{tibshirani1996lasso}. The loss function has the following form:
\begin{equation}\label{paralamda}
    \Xi = \arg\min ||\Theta \Xi-\dot{\bm{X}}||^2_2+\lambda\sum_{j=1}^p w_j|\xi_j|,
\end{equation}
in which $w_j=(|\xi_j|)^{-\delta}$ indicates the adaptive weight in loss function, and $\delta>0$.
It is clearly that the weights of $\xi_j$ keep changing with iterations of SINDy and employ a  continuous and smooth penalty mechanism, which is different from STLS in the original SINDy and 
is capable of handling more complex situations with greater precision.

Adalasso uses $L_1$ norm to penalise coefficient in the problem to avoid overfitting and achieve sparsity, which has more accurate results than STLS algorithm.
In the iterative training procedure, the parameter $\lambda$ is continually optimised, and the coefficient matrix is repeatedly updated until the error and the parsimony of equations reach our expectation.
To improve computational speed, least angle regression (LARS) algorithm\cite{efron2004least} is employed to accelerate the calculations of Adalasso.
The computational cost of the algorithm is $O(mp^2)$.
To alleviate the influence of features with disparate magnitudes on the computation results, we normalise the latent vectors before incorporating them into SINDy to ensure that the features are brought to the same scale.

\section{Data availability}
All simulation data are generated via \textit{Fluidity}, \href{http://fluidityproject.github.io/}{http://fluidityproject.github.io/}, which is an open source CFD model and is able to numerically solve Navier-Stokes equations. The data can be found at \href{https://github.com/CFD-NIROM/SAE-POD-SINDy}{https://github.com/CFD-NIROM/SAE-POD-SINDy}.

\section{Code availability}
The code is available at 
\href{https://github.com/CFD-NIROM/SAE-POD-SINDy}{https://github.com/CFD-NIROM/SAE-POD-SINDy}.

\vspace{-6pt}

\section*{Acknowledgments}
\noindent 
The authors would like to acknowledge the support of the Fundamental Research Funds for the Central Universities.  

\clearpage
\bibliographystyle{unsrt} 
\bibliography{bibliography}

\end{document}


\begin{frontmatter}
\renewcommand{\thefootnote}{\fnsymbol{footnotemark}}

\fancypagestyle{plain}{%
\fancyhf{} 
\fancyhead[RO,RE]{\thepage} 
}

\title{\textbf{Supplementary Information}
\\ 
Data discovery of low dimensional fluid dynamics of turbulent flows}
    \author[lab1,lab2]{X. Lin}
    \author[lab1,lab2]{D. Xiao\corref{cor1}} 
    \cortext[cor1]{Corresponding author}
    \ead{xiaodunhui@tongji.edu.cn}   
    \author[lab3]{F. Fang}
    \address[lab1]{School of Mathematical Sciences, Tongji University, Shanghai, P.R. China,200092}
    \address[lab2]{Key Laboratory of Intelligent Computing and Applications(Tongji University), Ministry of Education, China}
  \address[lab3]{Department of Earth Science and Engineering, Imperial College London, UK, SW7 2BP}  
 \end{frontmatter}

 \begin{table}[]
\centering
\begin{tabular}{cc}
   \toprule
   Notation & Description \\
   \midrule
   $\bm{u}$  &  The state variables of dynamics \\
   $t$  &  Time evalution of state variables and latent codes \\
   $\bm{x}$  & The coordinates of the high-dimensional computational domain \\
   $\mathcal{D}$  &  The whole computational domain, $\bm{x}\in \mathcal{D}$\\
   $\tilde{\bm{u}}$  & The approximation solution of $\bm{u}$ by SINDy-POD or other algorithms\\
   $\bm{u}_{ref}$  &  The reference state of $\bm{u}$\\
   $g_{enc}, g_{dec}$  &  The decoder and encoder of stacked autoencoders \\
   $\mathcal{P}$  & The principal component analysis transformation functions \\
   $r_i$  & The $i$-th vector of latent codes, in which $i=\dots,d$ \\
   $\mathcal{S}$  & The space consisting of the high-dimensional solution of latent codes decoded by an autoencoder decoder, \\
   & which means $\mathcal{S} = \{g_{dec}(\bm{r})|\bm{r}\in \mathbb{R}^d\}$ \\
   
   \bottomrule
\end{tabular}
\end{table}

\section{Methods} \label{sec:method}
The innovations presented in this work include three key ideas: (i) finding the low-dimensional representation of turbulent flows, (ii) selecting a number of terms in the libraries of candidate functions to represent the fluid dynamics in the manifold space. (iii) optimising the sparse coefficients of selected terms.


The original high-dimensional state space is defined as $\mathbb{R}^{N}$ and low-dimensional manifold space is $\mathbb{R}^d$. The spaces are finite dimensional due to the discrete data. Our purpose is to seek an approximate solution $\tilde{\bm{u}}$ of $\bm{u}$. Time evolution is $t \in \mathbb{R}^+$ and the set of coordinates of for computational domain is regarded as $\bm{x}\in \mathcal{D}$. $\mathcal{R}$ indicates the parameter space in the parametric PDEs.
$\tilde{\bm{u}}$ has the following form
\begin{equation}
    \bm{u}(t,\bm{x};\mu) \approx \tilde{\bm{u}}(t,\bm{x};\mu) = \bm{u}_{ref}(\bm{x};\mu) + g_{dec} (\mathcal{P}(r_1(t;\mu),\dots,r_d(t;\mu)))
\end{equation}
where $\tilde{\bm{u}}:\mathbb{R}^+ \times \mathcal{D} \times \mathcal{R} \rightarrow \bm{u}_{ref}(\bm{x};\mu) + \mathcal{S}$. $\bm{u}_{ref}(\bm{x};\mu)$ is a reference in high-dimensional space, usually generated by centralisation or normalisation and $\mathcal{S}$ is the variation space influencing the interpretation of ROM and is defined as 
\begin{equation}
    \mathcal{S}:=\{g_{dec}(\bm{r})|\bm{r}\in \mathbb{R}^d \}
\end{equation}
where $\bm{r}=(r_1,\dots,r_d):\mathbb{R}^+\times \mathcal{R} \rightarrow \mathbb{R}^d$ represents the latent generalised coordinates gained by $g_{enc}$ and reproduced by $g_{dec}$. $g_{enc}$ and $g_{dec}$ are trained in the stacked-autoencoder network structure with the ability to map between high and low dimensional spaces, and they are introduced in Section \ref{sec:sae}. 

It should be noted here that in order to obtain better interpretations in low dimensions, i.e., to obtain differential equations with both sparsity and accuracy, we have adopted the SINDy algorithm for the time derivative term of $\bm{r}$, which is explained in Section \ref{sec:SINDy}. In the offline process, a system of ODE is trained by SINDy and has a form of
\begin{equation}
    \frac{d\bm{r}}{dt} = f(\bm{r},t;\mu)
\end{equation}

\section{Autoencoder}\label{sec:sae}

The traditional autoencoder (AE) is a type of unsupervised feedforward neural network which intends for reducing the dimensional size and identify latent mapping. An autoencoder accomplishes its goals through two specific components: the encoder $h_{enc}$ and the decoder $h_{dec}$, such that:
\begin{equation}
    \begin{aligned}
        h_{enc}:\bm{u}\rightarrow \bm{r} \\
        h_{dec}:\bm{r}\rightarrow \tilde{\bm{u}} \\
    \end{aligned}
\end{equation}
The encoding progress can be generalised as follows:
\begin{equation}
    \bm{r}_i = h_{enc}(\bm{W}_e\bm{u}_i+\bm{b}_e)
\end{equation}
where $h_{enc}$ is the encoding progress, and $\bm{W}_e,\bm{b}_e$ stand for the weight and bias of encoder respectively.
In autoencoder network, taking the $i$-th AE an an example. The encoded result is $r_i$ and the input of autoencoder is $u_i$. Then the decoding progress can be generalised as follows:
\begin{equation}
    \widetilde{\bm{u}}_i = h_{dec}(\bm{W}_d\bm{r}_i+\bm{b}_d)
\end{equation}
where $h_{dec}$ is the decoding progress, and $\bm{W}_d,\bm{b}_d$ stand for the weight and bias of decoder respectively. The structure of the autoencoder is shown in the Figure \ref{Model:AE + SAE} \textbf{a}.

\begin{figure}[ht]
\centering
\begin{tabular}{cc}
\\
\textbf{a} &
\textbf{b}
\\
\begin{minipage}{0.4 \linewidth}
\includegraphics[width = \linewidth,angle=0,clip=true]{Model/ae model2.png} 
\end{minipage}
&
\begin{minipage}{0.4 \linewidth}
\includegraphics[width = \linewidth,angle=0,clip=true]{Model/stacked model.png} 
\end{minipage}
\end{tabular}
\caption{\textbf{The structure of the autoencoder.}
\textbf{a} outlines the approximate structure and functionality of a single autoencoder.
\textbf{b} shows the structure of stacked autoencoder.}
\label{Model:AE + SAE}
\end{figure}

In the structure, the aim of training is to ensure $\widetilde{\bm{u}_i}=h_{dec}\circ h_{enc}(\bm{u}_i) \approx \bm{u}_i$. For simplifying the discussion, the simplest case is considered, which owns only one hidden layer.
The error $l_2$ norm in Euclidean space has the following form
\begin{equation}
    \mathcal{L}_{AE}^{(i)} = \min\limits_{h_{dec} , h_{enc}} ||\bm{u_i}-h_{dec} \circ h_{enc}(\bm{u_i})||^2_2. 
\end{equation}
A non-linear activation function $\sigma$ is applied before the output layer to improve the capability of capturing larger non-linear information.
\begin{equation}
    q_i = \sigma(\bm{W}\bm{u_i}+\bm{b})
\end{equation}
The $q_i$ is so-called the code of the autoencoder.

In the framework of SAEs, $k$ autoencoders are stacked in the hidden layers within an unsupervised layer-wise learning algorithm, facilitated by a supervised fine-tuning methodology for parameter adjustment. The overall architecture is illustrated in Figure \ref{Model:AE + SAE} \textbf{b}. Suppose that
\begin{equation}
    h^{(i)}: \bm{u} \rightarrow h_{dec}^{(i)} \circ h_{enc}^{(i)} (\bm{u})
\end{equation}
and every output of $i$-th layer is $r^{(i)} \in \mathbb{R}^{d_{i}}$. $r^{(i-1)}$ represents the input of the $i$-th layer.
After layer-by-layer training, a SAE encoder with $k$ layers has a form of
\begin{equation}
    g_{enc}: \bm{u}  \rightarrow h_{enc}^{(k)}  \circ \dots \circ h_{enc}^{(1)} (\bm{u})
\end{equation}
where $h_{enc}^{(i)}:\mathbb{R}^{d_{i-1}} \rightarrow \mathbb{R}^{d_i}, i=1,\dots,k$ denotes the mapping relation represented by the $i$-th hidden layer. $d_i$ and $d_{i-1}$ are the output and input dimension of the $i$-th autoencoder with $d_i<d_{i-1}$ and $d_k=d$. $q$ is defined as $q:=h_{enc}(\bm{u}\in \mathbb{R}^d$.
A decoder with $k$ layers can also be depicted in a analogous manner:
\begin{equation}
    g_{dec}: \bm{u}  \rightarrow h_{dec}^{(k)}  \circ \dots \circ h_{dec}^{(1)} (\bm{u})
\end{equation}
where $h_{dec}^{(i)}:\mathbb{R}^{d_{i-1}} \rightarrow \mathbb{R}^{d_i}, i=1,\dots,k$.
Meanwhile the loss function of SAE can be expressed as
\begin{equation}
    \mathcal{L}_{SAE}=\min \frac{1}{n} \sum_{i=1}^n \mathcal{L}_{AE}^{(i)}.
\end{equation}

When $\mathcal{L}_{SAE}\rightarrow 0$, the information in the original space can be approximately represented by the latent manifold. The trained $q$ is the code of SAE and will serve as the input for SINDy approximation.

Notice that the traditional activation functions, such as hyperbolic tangent(tanh) and sigmoid activation function , often lead to a dramatic decrease in the gradients, causing training error during the propagation to forward layers. Accordingly, the rectified linear units activation function(ReLU) can avoid the problem, and resulting in computational efficiency throughout the entire process. The ReLU function has the following form:
\begin{equation}
    f_a(x) = max(0,x)
\end{equation}



\section{Proper Orthogonal Decomposition}\label{sec:pod}
In order to enhance the smoothness of low-dimensional manifolds and facilitate the searching for approximate polynomial estimations, the Proper Orthogonal Decomposition (POD) algorithm is performed to SAE before we use SINDy to search the governing equations in the manifold space. 

Consider a function $u(\bm{x};t): \Omega \times \mathbb{T} \rightarrow \mathbb{R}^l$, where $\mathbb{T}=[0,T]$ and $\Omega$ denote for time and the whole flow computational fields, respectively. The purpose of POD is to compute the reduced bases of $u(\bm{x};t)$ and project the high-dimensional dynamics into these bases, which should minimise the kneric energy $E(\cdot)$ of the flow field.
\begin{equation}
    \arg \min_{\{a_i,\phi_i\}} E(u(\bm{x};t)-\sum_{i=1}^{d} a_i(t)\phi_i(t))
\end{equation}
where $\{a_i(t) , a_i:[0,T]\rightarrow \mathbb{R}\}_{i=1}^d$ stands for POD coefficients and $\{\phi_i(t) , \phi_i: \Omega \rightarrow \mathbb{R}^l \}_{i=1}^d$. Usually $u$ is given by a series of snapshots $t_1,t_2,\dots,t_{N_t}$. To simplify the problem, we assume the time intervals are equal, and the computational field is based on a discrete point set $\bm{x}=(x_1,x_2,\dots,x_{N_s})$, meaning $\Omega := \mathbb{R}^{N_s}$. $E(\cdot)$ can be replaced by the norm $||\cdot||_{\Omega}$, which is induced by $L^2$ inner product $\langle \cdot, \cdot, \rangle$. The $u(\bm{x};t_i)$ is defined as $u_i$, $i=1,2,\dots,N_t$.

In POD, let $Y=[u_1,u_2,\dots,u_{N_t}]\in \mathbb{R}^{N_s\times N_t}$ be a real-valued matrix, with $rank(Y) = d \leq \min({N_s,N_t})$. Consequently set mode 0 of $Y$ as the average value of columns, 
\begin{equation}
    \overline{u} = \frac{1}{N_t} \sum_{j=1}^{N_t}u_j.
\end{equation}

The SVD of $Y$ can be represented as
\begin{equation}
    S=W^TYZ=
    \begin{pmatrix}
        D & 0 \\
        0 & 0
    \end{pmatrix}
    := \Sigma, 
\end{equation}
in which $D$ is a diagonal matrix and $D=diag(\sigma_1,\dots,\sigma_d)$, $\sigma_1\geq \sigma_2 \geq \dots \geq \sigma_d > 0$. $W\in R^{N_s\times N_s}$ and $Z \in R^{N_t\times N_t}$ are orthogonal matrix with vertical unit columns $\{w_i\}_{i=1}^{N_s}, \{z_i\}_{i=1}^{N_t}$ satisfying
\begin{equation}
    \begin{aligned}
         Yz_i &= \sigma_i w_i, \qquad Y^Tw_i = \sigma_i z_i \quad i = 1,\dots,d \\
         Yz_i &= 0 \quad  i = d+1,\dots,N_s ,\qquad Y^Tw_i = 0 \quad i = d+1,\dots,N_t \\
    \end{aligned}
\end{equation}\label{equ:svd}
In addition, $\{w_i\}_{i=1}^{d}, \{z_i\}_{i=1}^{d}$ can be viewed as the eigenvectors of $YY^T$ and $Y^TY$ corresponding with eigenvalues $\lambda_i = \sigma_i^2 >0$. 

The minimisation issue can be transformed into a maximization problem in the form of the corresponding inner product: for any $l\in \{1,\dots,d\}$, 
\begin{equation}\label{equ:max}
    \max_{\phi_1,\dots,\phi_l\in R^{N_s}}\sum_{i=1}^{l}\sum_{j=1}^{N_t} |\langle u_j,\phi_i\rangle_{\mathbb{R}^{N_s}}|^2 
    \quad  s.t. \quad \langle \phi_i,\phi_j \rangle_{\mathbb{R}^{N_s}} = \delta_{ij}, 
    \quad 1\leq i,j\leq l,
\end{equation}
where $\delta_{ij}$ denotes kronecker delta function.
Without prejudice to generality, we constrained $||\phi||_{\mathbb{R}^{N_s}}=1$ and introduce Largrangian functional of the issue:
\begin{equation}
    \mathcal{L}(\phi_1,\dots,\phi_l;\Lambda) = \sum_{i=1}^{l}\sum_{j=1}^{N_t} |\langle u_j,\phi_i\rangle_{\mathbb{R}^{N_s}}|^2 - \sum_{i=1}^{l} \lambda_{ij}(1-\langle \phi_i,\phi_j \rangle_{\mathbb{R}^{N_s}}),
\end{equation}
for $\Lambda=(\lambda_{ij})\in R^{l\times l}$ and $\phi_1,\dots,\phi_l\in \mathbb{R}^{N_s}$. We state that the singular vectors $\{w_i\}_{i=1}^l, l \leq d$ is one solution to the optimisation problem. To prove this, we consider the first-order necessary condition and second-order sufficient condition.
The first-order condition is given by
\begin{equation}\label{equ:1-order}
    \begin{aligned}
        0 & = \frac{\partial \mathcal{L}}{\partial \phi_k}(\phi_1,\dots,\phi_l;\Lambda)\delta\phi_k \\
        & = 2\sum_{i=1}^l \sum_{j=1}^{N_t} \langle u_j,\phi_i\rangle_{\mathbb{R}^{N_s}} \langle u_j,\phi_k \rangle_{\mathbb{R}^{N_s}} \delta_{ik}
        -\sum_{i,j=1}^l \lambda_{ij}\langle \phi_i, \delta\phi_k \rangle_{\mathbb{R}^{N_s}}
        -\sum_{i,j=1}^l \lambda_{ij}\langle \delta\phi_k, \phi_i \rangle_{\mathbb{R}^{N_s}}  \\
        & = \bigg \langle 2\sum_{j=1}^n \langle u_j,\phi_k \rangle_{\mathbb{R}^{N_s}}uj
        -\sum_{i=1}^l (\lambda_{ik}+\lambda_{ki})\phi_i,\delta\phi_k \bigg \rangle_{\mathbb{R}^{N_s}} \\
        0 & = \frac{\partial \mathcal{L}}{\partial \lambda_k}(\phi_1,\dots,\phi_l;\Lambda) = ||\phi_k||_{\mathbb{R}^{N_s}}-1,
    \end{aligned}
\end{equation}
holds $\forall \phi_k \in \mathbb{R}^{N_s}$ and $\forall k\in \{1,\dots,l\}$. 
The equation (\ref{equ:1-order}) holds if and only if $\forall k\in \{1,\dots,l\}$
\begin{equation}
    \begin{aligned}
        YY^T\phi & = \sum_{j=1}^{N_t}\langle u_j, \phi \rangle_{\mathbb{R}^{N_s}} u_i = \frac{1}{2}\sum_{i=1}^l(\lambda_{ik}+\lambda_{ki}) \phi_i    \\
        ||\phi_k||_{\mathbb{R}^{N_s}} &=1
    \end{aligned}
\end{equation}
When $l=1,k=1$, it is clear that $YY^Tw_1=\lambda_1 w_1$ holds with $\lambda_1=\lambda_{11}$. By induction in \cite{roth2021pod}, it follows from (\ref{equ:svd}) that
\begin{equation}
    YY^Tw_k=\lambda_k w_k, \quad \forall k\in \{1,\dots,l\}.
\end{equation}
This prove that $\{w_i\}_{i=1}^l$ is a solution to (\ref{equ:max}) and for $\forall i\in \{1,\dots,d\}$,
\begin{equation}\label{equ:lad}
    \arg\min\sum_{j=1}^{N_t}||u_j-\sum_{k=1}^l\langle u_j,\phi_k \rangle_{\mathbb{R}^{N_s}}\phi_k||^2_{\mathbb{R}^{N_s}}=\sum_{i=1}^l \sigma_i^2=\sum_{i=1}^l \lambda_i
\end{equation}
From Eckart-Young Theorem in \cite{higham1989matrix}, the criterion to choose $l$ can be decided heuristically, by finding $l \in \mathbb{N}^{+} $ as
\begin{equation}\label{equ:rate}
    \mathcal{E}(l):=\frac{\sum_{i=1}^{l}\sigma_i}{\sum_{i=1}^{d}\sigma_i} \leq \epsilon
\end{equation}
In practice, the definition of $\mathcal{E}(l))$ denotes the reproduced rate of POD. According to (\ref{equ:lad}), $\lambda_i$ stands for the essence of each POD basis function, meaning a reasonable tolerance $\epsilon$, like $95\%$, can promise that at least $95\%$ of kneric energy can be captured by $l$ POD basis and discard bases with smaller singular eigenvalues.

\begin{remark}
    For the dimension of POD, if we choose $l=d$, it is clear by (\ref{equ:rate}) that nearly 100\% rate of energy can be obtained by POD bases $\{w_i\}_{i=1}^l$. The only error will come from rounding-off errors during computing.
\end{remark}

In the other word, $\{z_i\}_{i=1}^{d}$ is rewrote as the eigenvectors of $Y^TY$, then the POD  basis and coefficient can be constructed as
\begin{equation}
    \begin{aligned}
        \phi_k(\bm{x}) &= w_k=\frac{1}{\sqrt{\lambda_k}}\sum_{j=1}^{N_t}(z_k)_j u_j \\
        a_i(t) &= \langle u_i,\phi_k \rangle_{\mathbb{R}^{N_s}}
    \end{aligned}
\end{equation}
where $(z_k)_j$ is the $j$-th element of $z_k$.

\section{Sparse identification of nonlinear dynamics (SINDy)}\label{sec:SINDy}

Brunton et al.\cite{2016SINDy} proposed a method for discovering equations from data by combining sparse-promoting techniques with symbolic regression, which is called sparse identification of nonlinear dynamics (SINDy). 
The state of a dynamical system at different times can be denoted as $\bm{x}(t)=[x_1(t),x_2(t),\dots,x_n(t)]\in R^n$, where $t$ represents the evolution of time and $n$ is the dimensionality of input features. $\bm{x}(t)$ can be represented in the form of the following system of differential equations:
\begin{equation}
    \dot{\bm{x}}(t) = f(\bm{x}(t)).
    \label{equ:xt=fxt}
\end{equation}
The function $f$, whose form stays unknown initially, represents the dynamic relationships within the dynamical system. The primary training target of SINDy is to discover a $f$ with sparsity and accuracy. 
The system states $\bm{x}(t)$ at $m$ time snapshots and the time derivatives $\dot{\bm{x}}(t)$ obtained through numerical differentiation are collected and arranged data matrices $\bm{X},\dot{\bm{X}}\in R^{m\times n}$:
\begin{equation}
\bm{X} = 
    \begin{pmatrix}
    \bm{x}^{T}(t_1) \\
    \bm{x}^{T}(t_2) \\
    \cdots \\
    \bm{x}^{T}(t_m)  \\
    \end{pmatrix}
=
    \begin{pmatrix}
    x_1(t_1) & x_2(t_1) & \cdots & x_n(t_1) \\
    x_1(t_2) & x_2(t_2) & \cdots & x_n(t_2) \\
    \vdots   & \vdots   & \ddots & \ddots    \\
    x_1(t_n) & x_2(t_n) & \cdots & x_n(t_n) \\
    \end{pmatrix}
\end{equation}
\begin{equation}
\dot{\bm{X}} = 
    \begin{pmatrix}
    \dot{\bm{x}}^{T}(t_1) \\
    \dot{\bm{x}}^{T}(t_2) \\
    \cdots \\
    \dot{\bm{x}}^{T}(t_m)  \\
    \end{pmatrix}
=
    \begin{pmatrix}
    \dot{x}_1(t_1) & \dot{x}_2(t_1) & \cdots & \dot{x}_n(t_1) \\
    \dot{x}_1(t_2) & \dot{x}_2(t_2) & \cdots & \dot{x}_n(t_2) \\
    \vdots   & \vdots   & \ddots & \vdots    \\
    \dot{x}_1(t_n) & \dot{x}_2(t_n) & \cdots & \dot{x}_n(t_n) \\
    \end{pmatrix}
\end{equation}

Time derivatives matrix $\dot{\bm{X}}$ is computed by second-order central difference scheme in our work. 
Without loss of generality, let's assume that the form of a linear combination of $\bm{x}$ and its nonlinear terms can approximate the governing function $f$. Subsequently, to solve the equation (\ref{equ:xt=fxt}), we construct a basis function library $\Theta(\bm{X})$, known as the SINDy library, to store the candidate function basis associated with the state $\bm{x}(t)$. The library $\Theta:=\Theta(\bm{X})$ can be wrote as
\begin{equation}
    \Theta = 
    \begin{pmatrix}
    |  &  |  &  |  &  |  &  \cdots  &  |  &  |  &  \cdots  \\
    1  & \bm{X} & \bm{X}^P_2 & \bm{X}^P_3 &  \cdots  & \sin(\bm{X}) & \cos(\bm{X}) & \cdots \\
    |  &  |  &  |  &  |  &  \cdots  &  |  &  |  &  \cdots
    \end{pmatrix}  
\end{equation}
where $\bm{X}^P_n$ stands for the n-order polynomials generated by $\bm{x}(t)$. Notice that $\Theta$ is data dependent. For example, $\bm{X}^P_2$ is
\begin{equation}
    \bm{X}^P_2 = 
    \begin{pmatrix}
    x_1^2(t_j) & x_1^2(t_j)x_2^2(t_j) & x_2^2(t_j) & \cdots & x_n^2(t_j) 
    \end{pmatrix}_{j=1,2,\cdots,m}
\end{equation}
According to SINDy library with nonlinear terms, the purpose of finding function $f$ is transformed into the process of training the sparse coefficient matrix $\Xi=[\xi_1,\xi_2,\dots,\xi_p]$ in
\begin{equation}
    \dot{\bm{X}}=\Theta(\bm{X})\Xi.
\end{equation}
Instead of using sequential thresholed least-squares (STLS) to determine the sparse coefficient matrix proposed by Brunton et al. in \cite{2016SINDy}, 
we transform this problem into a optimisation problem and utilise adaptive least absolute shrinkage and selection operator\cite{zou2006Alasso} (Adalasso), which would be discussed in Section \ref{sec:alasso}, as the loss function having the following form:
\begin{equation}\label{paralamda}
    \Xi = \arg\min ||\Theta \Xi-\dot{\bm{X}}||^2_2+\lambda\sum_{j=1}^p w_j|\xi_j|
\end{equation}
in which $w_j=(|\xi_j|)^{-\delta}$ indicates the adaptive weight in loss function, with $\delta>0$. Adalasso uses $L_1$ norm to penalise coefficient in the problem to avoid overfitting and achieve sparsity.

In the iterative training procedure, the parameter $\lambda$ is continually refined, and the coefficient matrix is repeatedly updated until the error and the parsimony of equations reach our expectation.
To alleviate the influence of features with disparate magnitudes on the computation results, we normalise the latent vectors before incorporating them into SINDy. This normalisation ensures that the features are brought to the same scale, thus enhancing the reliability and consistency of SINDy.


\section{Adalasso version 2}\label{sec:alasso}
Adaptive least absolute shrinkage and selection operator regression (Adalasso) initially was proposed by Zou in \cite{zou2006Alasso}. It is a more widely used sparse regression algorithm compared to Lasso regression\cite{tibshirani1996lasso}. It's worth noting that in the SINDy algorithm constructed by Brunton\cite{2016SINDy}, a hard thresholding algorithm was chosen for computational speed considerations.

\begin{figure}[htbp]
\centering 
\begin{minipage}{1.0 \linewidth}
\includegraphics[width = \linewidth,angle=0,clip=true]{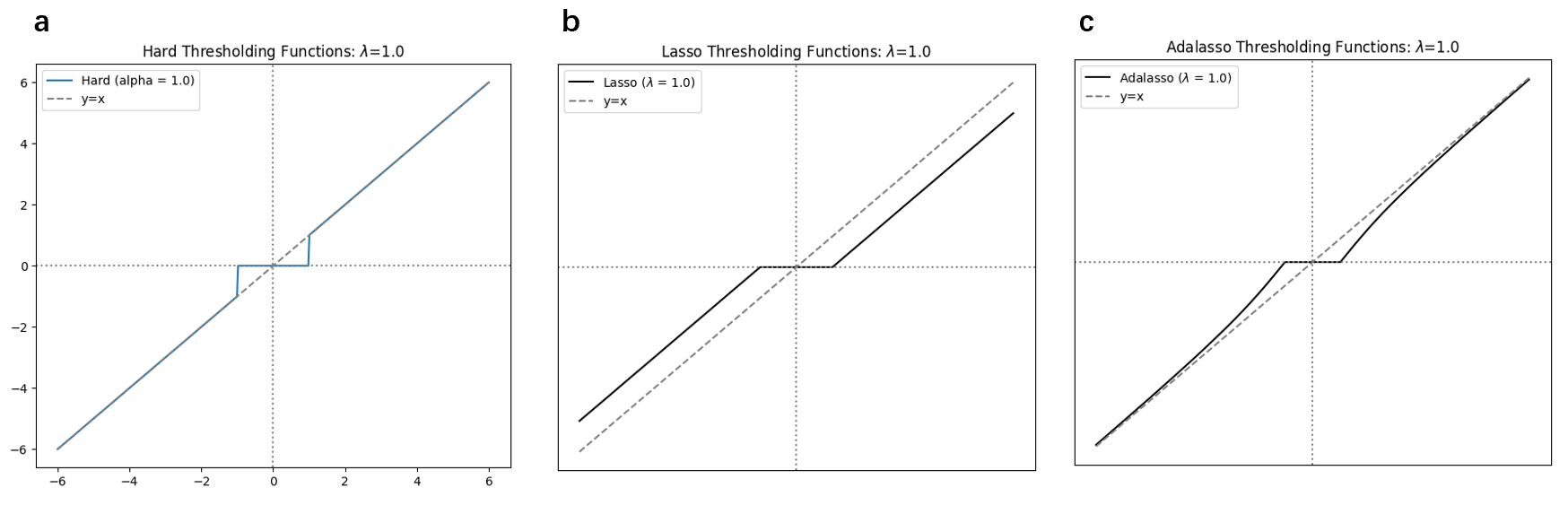} 
\end{minipage}
\caption{threshold functions compare}
\label{fig:thre-function}
\end{figure}

Hard thresholding regression and soft thresholding regression are two different sparse regression algorithms, and both aim at inducing sparsity by penalising coefficients. The key difference lies in the following: 
In hard thresholding regression, when coefficients fall below a predefined threshold, they are forcefully set to zero, thereby reducing the count of non-zero coefficients and effectively eliminating all less important features. 
In contrast, soft thresholding regression employs a continuous and smooth penalty mechanism. Unlike the direct zeroing of coefficients, it adjusts coefficients smaller than a certain threshold towards zero without entirely removing unimportant features. Consequently, it finds applicability in a broader range of scenarios but may introduce some computational overhead.
The specific comparison of their threshold functions near the origin can be checked in Figure \ref{fig:thre-function}. However, least angle regression in \cite{efron2004least} effectively reduce the computational consumption of LASSO, making it possible to be used in SINDy.

Recently, soft thresholding regression is more widely used as it balances the feature selection and information preservation. 
In addition, from Figure \ref{fig:thre-function}, it can be observed that Adalasso is asymptotically unbiased, whereas Lasso is biased and exhibits lower accuracy than Adalasso in many scenarios. 

It is clear that the form of SINDy can be simplified as a variable selection and model estimating in linear regression processes.
Suppose $\bm{y}=[y_1,\dots,y_m]^T$ displays the target vector, which is $\bm{\dot{X}}$ as well as $\frac{dr}{dt}$ in Section \ref{sec:SINDy}. For notation simplicity, we next uniformly denote this as $y$. $\bm{x_j}=(x_{1j},\dots,x_{mj})^T, j=1,\dots,p$ is the predictors, often being regarded as linearly independent, which is related to library in Section \ref{sec:SINDy} as well. The predictor matrix is defined as $\bm{X}:=[\bm{x_1},\dots,\bm{x}_p]$. Assume the relation between $\bm{x}$ and $y$ as
\begin{equation}\label{equ:ols}
    y = \sum_{j=1}^p x_j\xi_j + \epsilon \in \mathbb{R}
\end{equation}
where $\epsilon \sim \mathcal{N}(0, \sigma^2)$ and data is centered.
Set $\mathcal{A}=\{j:\xi_j \neq \}$ and presume $|\mathcal{A}|=p_0<p$ for sparsity considerations. Denote $\bm{\xi}(\delta)$ as the coefficient estimation of Equation \ref{equ:ols} by algorithm $\delta$.

Each iteration of STLS is done by following steps:
(a) Solving matrix equations directly; 
(b) setting $\xi_j$ less than the threshold $\lambda$ directly to 0; 
(c) considering it again in the next round of iterations. 
This is actually a hard-thersholding method, which is intermittent at certain points near the origin, and is manifested in the following:
\begin{equation}
    \xi^*_j(STLS) = 
    \begin{cases}
        \xi_j, \quad |\xi_j|\geq \lambda  \\
        0  ,\quad |\xi_j| < \lambda \\
    \end{cases}
    \quad j=1,\dots,p
\end{equation}
in which $\xi=(\xi_1,\dots,\xi_p)$ is obtained by solving linear matrix equation $\bm{y}=\bm{X}\bm{\xi}$ by direct inverse matrix method.  
This method was adopted for computational speed considerations and it is also unbiased. However, when $\bm{y}=\bm{X}\bm{\xi}$ is an over-determined system of equation or rank of $X$ stays low, a great error occurs. In addition, STLS is a hard thresholding algorithm and is discontinuous, resulting in a non-convex problem\cite{donoho1994ideal}. This means STLS could be affected by local minima easily when applying on a more complex fluid problems, making the final result may not be optimal.

Instead, soft shresholding algorithm such as LASSO is more popular recently \cite{tibshirani1996lasso} as it turns the problem into a continuously optimised and convex problem. 
Lasso has the following form
\begin{equation}
    \bm{\xi}^*(lasso)=\arg\min_{\bm{\xi}} || \bm{y} - \sum_{j=1}^p \bm{x_j}\xi_j ||^2 +\lambda \sum_{j=1}^p |\xi_j|,
\end{equation}
$\lambda$ is non-negative regular parameter and  not data-dependent. Every coefficient $\xi_j$ is equally penalised. As $\lambda$ increases, $\bm{\xi}$ shrinks continuously and some of $\xi_j$ decrease into exactly zero when $\lambda$ is a special value, which improves prediction accuracy and balances the sparsity and fidelity of the model. Notice that soft shresholding possesses near-minimax optimality according to \cite{donoho1995wavelet}, ensuring a wider range of adaptability and robustness. In addition, LASSO transforms the equations into a convex problem, allowing the optimal point to be obtained by iteration, which is an superiority over the STLS.

To overcome some of the issues of LASSO, we introduce an adaptive lasso (Adalasso), which can be expressed as
\begin{equation}
    \bm{\xi}^*(adalasso)=\arg\min_{\bm{\xi}} || \bm{y} - \sum_{j=1}^p \bm{x_j}\xi_j ||^2 +\lambda \sum_{j=1}^p w_j |\xi_j|,
\end{equation}
where $w_j=|\xi_j|^{\gamma}$ is known as weights vector and often data-dependent. 
Adalasso and Lasso have some of the same properties, such as sparsity, convexity, etc.
Unlike lasso, Adalasso does not penalise each parameter equally, but introduces weight values that will be constantly updated. 
 $w_j$ represents the newly added weights, which will change with the variation of SINDy coefficients. Each $w_j$ depends on the corresponding $\xi_j$. If $\xi_j$ is very small (approaching zero in absolute value), the coefficient $w_j$ corresponding to $\xi_j$ will tend toward infinity. If $\xi_j$ is relatively large, $w_j$ will converge towards a fixed finite constant. 

The key to establish Adalasso is to set the appropriate weights for each coefficient $\xi_j$. It guarantees the consistency and unbiasedness of Adalasso. We can also observe unbiasedness from the Figure \ref{fig:thre-function}, in which Adalasso realizes unbiased when the coefficients are sufficiently large. Finally, we compute the standard errors of the adaptive lasso estimates. According to \cite{tibshirani1996lasso,fan2001oracle}, the covariance of estimates of non-zero parts can be computed by local quadratic approximation. A sandwich formula is provided and proven to be consistent in \cite{zou2008one}. For non-zeros components in $\bm{\xi(adalasso)}$, the local quadratic approximation of penility is given as
\begin{equation}
    w_j |\xi^*_j| \approx w_j|\xi_{j0}| +\frac{1}{2} \frac{w_j}{|\xi_{j0}|}(\xi_j^{*2}-\xi_{j0}^2), 
\end{equation}
where $\bm{\xi_0}=(\xi_{10},\dots,\xi_{p0})$ represents the initialisation. Assume $\Sigma(\bm{xi})=diag(\frac{w_1}{|\xi^*_1|},\dots,\frac{w_p}{|\xi^*_p|})$, and $\bm{X}_p$ stands for the first $p$ columns of $\bm{X}$. Then the estimates of Adalasso can be solved as 
\begin{equation}
    (\xi^*_1,\dots,\xi^*_p)^T=(\bm{X}_p^T \bm{X}_p + \lambda\Sigma(\bm{\xi_0}))^{-1}\bm{X}_d^T\bm{y}.
\end{equation}
And the covariance matrix of nonzeros estimates is
\begin{equation}
    cov(\xi^*_{\mathcal{A}_n^*}) = \sigma^2(\bm{X}_{\mathcal{A}_n^*}^T \bm{X}_{\mathcal{A}_n^*}+\lambda\Sigma(\xi^*_{\mathcal{A}_n^*}))^{-1} \bm{X}^T_{\mathcal{A}_n^*} \bm{X}_{\mathcal{A}_n^*} (\bm{X}_{\mathcal{A}_n^*}^T \bm{X}_{\mathcal{A}_n^*}+\lambda\Sigma(\xi^*_{\mathcal{A}_n^*}))^{-1}. 
\end{equation}
For coefficient values with zero, there is no need to consider them in the covariance matrix as the standard errors remain zero.

least angle regression (LARS) is able to enhance the computational efficiency \cite{efron2004least}.
The whole process of Adalasso is shown in Algorithm \ref{alg:lars}.

\begin{algorithm}[H]\label{alg:lars}
    \SetAlgoLined
    \KwIn{$\bm{X},\bm{y}$;}
    Set a certain $\lambda$\;
    Initialise weights vector $\bm{w}$ if necessary\;
    Define $x^{**}_j=x_j / w_j$\;
    Using LARS to solve the lasso problem: $\bm{\xi}^{**}=\arg\min_{\bm{xi}^{**}}||\bm{y}-\sum_{j=1}^p\bm{x}_j^{**}\xi_j||^2+\lambda\sum_{j=1}^p|\xi_j|w_j$\;
    Update weights vector $\bm{w}$ and go into next iteration: $w_j \leftarrow (|\xi_j|)^{-\delta}$ \;
    \KwOut{$\bm{\xi}$: $\xi_j \leftarrow \xi^{**}_j / w_j$.}
\caption{Adalasso with least angle regression}
\end{algorithm}

The computational cost of the algorithm is $O(mp^2)$. The utilisation of weight values $w_j$ mitigates the issue of multiple local minima during the optimisation process, compared with STLS.  

The parameter set is defined as $\Lambda=\{\lambda_v|v\in \mathcal{F}\}$, where $\mathcal{F}$ is a finite set of indicators. The whole SINDy progress with Adalasso is shown in Algorithm \ref{alg:sindy}.

\begin{algorithm}[H]\label{alg:sindy}
    \SetAlgoLined
    \KwIn{$\bm{X},\bm{y}$,$\Lambda$,the max iteration $max\_i$;}
    \For{$\lambda \in \Lambda$}{
    Initialise weights vector: $\bm{w} \leftarrow 1$\:
    $x^{**}_j \leftarrow x_j / w_j$\;
        \While{Number of iterations$i\leq max\_i$ || $\xi_j$ not change anymore}{
        Using LARS to solve the lasso problem: $\bm{\xi}^{**}=\arg\min_{\bm{xi}^{**}}||\bm{y}-\sum_{j=1}^p\bm{x}_j^{**}\xi_j||^2+\lambda\sum_{j=1}^p|\xi_j|w_j$\;
        Update weights vector $\bm{w}$ and go into next iteration: $w_j \leftarrow (|\xi_j|)^{-\delta}$ \;
        Update $\bm{\xi}$: $\xi_j \leftarrow \xi^{**}_j / w_j$\;
        }
        Solve the system of linear equations by the Runge-Kutta method:   $\hat{\bm{y}}=\bm{\Theta}\bm{\xi}$\;
        Get SINDy's solution manifold $\hat{\bm{y}}$\;
        Assess $\hat{\bm{y}}$ with the original latent manifold $\bm{y}$ and output a score to represent the effect of the algorithm with this $\lambda$\;
    }
    Compare the scores and pick one $\lambda$ as final parameter in Adalasso\;
    Solve the lasso problem: $\bm{\xi}^{**}=\arg\min_{\bm{xi}^{**}}||\bm{y}-\sum_{j=1}^p\bm{x}_j^{**}\xi_j||^2+\lambda\sum_{j=1}^p|\xi_j|w_j$\;
    $\bm{\xi}$: $\xi_j \leftarrow \xi^{**}_j / w_j$\;
    Solve the system of linear equations: $\hat{\bm{y}}=\bm{\Theta}\bm{\xi}$\;
    \KwOut{$\hat{\bm{y}}$,$\bm{\xi}$}
\caption{SINDy with Adalasso}
\end{algorithm}

\vspace{-6pt}
 
 \bibliographystyle{unsrt} 
\bibliography{bibliography}